\documentclass{article}
\usepackage{arxiv}
\usepackage[utf8]{inputenc} 
\usepackage[T1]{fontenc}    
\usepackage[english]{babel}
\usepackage{amsmath}
\usepackage{hyperref}
\usepackage{amssymb}
\usepackage{multirow}
\usepackage{float}
\usepackage[export]{adjustbox}
\usepackage{algorithm}
\usepackage{algorithmic}

\usepackage{tikz}
\usepackage{lipsum}
\usepackage{bm}

\DeclareMathOperator*{\argmin}{argmin}

\usepackage[numbers,sort&compress]{natbib}

\begin{document}

\title{Preconditioning Natural and Second Order Gradient Descent in Quantum Optimization: A Performance Benchmark}

\author{
Théo Lisart-Liebermann \\
\texttt{theo.lisart@itwm.fraunhofer.de} \\
\and
Arcesio Castañeda Medina \\
\texttt{arcesio.castaneda.medina@itwm.fraunhofer.de} \\
\and
Fraunhofer ITWM, 67655 Kaiserslautern, Germany
}

\onecolumn
\maketitle

\begin{abstract}
  The optimization of parametric quantum circuits is technically hindered by three major obstacles: the non-convex nature of the
  objective function, noisy gradient evaluations, and the presence of
  barren plateaus. As a result, the selection of classical optimizer
  becomes a critical factor in assessing and exploiting quantum-classical applications. One promising approach to tackle
  these challenges involves incorporating curvature information into
  the parameter update. The most prominent methods in this field
  are quasi-Newton and quantum natural gradient methods, which can
  facilitate faster convergence compared to first-order
  approaches. Second order methods however exhibit a
  significant trade-off between computational cost and accuracy, as
  well as heightened sensitivity to noise. This study evaluates the
  performance of three families of optimizers on synthetically
  generated MaxCut problems on a shallow QAOA algorithm. To address noise sensitivity and iteration cost, we demonstrate that incorporating
  secant-penalization in the BFGS update rule (SP-BFGS) yields
  improved outcomes for QAOA optimization problems, introducing a
  novel approach to stabilizing BFGS updates against gradient noise.
\end{abstract}

\twocolumn

\section{Introduction}
Variational Quantum Algorithms (VQAs) are considered the main
candidate for near-term quantum advantage~\cite{farhi2019quantum,
  Rudolph_2023, Benedetti_2019, yu2023provable, Cerezo_2021}. These
algorithms employ an iterative process that leverages the strengths of
both quantum and classical computing. The approach involves evaluating
an objective function using a parameterized quantum circuit on quantum
hardware, and then utilizing a classical optimization routine to
refine the results. This hybrid strategy is particularly valuable when
the objective function can be efficiently computed on quantum
hardware, potentially offering significant performance advantages across various
applications: quantum chemistry~\cite{McArdle_2020, Guo_2024,
  Liu_2022, Peruzzo_2014}, combinatorial
problems~\cite{Pirnay2022AnIS, farhi2014quantum,
  farhi2000quantumcomputationadiabaticevolution,
  abbas2023quantumoptimizationpotentialchallenges},
simulation~\cite{feyn_1982, Fauseweh_2024, Georgescu_2014},
control~\cite{Choquette_2021, Egger_2014, magann_2021}, quantum
machine learning (QML)~\cite{Zoufal_2019, Abbas_2021}, numerical
methods subroutines~\cite{Bravo_Prieto_2023}.

The performance of these applications is heavily influenced by the
classical routine's ability to identify a satisfactory approximation
of the global optimum. This capability depends on the selection of an
optimal optimization algorithm and the underlying optimization
landscape associated with the objective function, which can be
compromised by various factors. In particular, the so-called barren
plateaus (BP) lead to entire regions of vanishing gradients as the
result of sources such as noise~\cite{Wang_2021}, type of
cost-function~\cite{Cerezo_2021_shallow}, and ansatz design and
presence of entanglement~\cite{holmes_2023, McClean_2018}. Although
recent research has shown that the origin of BP from the ansatz
structure can be avoided for a large class of
problems~\cite{letcher2024tightefficientgradientbounds,mele2024noiseinduced,
  sharma_22, Pesah_2021}, poor gradient quality remains a practical
concern in many cases~\cite{Larocca_2022}.

Optimization of VQA models in realistic settings is hampered by the
inherent non-convexity of their objective functions~\cite{Cerezo_2021,
  Huembeli_2021, Shaydulin_2019} and noisy gradient
estimates~\cite{Preskill_2018}. This dual challenge undermines
theoretical guarantees of convergence for traditional optimization
methods~\cite{chouzenoux2023kurdykalojasiewiczpropertystochasticoptimization,
  Jain_2017, de2018convergence}, and the classical
optimizer must be tailored to the specific problem.

Success in optimizing parametric quantum circuits relies on tradeoffs
between minimizing quantum hadware calls, number of iterations and
performance in noisy function calls. Previous efforts focused on
benchmarking gradient-less (zero-order order) and first order
methods. Thus, for example, authors
in~\cite{gidi2023stochasticoptimizationalgorithmsquantum} approach
noisy objective function evaluation using stochastic methods, in
particular the Simultaneous Perturbation Stochastic Approximation
(SPSA)~\cite{jspall_spsa87} and second order
extention~\cite{SPALL1997109}. On the other hand, in
~\cite{lockwood2022empiricalreviewoptimizationtechniques} an extensive
study of \textit{SciPy}~\cite{Virtanen_2020} implementations of
popular optimizers is provided, covering stochastic gradient methods,
zero-order and first order methods. This last work extends
on~\cite{Lavrijsen_2020}, which focused on establishing best practice
for gradient-free methods in controlled noisy emulators. The latter
argues that specifically tuned optimizers are crutial and will be
needed even in the presence of usable error correction schemes,
justifying the need for improvements using hyper-parameter fine tuning
using more advanced surrogate models~\cite{Sung_2020,
  Shaffer_2023}. Within those studies, particular attention was
brought to the definition of relevant cost models designed to balance
empirical efficiency (or direct runtime cost) and solution accuracies
relevant for the specific needs of quantum computing.

A rich literature of methods from classical optimization has emerged in the context of VQA-based quantum
optimization~\cite{leng2019robustefficientalgorithmshighdimensional,
  yao2020policygradientbasedquantum, Nannicini_2019,
  romero2018strategiesquantumcomputingmolecular}. For reference, we
can list the Nelder-Mead simplex algorithm~\cite{Nelder1965ASM},
SPSA~\cite{spall_spsa92}, collective and swarm optimization
(ant-based, particle-swarm), Bayesian optimization, gradient-based
reinforcement learning, Powell's method, Conjugate Gradient,
Constrained Optimization by Linear Approximation (COPYLA), Sequential
Least Squares Programming (SLSQP), Broyden–Fletcher–Goldfarb–Shanno
(BFGS), and Byrd-Omojokun Trust Region Sequential Quadratic
Programming (trust-constr)~\cite{NoceWrig06}. Another family of
promising methods comes from the field of machine learning, which has
seen significant success over the last two decades. Namely gradient-based methods such as Stochastic Gradient
Descent (SGD), Adaptive Gradient Algorithm
(AdaGrad)~\cite{durchiadagrad}, Root Mean Square Propagation
(RMSProp)~\cite{graves2014generatingsequencesrecurrentneural}, Follow
the Regularized Leader (FTRL)~\cite{adclickprediction_mcmahan}, Adam,
Adamax, Nadam, and AMSGrad~\cite{reddi2019convergenceadam,Dozat2016IncorporatingNM,
kingma2017adammethodstochasticoptimization}. Machine-learning centric methods have been ranked for
quantum optimization problems in~\cite{lockwood2022empiricalreviewoptimizationtechniques}.

Systematic testing of cutting-edge quasi-Newton methods has not been
as thoroughly examined as zero-order and first-order methods. Benchmarking
different variants of these methods, using distinct Hessian matrix
approximations and alternative shot noise mitigation strategies, can
provide better alternatives for VQAs. Thus, for example, although the
Davidon-Fletcher-Powell (DFP) method~\cite{Fletcher1963ARC}, which
predates BFGS and its limited memory
variant~\cite{Shanno1970ConditioningOQ}, has been shown to be less
accurate and stable than its successor, fine-tuned DFP could
potentially lead to faster convergence due to its more aggressive
Hessian approximation. Other methods, like the The Symmetric Rank One
(SR1) method~\cite{Conn1991ConvergenceOQ}, offer a flexible Hessian
matrix approximation, particularly in situations where well-posedness
cannot be guaranteed. Another overlooked approach is the quasi-Newton
update based on the Conjugate Gradient (NCG) Hessian
approximation~\cite{SHERALI1990361}, which inherits the advantages of
CG for quasi-Newton updates while offering improved scaling and memory
usage, although tends to perform poorly in non-convex
geometries. An alternative strategy for leveraging
second-order information at key points in parameter space involves
applying a contemporary approach that incorporates a gradient quality
metric when updating the BFGS algorithm
(SP-BFGS)~\cite{Irwin_2023}. By penalizing the update, this method has
a stabilizing effect on the Hessian approximation, enhancing overall
robustness and dynamical properties. The technique seamlessly
interpolates between first- and second-order methods, thereby
preventing accumulation of errors in the Hessian approximation, and
generalizes BFGS updates to noisy function evaluations.

Second order methods scales the step size of the Newton update
according to the Hessian matrix information. It is known however that
local gradients evaluations are not the best possible update
direction~\cite{amari1993, amari}. Natural Gradient Descent (NGD) use the
Fisher information matrix (FIM) to precondition the update step, as
the natural direction of greatest change. Under this conceptual
framework, the Quantum Natural Gradient (QNG) updates the parameters in the
direction of biggest change according to Quantum Information
Geometry~\cite{Stokes2020} through a block approximation of the
Fubini-Study metric tensor, also known as the Quantum Fisher
Information Matrix (QFIM). Contrary to the classical natural gradient
method, the QFIM is not explicitly linked to the Hessian of the
objective function, exploring the geometry of the parameter space in
probability space. This approach can be considered as a
pre-conditioned gradient update.

Building on the principles of Quantum Natural Gradient (QNG), various
methods have been proposed to enhance the efficiency of Quantum Fisher
Information Matrix (QFIM) evaluations. Notably, the introduction of
Quantum-Natural Stochastic Perturbation Synthesis Algorithm (QNSPSA)
in~\cite{Gacon_2021} represents an initial step in this
direction. More recently, momentum-based and Broyden-type methods have
been developed to offer gradient-based QFIM approximations,
exemplified by the qBroyden and qBang
algorithms~\cite{Fitzek_2024}. The
inclusion of stochastic methods such as SPSA, 2SPSA, and QNSPSA
provides valuable insight into the impact of Hessian and QFIM
approximations on optimizing quantum circuits, with experimental
relevance established ~\cite{Escudero_2023, Kandala_2017, D_ez_Valle_2021}. Furthermore, other the Random Coordinate Descent (RCD) algorithm
~\cite{Necoara2013ARC, ding2023random}, have demonstrated
improved performance for PQCs optimization, which we verify. We use the stochasic methods
as a reference point for comparing methods.

The three main categories of methods - quasi-Newton, natural gradient,
and stochastic techniques - serve as
the foundation for our examination of second-order methods in quantum
optimization, which are summarized in Tab.~\ref{tab:table_available_methods}.

\begin{table}[H]
    \centering
    \begin{tabular}{ |c|c| }
    \hline
    Family & Method \\
    \hline
    \multirow{3}{4em}{Second order quasi-Newton} & NCG \\
    &BFGS\\
    &DFP\\
    &SR1\\
    &SP-BFGS\\
    \hline
    \multirow{3}{4em}{Quantum Natural Gradient} & QNG (block-diag) \\
    &qBroyden\\
    &qBang\\
    &m-QNG\\
    \hline
    \multirow{3}{4em}{Stochastic} & SPSA \\
    &2SPSA \\
    &QNSPSA\\
    &RCD\\
    \hline
    \end{tabular}
    \caption{Scope of studied methods within this work}
    \label{tab:table_available_methods}
\end{table}

Most of these methods, excluding pure
QNG, demand meticulous tuning of hyperparameters to achieve
optimal results. The literature on hyper-parameter optimization is
extensive, with numerous techniques available, including genetic
algorithms~\cite{YOO201975},
bandit-based approaches~\cite{li2018hyperbandnovelbanditbasedapproach}, random
search~\cite{liashchynskyi2019gridsearchrandomsearch}, and
autoML~\cite{NIPS2015}. One method that stands out for its ability to
handle low-dimensional spaces is Bayesian
optimization~\cite{snoek2012practicalbayesianoptimizationmachine,
  7352306}, which proves particularly effective when each model
evaluation is computationally intensive. In this study, the
hyper-parameter space is explored through successive iterations of
Bayesian optimization, resulting in a comprehensive coverage of both
randomly sampled and optimal scenarios for each hyperparameters
optimization run.

The primary concern for evaluating combinatorial solvers is not their
inherent ability to find optimal solutions, but rather their capacity
to identify solutions that are satisfactory for specific applications
or domains. The ultimate goal is to determine how efficiently and
reliably a solver can locate an acceptable solution. Consequently, the
performance of these optimizers is typically measured against
quality-to-cost metrics. Within this study, our focus lies on
understanding solver dynamics and solution quality, which can be
quantified through various indicators. Specifically, we consider the
pairwise Hamming distance associated with bitstring
solutions~\cite{hamming_dist}, as well as overall convergence ratios
that estimate the success probability of the methods
employed. Additionally, standard metrics such as objective function
values, quantum calls, overall walltime, and average iteration to
convergence are also taken into account.

For the second-order quasi-Newton family, our results suggest that the
choice of Hessian approximation is crucial for quasi-Newton methods'
ability to achieve acceptable solutions. Specifically, DFP methods
offer higher convergence probabilities and solution quality at the
cost of numerical instability. In contrast, more dynamic rank-1
methods like SR1 do not provide significant improvements. However,
SP-BFGS offers reasonable convergence ratios while also stabilizing
BFGS and DFP Broyden updates, although this comes with the added
complexity of two additional hyperparameters to fine-tune.

For quantum natural gradient (QNG) methods, in general, principal appeal is
the absence of hyper-parameter tuning in general, this is however not the case for its approximate variants,
which often require challenging hyper-parameter tuning. However, the addition of momentum in
QNG does improve convergence speed drastically, but reduces significantly convergence rates particularly at higher problem sizes. Notably, QNG-type methods exhibit better overall
convergence consistency compared to quasi-Newton methods, suggesting
that cheaper QFIM estimation remains a worthwhile area of
investigation.

Stochastic methods overall require parameter tuning for an appreciable performance, overall we find that
QNSPSA helps at low dimensions convergence rates. 2SPSA however requires intensive tuning to scale, which is not necessarily possible. At convergence, the introduction of second order estimators do drastically increases convergence speed, but at the cost of drastic increase in variability in performances.

This paper is organized as follows: In
Sec.~\ref{sec:theoretical_framework}, we provide a detailed
introduction to the theoretical framework underlying the studied
methods, including an overview of the standard MaxCut problem and its
Quantum Approximate Optimization Algorithm (QAOA) implementation. We
then devote Sec.~\ref{sec:methods} to an in-depth examination of
the methods listed in Tab~\ref{tab:table_available_methods},
followed by a presentation of Bayesian hyper-parameter fine tuning
techniques in Sec.~\ref{sec:bayes}. The performance metrics and
context for hyperparameters fine tuning are specified in
Sec.~\ref{sec:performance_benchmarks}. Finally, the obtained
results are discussed in full detail in
Sec.~\ref{sec:results_discussion}, while our key observations and
suggestions for future work on applying second-order methods for
quantum circuit optimization are presented in
Sec.~\ref{sec:Conclusion}.

\section{Theoretical framework}
\label{sec:theoretical_framework}
The quantum-classical optimization loop, appearing in quantum
algorithms such as QAOA, Variational Quantum Eigensolvers
(VQE)~\cite{Peruzzo_2014}, Harrow–Hassidim–Lloyd
(HHL)~\cite{Harrow_2009}, sensing~\cite{Degen_2017, Faist_2023}, and
quantum machine learning (QML)~\cite{wang2024comprehensive}, rely on
the definition of quantum ansatzs, whose tunable parameters are
defined on a parameterized quantum circuit~\cite{Benedetti_2019,
  Peruzzo_2014, Ostaszewski_2021}. In this section, we introduce the
MaxCut problem and the associated QAOA circuit forming the basis of
the numerical investigation, we then go into details of the second
order quasi-Newton methods and quantum natural gradient methods.

\subsection{MaxCut and the QAOA algorithm}
\label{secsec:MaxCut_qaoa}

Given a weighted graph \( G = (V, E, w) \), where \( V \) represents
the set of vertices, \( E \) denotes the set of edges, and \( w \) is
a weight function mapping each edge to a positive real number, the
MaxCut problem seeks to partition the vertex set \( V \) into two
non-overlapping subsets \( S \) and its complement
\( \bar{S} = V \setminus S \), with the goal of maximizing the total
weight of edges between these two subsets. By defining binary
variables \( z_i \in \{0, 1\} \) for each vertex \( i \in V \), the
problem's objective function can be formally expressed as:

\begin{equation}
\label{eq:classical_cost_function}
\text{MaxCut}(G) = \max_{\mathbf{z} \in \{0, 1\}^n} \sum_{(i,j) \in E} w_{ij} \cdot \frac{1}{2} (1 - z_i z_j)\,,
\end{equation}

where \( n = |V| \) is the number of vertices in the graph. To solve
this problem using a quantum-classical approach, the cost function
\eqref{eq:classical_cost_function} is re-expressed as a minimization
problem and the Quantum Approximate Optimization Algorithm (QAOA)
method is employed as the solver. QAOA represents an $n$-qubit Ansatz
Trotterization of the adiabatic evolution as described in
\cite{farhi2000quantumcomputationadiabaticevolution}, with its
precision controlled by a positive parameter $p$, which determines the
number of mixer-problem unitary applications. To establish the
problem's unitary operator, we directly translate the cost function
into Pauli-Z operators acting on each qubit:

\begin{equation}
  \label{eq:quantum_cost_function}
  \hat{P} = \sum_{(i,j) \in E} w_{ij} \cdot \frac{1}{2} (\hat{Z}_i \hat{Z}_j - I)\,.
\end{equation}

The Quantum Approximate Optimization Algorithm (QAOA) Ansatz is
defined as:

\begin{equation}
  \label{eq:quantum_state}
  |\bm{\gamma}, \bm{\beta}\rangle = U_B(\beta_p)U_P(\gamma_p) ... U_B(\beta_1)U_P(\gamma_1)|+\rangle^{\otimes n}\,,
\end{equation}

where the problem and mixer generators are given by
$U_P(\bm{\gamma}) = e^{-i\bm{\gamma}P}$ and
$U_B(\bm{\beta}) = e^{-i\bm{\beta}B}$, with the mixer Hamiltonian and
$B= \sum_{i=1}^n X_i$, respectively. This establishes a parameter
space that spans $2p$ real numbers,
$\bm{\gamma} = (\gamma_1,..., \gamma_p)$ and
$\bm{\beta} = (\beta_1,..., \beta_p)$.

With $p$ fixed, the QAOA algorithm estimates the solution by employing
a classical optimizer to identify the optimal set of parameters
$\bm{\theta} \equiv \{\gamma, \beta\}$ that minimizes:

\begin{equation}
  \label{eq:unconstrained_problem}
  f(\bm{\theta}) = \min_{\bm{\theta}} \langle \bm{\theta}| P(z) | \bm{\theta} \rangle\,,
\end{equation}

utilizing the Pauli-strings expectation values computed on a quantum
processor or emulator.

\subsection{Methods}
\label{sec:methods}
For functions with non-exponential decay or complex landscapes
featuring multiple critical points, first-order methods often struggle
to converge effectively. In contrast, second-order methods utilize the
Hessian matrix to scale the step size according to the landscape's
geometric characteristics and adjust the direction of the gradient
based on curvature information. This allows for improved convergence
in flat areas or heavily bumpy surfaces with numerous critical points.

Starting from the Taylor expansion
$f(\theta_i)\approx f(\theta_i) + (\theta_{i+1} -
\theta_i)^T\nabla_\theta f(\theta_i) + \frac{1}{2}(\theta_{i + 1} -
\theta_i)^T H(\theta_{i + 1} - \theta_i)$, the second-order Newton
update rule with step size $\alpha$ can be derived as follows:

\begin{equation}
  \label{eq:second_order_newton}
  \theta_{i+1} = \theta_i - \alpha H^{-1}\nabla_\theta f(\theta_i)\,,
\end{equation}

By solving for the critical point using the inverse Hessian $H^{-1}$,
the optimization step can be performed through a standard line search
along the search direction $p_i=-H^{-1}_{i}\nabla_\theta
f(\theta_i)$. Further improvements can be achieved by incorporating
problem-adapted linear updates or more efficient search methods, such
as Trust Region
techniques~\cite{fridovichkeil2022approximately,tankaria2021regularized}. However,
if the function is not strongly convex, enforcing the Wolfe condition
becomes essential to ensure stability of the line search
procedure. The sub-optimization fulfills this condition when the Amijo
rule and the curvature
condition are met:

\begin{align}
  \label{eq:amijo}
  f(\theta_i - \alpha_ip_i) &\leq f(\theta_i) + c_1\alpha_ip_i^T\nabla_\theta f(\nabla_i) \\
  -p_i^T\nabla_\theta f(\theta_i + \alpha_ip_i) &\leq -c_2p_i^T\nabla_\theta f(\theta_i)
                                                  \label{eq_curvature_cond}
\end{align}

These conditions provide heuristical upper and lower bounds to the
optimal choice of step size in the line search. Typical values for
hyperparameters $c_1$ and $c_2$ are $10^{-4}$ and $0.9$, respectively,
although they should ideally be tuned empirically.

Moreover, the positiveness of the Hessian imposes additional
constraints on the direction of search, such as the curvature
condition $p_i^Ty_i > 0$, where
$y_i = \nabla_{\theta}f(\theta_{i+1}) - \nabla_\theta f(\theta_i)$. In
this study, we employ a simple line search and backtracking with a
maximum number of iterations. The conditions~\eqref{eq:amijo}
and~\eqref{eq_curvature_cond} are iteratively checked to reduce the
step size $\alpha$ by $\beta$ for a given $(c_1, c_2)$ pair until
either the maximum backtracking iteration is reached or the conditions
are satisfied.

In practice, direct use of Newton methods can be computationally
expensive due to the need to compute the full Hessian at each
iteration and handle instabilities with aggressive
regularizations. For large matrices, explicit inversion of $H$ can
also be challenging if the matrix is dense~\cite{Tan_2019}. In the
context of VQAs, gradient and higher-order derivative estimations can
be achieved using parameter-shift routines~\cite{maria19, Mari_2021,
  Wierichs_2022}, introducing auto-differentiation approaches to
gradient estimation in parametric quantum circuits.

\subsection{Quasi-newton second order methods}
\label{sec:appendix_secondorder_approx}
Quasi-Newton methods address the limitations of Newton's method by
introducing inverse Hessian approximations, $B_i = \tilde{H}^{-1}_i$,
that are based on the history of gradient evaluations. This results in
more stable and efficient algorithms, with reduced computational costs
from $\mathcal{O}(n^3)$ to $\mathcal{O}(n^2)$ for an $n \times n$
Hessian matrix, irrespective of the specific Hessian approximation
employed~\cite{tankaria2021regularized, dennis_quasi_newton}.

These methods have been extensively studied in various fields and are
represented by the well-known (L-)BFGS
method~\cite{berahas2019robustmultibatchlbfgsmethod}, which has been
applied to diverse quantum optimization
problems~\cite{lockwood2022empiricalreviewoptimizationtechniques,
  Lavrijsen_2020, sohldickstein2014fastlargescaleoptimizationunifying,
  NoceWrig06}. Other quasi-Newton methods include the
Davidon-Fletcher-Powell (DFP) method, which is less robust but more
efficient per iteration; the Symmetric Rank One (SR1)
method~\cite{Conn1991ConvergenceOQ}, which improves upon the dynamical
scaling of Hessian approximations; and the Quasi-Newton Conjugate
Gradient (NCG)~\cite{SHERALI1990359}, which employs the conjugate
gradient method for the Hessian approximation. Broyden's method
families~\cite{Broyden1965ACO} are a generalization of these methods
and will be discussed in Sec.~\ref{secsec:qng} in the context of
quantum natural gradient methods.

Quasi-Newton methods typically begin with an initial guess, often the
identity matrix, and iteratively build upon it to develop a Hessian
approximation. The accuracy of these approximations relies heavily on
the quality of the Hessian update rule. Below, we provide a brief
overview of some standard quasi-Newton approximations. For more
detailed descriptions, please refer to the references~\cite{berahas2019robustmultibatchlbfgsmethod, NoceWrig06,
  Shanno1970ConditioningOQ}.

\subsubsection{Davidon-Fletcher-Powell (DFP)}
Introduced by Davidon in 1959, the DFP update rule is a seminal
contribution to quasi-Newton methods. The inverse Hessian
approximation, where the update direction is given as
$s_{i}=\alpha p_{i}$, is derived from the Sherman–Morrison–Woodbury
formula~\cite{NoceWrig06}:

\begin{equation}
B_{i+1} = B_i + \frac{s_i s_i^T}{s_i^T y_i} - \frac{B_i y_i y_i^T B_i}{y_i^T B_i y_i}\,,
\end{equation}

Although the DFP method was largely superseded by subsequent methods,
it can outperform the BFGS method in specific cases of objective
function non-linearities. This is particularly evident in situations
involving noisy gradient evaluations, as the DFP method exhibits
reduced sensitivity to initial conditions and improved recovery from
poor Hessian estimation~\cite{NoceWrig06}.

\subsubsection{Broyden-Fletcher-Goldfarb-Shanno (BFGS)}
Developed in an effort to find a better inverse Hessian approximation from the
DFP method, the BFGS method has been empirically shown to be the best
known quasi-Newton update formula~\cite{NoceWrig06}. The rank-2 update
method takes the form:

\begin{equation}
  \label{eq:bfgs_update}
  B_{i+1} = B_i + \left(1 + \frac{y_i^T B_i y_i}{s_i^T y_i}\right) \frac{s_i s_i^T}{s_i^T y_i} - \frac{s_i y_i^T B_i + B_i y_i s_i^T}{s_i^T y_i}\,.
\end{equation}

The BFGS update is derived similarly to the DFP method, introducing a
quadratic model of the Hessian. For a displacement $s_i$ and variation
in gradients $y_i$, the secant condition requires that the new
approximation maps $s_i$ and $y_i$ together: $B_{i+1}s_i=y_{i}$, which
is known as the secant condition.

\subsubsection{Symmetric Rank One (SR1)}
In SR1 method the inverse Hessian is given by the rank-1 symmetric update:

\begin{equation}
B_{i+1} = B_i + \frac{(s_i - B_i y_i)(s_i - B_i y_i)^T}{(s_i - B_i y_i)^T y_i}\,,
\end{equation}

which maintains matrix symmetry and satisfies
the secant condition. However, it does not guarantee positive
definiteness, implying that a line-search or trust region method
should be employed in conjunction with this update to ensure
stability~\cite{NoceWrig06}.

\subsubsection{Quasi-Newton Conjugate Gradient (NCG)}
The Conjugate Gradient (CG) algorithm occupies a middle ground between
steepest descent methods and quasi-Newton approaches~\cite{SHERALI1990359}. A significant
advantage of CG methods is their ability to operate without explicit
storage of matrices, thus enhancing both memory efficiency and
computational speed.

Mathematically, the update rules for conjugate gradient methods are
defined by:

\begin{equation}
    x_{i+1} = x_i +\alpha p_i\,,
\end{equation}

where the direction of the line search is given by
$p_i = -\nabla_\theta f(\theta_i) + \beta_ip_{i-1}$.

Variations among different CG methods arise from the selection of the
multiplier $\beta_i$ at each iteration, which forms the core of the
line-search algorithm. The quasi-Newton Conjugate Gradient (NCG)
method builds upon standard CG approaches by incorporating an inexact
line search and leveraging the quasi-Newton condition instead of the
conjugacy condition in the update rule. Furthermore, NCG methods
utilize second-order approximations to derive both step sizes and
search directions. In the context of quantum optimization, NCG
approaches permit the relaxation of the quadratic condition through
the use of reset rules.

In an inexact line-search scenario, the Quasi-Newton condition can be
leveraged to derive a $\beta_i$ with $\mathbf{p}_i = -\tilde{H}^{-1}_i {g}_i$, where
$g_i=\nabla_\theta f(\theta_i)$.

Scaled quasi-Newton update with inexact line search~\cite{SHERALI1990361}:
\begin{equation}
    \beta_i = \frac{y_i^Tg_i - (1/s_i)p_i^Tg_i}{y_i^Td_{k-1}}\,,
\end{equation}

Similarely to SP-BFGS method, local evaluation of gradient quality interpolates continuously between the most relevant method depending on second order metrics, in the case of CG updates, Perry's strategy as the factor $(s_i\to 1)$ and coincides with the Hestenes and Stiefel's strategy when $(s_i\to \infty)$. Various quasi-Newton approximations can be used in evaluating $\beta_i$, standards ones can be found in the original paper~\cite{SHERALI1990359}

\subsubsection{The Secant-Penalized BFGS (SP-BFGS)}
This approach endeavors to mitigate the influence of local dependence
on the accuracy of gradient estimation by introducing an adaptive
parameter $\beta_{i}$, derived from the curvature condition
($-\beta_{i}^{-1}<s_i^Ty_i$), which smoothly transitions between a
full BFGS step (as $\beta_i\to\infty$) and a standard gradient descent
update (as $\beta_i\to 0$)~\cite{Irwin_2023,Freund2004PenaltyAB}.

The update rule for the SP-BFSG method is derived from the integration
of penalty terms into the BFGS update rule~\cite{Irwin_2023} and can
be expressed as follows:

\begin{align}
  B_{i+1} =&\Big(I - \omega_is_iy_i^T\Big)B_i\Big(I - \omega_iy_is_i^T\Big) + \\&\omega_i\Big[\frac{\gamma_i}{\omega_i} + (\lambda_i - \omega_i)y_i^TB_iy_i\Big]s_is_i^T\,,
\end{align}

where the parameters $\gamma_i$ and $\omega_i$ are defined by:

\begin{equation}
  \gamma_i = \frac{1}{s_i^Ty_i + \frac{1}{\beta_i}}, \quad \omega_i = \frac{1}{(s_i^Ty_i + \frac{2}{\beta_i})}\,.
\end{equation}.

\begin{algorithm}[t]
  \begin{algorithmic}[1]
      \STATE \textbf{Input:} $f$, $\beta\_compute\_linear$, $x_\text{init}$, $\alpha$, $\beta_{reduce}$, $c_0$, $c_1$, $N_0$, $N_s$, MAXIT
      \STATE \textbf{Output:} $x$, best\_value

      \STATE $x \gets x_{\text{init}}$
      \STATE $i \gets 0$
      \STATE $H_i \gets \text{Identity Matrix}$

      \WHILE{$i < \text{MAXIT}$}
          \STATE $\nabla_i \gets \nabla_\theta f(x)$
          \IF{$\lVert \nabla_i \rVert < \text{tol}$}
              \STATE \textbf{break}
          \ENDIF

          \STATE $p_i \gets -H_i \cdot \nabla_i$

          \WHILE{not satisfying Armijo-Wolfe condition}
              \IF{$f(x + \alpha p_i) \leq f(x) + c_0 \alpha \cdot \nabla_i^T p_i$}
                  \STATE \textbf{break}
              \ENDIF
              \IF{$-p_i^T \nabla_\theta f(x + \alpha p_i) \leq -c_1 p_i^T \nabla_i$}
                  \STATE \textbf{break}
              \ENDIF
              \STATE $\alpha \gets \beta_{reduce} \cdot \alpha$
          \ENDWHILE

          \STATE $x_i \gets x + \alpha p_i$
          \STATE $s_i \gets x_i - x$
          \STATE $\nabla_{i+} \gets \nabla f(x_i)$
          \STATE $y_i \gets \nabla_{i+} - \nabla_i$

          \STATE $\beta_i \gets \beta\_compute\_linear(N_0, N_s, s_i)$
          \STATE proj $\gets s_i^T \cdot y_i$
          \STATE $\gamma_i \gets \frac{1}{proj + \frac{1}{\beta_i}}, \quad \omega_i \gets \frac{1}{proj + \frac{2}{\beta_i}}$
          \STATE $H_{ns} \gets (I - \omega_i (s_i \odot y_i)) \cdot H_i (I - \omega_i (y_i \odot s_i))$
          \STATE $H_{nc} \gets \omega_i \cdot \left(\frac{\gamma_i}{\omega_i} + (\gamma_i - \omega_i) \cdot y_i^T \cdot H_i y_i\right) \cdot s_i s_i^T$
          \STATE $H_n \gets H_{ns} + H_{nc}$

          \STATE $x \gets x_i$
          \STATE $H_i \gets H_n$
          \STATE $i \gets i + 1$
      \ENDWHILE

      \RETURN {$x$}
  \end{algorithmic}
  \caption{SP-BFGS implementation, with the Amijo-Wolf backtracking. \textit{compute\_linear} function is the expression in adapting $\beta_i$ with interception or noise model in Eq.~\ref{eq:mitigation_interception}}
  \label{alg:sp_bfgs_implementation}
\end{algorithm}

Adaptation of the parameter $\beta_i$ relies on assumptions regarding
the noise distribution that influences the gradient measurements. The
original work proposed a rectified version of the Uniform Gradient
Noise Bound, which schedules the gradient measure linearly with
respect to the continuous update parameter:
($\beta_i \propto ||s_i||_2$)~\cite{Irwin_2023}. To enhance update
stability, an interception rule can be implemented when gradient
information is insufficient for a proper update.

\begin{equation}
  \label{eq:mitigation_interception}
  \beta_i = \max\Big\{N_s||s_i||_2 - N_0, 0\Big\}
\end{equation}

We implement in~Alg.\ref{alg:sp_bfgs_implementation} the SP-BFGS procedure.

The phenomenological parameters $N_0$ and $N_s$ improve the stationary
point dynamics by introducing a minimum gradient value beyond which
noise dominates and second-order information is destroyed. They also
scale the effects of noise. In this work, both hyperparameters are
learned using Bayesian optimization.

\subsection{Quantum natural gradient methods}
\label{secsec:qng}

Natural Gradient (NG) methods have been developed from the observation
that the Euclidean metric is not always the most optimal description
for a large class of objective functions. The exact form of the Newton
update rule, as given in Eq.~\eqref{eq:exact_natural_gradient},
implies an arbitrary choice of the $l_2$ norm:

\begin{equation}
  \label{eq:exact_natural_gradient}
  \theta_{i+1} = \argmin_{\theta\in \mathbb{R}^n}\Big[\langle\theta - \theta_i, \nabla_\theta f(\theta_i)\rangle + \frac{1}{2\alpha_i} ||\theta - \theta_i||^2_2\Big]\,.
\end{equation}

Then, by introducing a metric change,
$||\cdot||_2 \to ||\cdot||_{g_{\theta}}$, convergence can be improved by
exploring changes in the model space instead of changes in the
parameter space. The NG update
rule, adapted from Eq.~\eqref{eq:second_order_newton}, is given by:

\begin{equation}
  \label{eq:quantum_natural_gradient_update}
  \theta_{i + 1} = \theta_i - \alpha_i g^{-1}(\theta_i)\nabla_\theta f(\theta_i)\,.
\end{equation}

This natural preconditioning is invariant under
parameter re-scaling and allows the gradient to always point towards
the direction of greatest descent relative to the information
geometry~\cite{amari1993, Stokes2020, Gacon_2021}.

Within the quantum computing context, the Fubini-Study metric tensor,
$g^F$, offers a natural metric corresponding to the variation of the
model in probability space~\cite{Lambert_2023}. Inspired by Amari's
NG descent~\cite{amari1993} and the Berry
connection between $g^F$ and the Quantum Fisher Information Matrix
(QFIM), namely $F(=4g^F)$~\cite{Meyer_2021, Lambert_2023}, the Quantum Natural
Gradient (QNG) utilizes quantum information geometry to improve
VQA optimization~\cite{Stokes2020}.

The QFIM Riemannian tensor, measuring the sensitivity of the model
from the parametric space to the probability space $p_\theta(x)$, is
given by:

\begin{equation}
  \label{eq:quantum_fisher}
  F_{ij} = \sum_{x\in [N]}p_\theta\frac{\partial \log p_\theta(x)}{\partial \theta_i}\frac{\partial \log p_\theta(x)}{\partial \theta_j}\,,
\end{equation}

where the sum extends over all possible $x$ outputs. The corresponding The Fubini-Study
metric is then given by:

\begin{equation}
  \label{equ:study_tensor}
  g_{ij}^F(\theta) = \Re\Biggl \{ \Biggl \langle \frac{\partial\psi}{\partial\theta_i}\Biggl | \frac{\partial\psi}{\partial\theta_j}\Biggl \rangle - \Biggl \langle \frac{\partial\psi}{\partial\theta_i}\Biggl | \psi\Biggl \rangle \Biggl \langle \psi \Biggl | \frac{\partial\psi}{\partial\theta_j}\Biggl \rangle \Biggl \}\,.
\end{equation}

Full estimation of the QFIM at every optimizer iteration is
prohibitively expensive in quantum resources
($\mathcal{O}(n^2_\theta)$ for $n_\theta$ parameters), but through
QNG's quantum circuit representation of the QFIM, $g^F$ can be
efficiently evaluated ($\mathcal{O}(n_\theta)$) by sub-circuit
sampling, particularly within block-matrix and diagonal matrix
approximations. These improvements can be, however, hindered by
discarded inter-system correlations~\cite{Wierichs2020}. On the other
hand, alternative methods and further approximations can also be
introduced for avoiding QFIM explicit calculation
altogether~\cite{Beckey_2022, Gacon_2024}. In the following sections,
we expose cheaper alternatives aiming at exploiting the QFIM by
quasi-Newton inspired generalizations.

\subsubsection{Momentum QNG methods}
QNG improvements in convergence brought about by quantum information
geometry are expected to be complementary to adaptive optimizers, such
as Adam~\cite{Stokes2020}. In moment-based techniques, preconditioning
is evaluated based on a temporal average of the optimization
function's behavior, rather than the local geometry of the quantum
state space. A collection of QNG variants, including momentum
averaging have been introduced
recently~\cite{Fitzek2024optimizing}. We first revisit the conceptual
framework underlying momentum optimization, before explaining some of
these implementations.

The classic momentum update is a modern adaptation of
Polyak's momentum method \cite{Polyak1964SomeMO,sutskever2013}, which
modifies the Newton update rule as follows:

\begin{align}
  \label{eq:momentum_update}
  v_{i+1} &= mv_i - \alpha H^{-1}\nabla_\theta f(\theta_i)\,,\\
  \theta_{i + 1} &= \theta_i + v_{i+1}\,.
\end{align}

Here, $m \in [0, 1]$ is the momentum coefficient and $v_i$ is the
corrected update at iteration $i$, or velocity vector. Incorporating this
physics-inspired momentum into parameter dynamics has a stabilizing
effect in noisy and non-convex landscapes with large amounts of
curvature, as it exponentially damps the oscillatory nature of
gradient corrections. However, selecting an optimal momentum parameter,
or introducing scheduling dynamics, is by itself another hyperparameter
fine-tuning task \cite{tao2021role, nakerst2020gradient}, which is
computationally expensive and heavily problem-dependent.

Adaptive methods based on gradient history can approximate the Hessian
and adapt multi-dimensional learning rates by naturally assigning
higher rates to parameters with low-frequency geometrical features,
while applying lower rates in directions with high curvature
features. Thus, for example, in ADAGRAD, the momentum update
\eqref{eq:momentum_update} follows the rule:

\begin{equation}
  v_{i+1} = mv_i - \frac{\alpha}{\sqrt{\sum_{\tau=1}^i |\nabla_\theta f(\theta)|_\tau^2}}\nabla_\theta f(\theta)\,,
\end{equation}

where the denominator computes the $l_2$ norm of all gradients at all
previous iterations over a window of length $\tau$. This approach
factors out gradient amplitudes into learning rates and exhibits an
annealing effect. Additionally, it allows large magnitude variations
in gradients while inheriting second-order update features as a
first-order method. However, due to the sensitivity to initial
conditions, it can be particularly sensitive to setting up initial
learning rates at early iterations of the method.

\subsubsection{qBang, qBroyden}
These methods are inspired by the extension of the classical natural
gradient~\cite{amari} through adaptive techniques~\cite{PARK2000755},
introducing both momentum and adaptive learning rates with moving
averages, and assumming the approximated QFIM is slowly varying as the
parameter space is explored. At each iteration, the metric tensor is
characterized by a low-pass filter of learning rate $\epsilon_i$, such
the inverse Hessian is given by:

\begin{equation}
  B_{i+1} = (1 - \epsilon_i)B_i + \epsilon_i\nabla f_i\nabla f_i^\dagger.
\end{equation}

This update leverages Hessian information to update the
QFIM~\cite{dash2024efficiencyneuralquantumstates}. Consequently, the
initial approximation of the algorithm dictates the nature of qBang
and qBroyden. At the first iteration, one computes either a full or
block approximation of the QFIM, followed by an adaptive Broyden
method using an averaging window to approximate the QFIM at subsequent
optimization steps. At each iteration, the inverse Hessian matrix is
given by rank-1 approximation: :

\begin{equation}
  B^{-1}_{i+1} = \left(I - \frac{\epsilon_i B_i^{-1}\nabla f_i\nabla f_i^\dagger}{1 - \epsilon_i(1 - \nabla f_i^\dagger B_i^{-1}\nabla f_i)}\right)\frac{B_i^{-1}}{1 - \epsilon_i}
\end{equation}

Similarly to the Adam algorithm~\cite{kingma2017adam}, qBang
introduces updates of both variance and momentum, with the variance
used to implement a trust region. The incorporation of momentum and
moving averages improves stability by shortening unreliable
steps. qBroyden, on the other hand, extends a Broyden method (a more
general version of BGFS), to the same QFIM approximation without
incorporating variance update rules or momentum. It is then a rank-1
approximation of the QFIM using low-pass learning rates and
quasi-Newton update rules.

An alternative momentum version of QNG was also introduced in
~\cite{Fitzek_2024}, which we will incorporate within this work as a
test of the efficacy of adding momentum and sliding averages to
QNG-type methods for quantum optimization.

\subsection{Bayesian optimization for hyperparameters fine-tuning}
\label{sec:bayes}

For the particular case of combinatorial problems involving QAOA the
complexity of the objective function implies that sensitivity to
optimizer settings will have a substantial impact on the solver's
ability to find solutions. In general, hyperparameter fine-tuning of
optimizers is a significant challenge in contemporary computer
science~\cite{yu2020hyperparameteroptimizationreviewalgorithms}. In
many instances, parameters are selected manually through trial and
error by the user, which can be a viable strategy in simple cases but
is often time-consuming and
ineffective~\cite{bischl2021hyperparameteroptimizationfoundationsalgorithms}. When
the number of parameters is low, and objective function evaluation is
computationally expensive, Bayesian optimization presents itself as an
ideal candidate, offering a reasonable cost-to-benefit
ratio~\cite{7352306}.

\begin{figure}[t]
    \hspace{-0.6cm}
    \centering \includegraphics[scale=0.33]{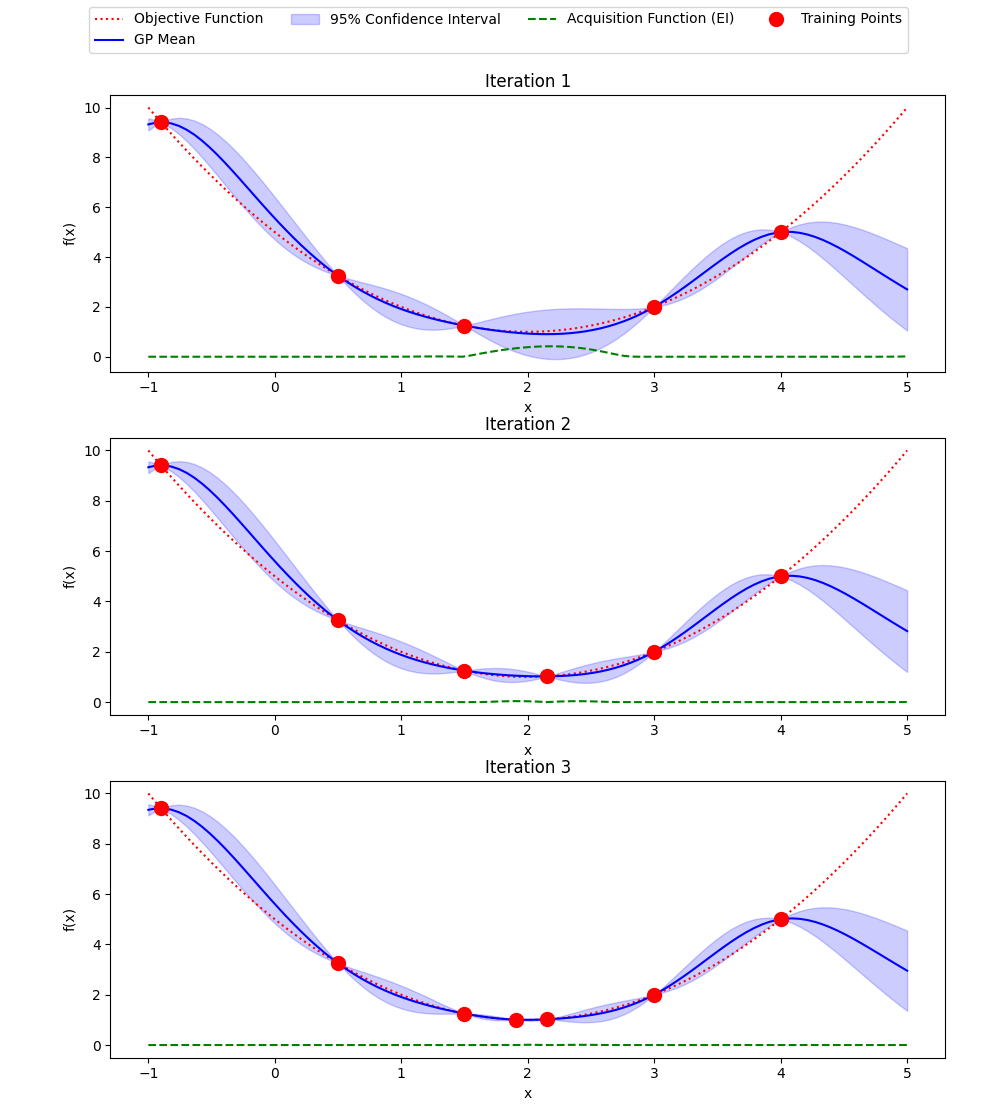}
    \caption{Example of iterative Bayesian process using a Gaussian kernel. Objective function of the form $f(x) =(x - 2)^2 + 1$. In practice the objective function is not known, it is only displayed for illustration. Edge acquisition policy and GP + simple Matèrn kernel}
    \label{fig:bayesian_opt}
\end{figure}

Bayesian optimization relies on constructing a probabilistic model of
the objective function to evaluate the most likely next step to take
in order to improve an overall performance metric of interest (see
Fig.~\ref{fig:bayesian_opt}). This method is based on two fundamental
concepts: surrogate models and acquisition functions.

\subsubsection{Gaussian Processes (GP)}
In Bayesian regression models, the distinction is made between
parametric and non-parametric models. Parametric models have a finite
number of parameters to tune in the regression problem, such as
Bayesian linear regression, while non-parametric models can be thought
of as an infinite-parameter fit to the objective function. The latter
approach can be derived from the former using a kernelization
trick. The advantage of this approach lies in its ability to introduce
a Gaussian Process (GP), which can integrate both simple and complex
data relationships as more data is provided to the model for updating.

A non-parametric Gaussian Process (GP) model, denoted as
$GP(\mu_0, k)$, is fully characterized by its prior mean function
$\mu_0$ and its positive definite kernel $K$. Given a collection of
$n$ measurements, dictated by a set of hyperparameters $\phi$,
the set of observations $D_n = \{(\bm{x}_i, \bm{y}_i)\}_{i=1}^n$,
according to Bayesian inference principles, the random variable
$\mathbb{E}(\bm{x})$, conditioned on the cumulative knowledge of $D_n$,
follows a normal distribution with the posterior mean and variance:

\begin{align}
  \label{eq:posterior}
  \mu_n(\bm{x}) &= \mu_0(\bm{x}) + \kappa(\bm{x})^T(\bm{K} + \sigma^2\bm{I})^{-1}(\bm{y} - \bm{m})\\
  \label{eq:posterior_two}
  \sigma_n^2(\bm{x})&= \kappa(\bm{x}, \bm{x}) - \kappa(\bm{x})^T(\bm{K} + \sigma^2\bm{I})^{-1}\kappa(\bm{x})
\end{align}

where $\kappa$ represents the vector of covariance terms between point
measurements, $\bm{K}$ is the kernel matrix formed from all pairs of
$\kappa$. A commonly used kernel choice is
the Matèrn kernel, which belongs to a class of parameterized stationary
kernels~\cite{porcu2023maternmodeljourneystatistics}. We employed this
kernel through the automatic fine-tuning feature in \verb|scikit-optimize|.

\subsubsection{Acquisition function}
The second element of Bayesian hyper-parameter estimation involves
designing a policy for updating $\phi$ based
on information obtained from the improving surrogate model. Various
options are available in the literature regarding such policy,
including improvement-based policies~\cite{kushner_64}, optimistic
policies~\cite{LAI19854, peter2003}, and information-based policies,
as well as the use of a portfolio of multiple acquisition functions~\cite{brochu2011portfolioallocationbayesianoptimization}. In this
study, we employ the default \verb|gp_hedge| update policy in the
\verb|scikit-optimize| function, which is a portfolio that scrambles and
randomly selects from lower confidence bound, negative expected
improvement, and negative probability of improvement. Utilizing a
portfolio-based update policy helps to mitigate biases and leverage
the strengths and weaknesses of multiple policies simultaneously~\cite{VASCONCELOS2022115847}, making it an appealing choice for
benchmarking optimization methods.

\subsection{Performance benchmarks}
\label{sec:performance_benchmarks}
The efficacy of the QAOA's ansatz-optimizer couple is evaluated
through a comprehensive assessment of its overall performance
capabilities, specifically its ability to provide efficient solutions
at reasonable computational costs. The key metrics employed in this
work are explained below.

\subsubsection{Convergence criteria}
\label{sec:convergence_criteria}
As reference point, we solve the synthetic MaxCut problem using a
brute force algorithm, and define convergence in terms of the set of
solutions that yield an objective function evaluation within a
predetermined margin from the optimal solution. This proximity is
measured in terms of relative or absolute tolerance. The choice of
tolerance typically depends on the scale and specificity of the
problem, as well as industry standards. In particular, for logistic
problems, including the Traveling Salesman Problem (TSP) and various
Vehicle Routing Problems (VRPs), research communities often adhere to
established tolerances set by algorithmic competitions and
standards. Generally, these tolerances range from 1 to 3 percent for
small-scale logistic problems in DIMACS, and 1 to 5 percent for the
same types of problems in the ROADEF competition. For large-scale
problems, tolerances may be set between 5 and 10 percent for ROADEF
and between 3 to 10 percent for DIMACS~\cite{dimacs_competition,
  roadec_competition}. In this work, a tolerance of $3\%$ is chosen,
and consider a specific run converged if this tolerance is reached at
least once.

\subsubsection{Local Lipschitz constant}
\label{sec:local_lip_theory}
In the context of continuous parameter optimization problems, the Lipschitz constant
provides a bound on the rate at which a function can
change. Specifically, a function
$f: \mathbb{R}^n \rightarrow \mathbb{R}$ is said to be Lipschitz
continuous, with a Lipschitz constant $L$, if
$L = \sup_{\mathbf{x}, \mathbf{y} \in D, \mathbf{x} \neq \mathbf{y}} \frac{|f(\mathbf{x}) - f(\mathbf{y})|}{\|\mathbf{x} - \mathbf{y}\|}$ for
$D$ in the acceptable parameter model domain. Due to the computational costs involved in precisely
calculating the Lipschitz constant, many algorithms instead estimate
its value~\cite{fazlyab2023efficientaccurateestimationlipschitz,
xu2024compositionalestimationlipschitzconstants}. Rates of convergence of gradient-based methods
are bound by the LP constant which sets the optimal step sizes~\cite{herrera2023locallipschitzboundsdeep}, for
a given optimization landscape, various gradient based methods can be compared by finding a local bound on the LP constant
for a given set of conditions. Equivalently to~\cite{herrera2023locallipschitzboundsdeep}, we focus on approximating $L$ along optimization ``trajectories'', multiple runs creating an overall stochastic form of each methods (mutliple starts of the same methods on the same problem, averaged over), computing different $L$-metrics as it is shown in
the Tab.~\ref{tab:lips_metrics}.

By gathering statistics from numerous optimization runs, we aim to observe the difference in
behavior between various optimizers when adapting to the landscape
geometry, checking the sensitivity to sudden changes and the ability
to handle them effectively. Overall, the largest the LP, the smallest the step-size the optimizer need to accept
to handle large local variations.

Increasing problem complexity, characterized by high
dimensionality and non-convexity, suggests that tightly packed
distributions of $L$ with low median and interquartile range (IQR),
coupled with low deviation and averages, do not necessarily
guarantee the best convergence behavior. Rather, a trade-off
between speed and accuracy should incorporate well-behaved
distributions, where statistical values are commensurate in
magnitude. To maximize coverage, employing convergence filtering
(i.e., performing all estimations on converged results)
necessitates higher averages, accompanied by increased
dispersion, albeit with median values tightly clustered around the
mean value for optimal performance. Conversely, very stable runs
(i.e., non-skewed distributions, low averages, deviations, and
IQR) will likely yield strong convergence at low dimensions but may
poorly scale due to overfitting or limited adaptivity.

\begin{table}[htpb]
  \centering
  \small
  \begin{tabular}{|p{0.23\columnwidth}|p{0.70\columnwidth}|}
    \hline
    Metric & Interpretation \\
    \hline
    Average & Low: stable updates in smooth regions, allowing larger step sizes without performance loss or instability. \\
    Variance & Low: consistent and robust across runs, but less adaptable to skewed landscapes. Lower STD enhances predictability upon convergence. \\
    Median & Low: Frequent low Lipschitz constants indicate stable iterations and effectiveness in less sensitive optimizations. \\
    Interquartile range (IQR) & Low: clustered Lipschitz constants show consistent performance, suggesting stability in similar landscapes but limited adaptability. \\
    \hline
  \end{tabular}
  \caption[lips_metrics]{Lipschitz constant statistical metrics used in the benchmark. Very small LP averages indicate low sensitivity, and tendencies to get stuck in local minimas. Target overall for a well behaved optimizer is non-skewed distributions, in most cases, $IQR > STD$ describes leptokurtic distributions (peaked distributions), when the opposite $IQR < STD$ can indicate platykurtic distributions (flat distributions)~\cite{walck1996hand}. The LP being a measure of model variation, this means we also desire for optimization purposes rather a peaked distribution, hence a clustered $IQR > STD$. Finally, the STD being a measure of overall spread, well-behaved, relatively higher STD is wanted for adaptibility of the procedure.}
  \label{tab:lips_metrics}
\end{table}




\subsubsection{Hyperparameters tuning}
\label{sec:hyper-param-tuning}
Exploring the behavior of methods within the hyper-parameter space
requires a substantial number of function evaluations, grid-division
approaches, for example, search for discretized configurations of
individual parameters. Using Bayesian optimization, on the other hand,
allows for an efficient hyper-parameter space exploration. This
approach evaluates what can be expected in terms of overall behavior
from each optimizer, even for small numbers of hyperparameters.  With
Bayesian fine-tuning, one can reasonably expect to observe multiple
optimizers converging with parameter selections at least along points
of interest, provided that each parameter's range is defined within
practical bounds.

An overview of the hyper-parameter space per method is presented in
Tab.\ref{tab:table_hyper_parameters}, while additional information
about the selected search space for each parameter is provided in
App.\ref{app:default_parameters}.

\begin{table}
    \centering
    \begin{tabular}{l c c}
        \hline
        \textbf{Methods} & \textbf{Params} & \textbf{Desc}\\
        \hline
        BFGS,DFP,NCG&$\alpha$&lr\\
        SR1&$\beta$&ls\\
         &$c_1 $&Amijo\\
         &$c_2$&Wolfe\\
         \hline
         SP-BFGS&$\alpha$&lr\\
         &$\beta$&ls\\
         &$N_s$&slope p.\\
         &$N_0$&inter\\
         &$c_1$&Amijo\\
         &$c_2$&wolfe\\
         \hline
         \hline
         QNG (block, diag)&$\alpha$&lr\\
         \hline
         qBroyden&$\alpha$&lr\\
         &$\epsilon$&filter lr\\
         \hline
         qBang&$\alpha$&lr\\
         &$\epsilon$&filter\\
         &$\beta_1$&decay\\
         &$\beta_2$&decay\\
         \hline
         m-QNG&$\alpha$&lr\\
         &$\epsilon$&filter lr\\
         &$\beta_1$&decay\\
         &$\beta_2$&decay\\
         \hline
         \hline
         SPSA&$\alpha$&decay lr\\
         &$a_{init}$&init lr\\
         &$c_{init}$&init pert\\
         &$\gamma$&decay pert\\
         &$A$&amplitude decay\\
         \hline
         2SPSA&$\alpha$&decay lr\\
         &$a_{init}$&init lr\\
         &$c_{init}$&init pert\\
         &$a^H_{init}$&init lr\\
         &$c^H_{init}$&init pert\\
         &$\gamma$&decay pert\\
         &$A$&amplitude decay\\
         \hline
         QNSPSA&$\alpha$&lr\\
         &$\epsilon$&amplitude decay\\

         \hline
         RCD&$\alpha$&lr\\
         &$\gamma$&decay\\
         \hline
    \end{tabular}
    \caption{The following hyperparameters are selected for tuning on
      each method: Amijo and Wolfe's line backtracking algorithm
      parameters, which include curvature and improvement; ls,
      denoting the line search penalty term associated with the
      learning rate update; perturbations of SPSA type, referred to as
      pert; and lr, representing any learning rate. Additionally, the
      exponent $H$ represents an approximation of the Hessian matrix.}
    \label{tab:table_hyper_parameters}
\end{table}

\section{Results}
\label{sec:results_discussion}
\begin{figure*}[t]
  \centering
  \includegraphics[scale=0.42]{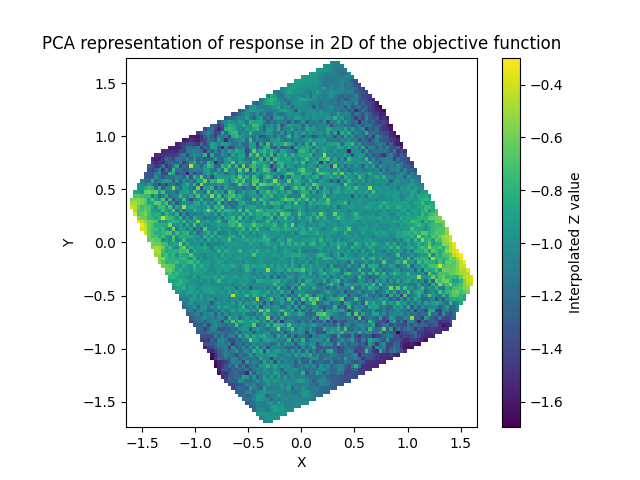}
  \includegraphics[scale=0.42]{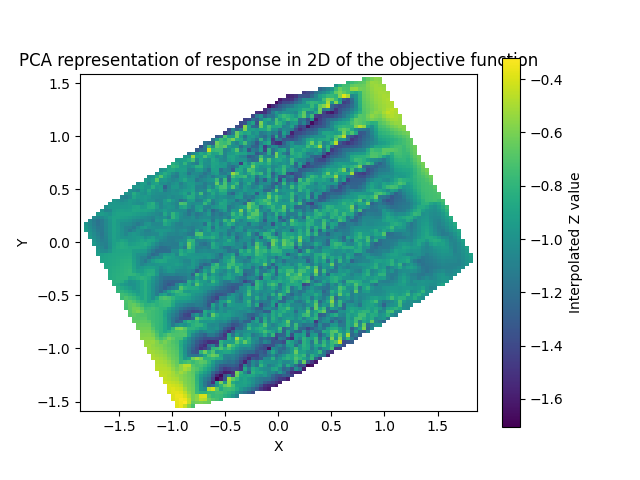}
  \caption{3 nodes problem (left) and 7 node problem (right) PCA analysis in a grid exploration. Increasing the problem size drastically increases rugosity and local minima, central symmetry due to symmetries in parametrization of Pauli gates.}
  \label{fig:space_analysis_PCA}
\end{figure*}

First, we conducted a series of exploratory runs for each optimizer on weighted graphs with 3 and 5 nodes, addressing the MaxCut problems using a one-layer QAOA ansatz (as detailed in Sec.\ref{sec:problem_def}). The initial conditions were fully randomized but can be reproduced through the \verb|seed| settings. We first present a high-dimensional function visualization, utilizing Principal Component Analysis (PCA) and a random projection onto a lower-dimensional plane. This illustrates the highly complex structures of the parameter space, characterized by non-convexities that scale with problem dimensions. We also performed an initial Bayesian hyperparameter estimation to identify a well-suited parameter set, as shown in Tab.\ref{tab:table_hyper_parameters}.

Subsequently, we present three benchmarks: Firstly, a small benchmark involving 3 nodes to assess convergence robustness by analyzing local Lipschitz constants discussed in Sec.\ref{sec:local_lip_theory}, followed by an examination of convergence behavior. Secondly, we conducted a benchmark with a larger set of properties through multiple Bayesian sweeps, as described in Sec.\ref{sec:hyper-param-tuning}. This latter benchmark was explored by gathering statistics on the averaged behaviors across various optimization initial conditions, allowing for a systematic discussion of what users can expect from these different optimization methods.

\subsection{Landscape analysis}
\label{sec:landscape_analysis}

\begin{figure*}[t]
  \centering
  \includegraphics[scale=0.37]{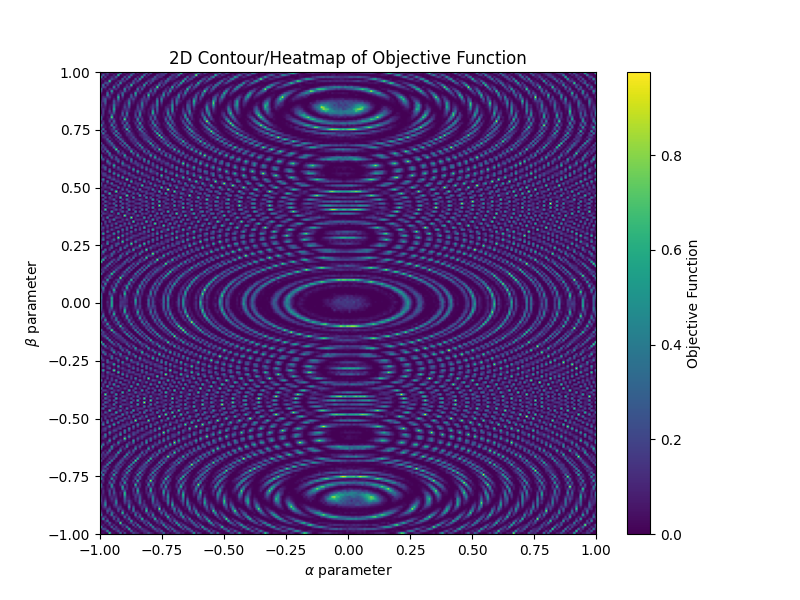}
  \includegraphics[scale=0.37]{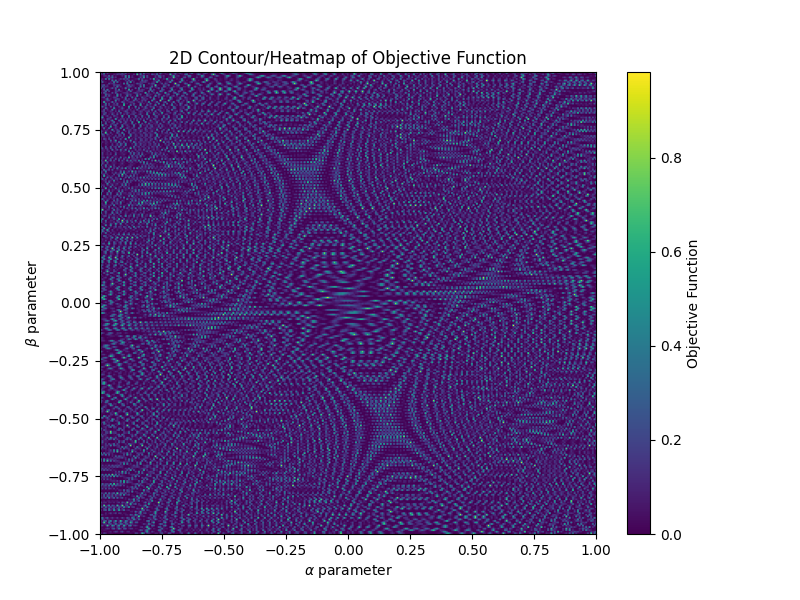}
  \caption{For 300 grid discretization, examples of two randomly picked directions in 5 nodes problem App.~\ref{sec:problem_def}, defining two random parameterized lines. Two random directions define a plan in the hyper-space in which the objective function is defined, parameters $\alpha$ and $\beta$ are unique parameterization of each random lines.}
  \label{fig:space_analysis_random_direction}
\end{figure*}

To motivate the following discussion, we investigate the optimization landscape of the QAOA ansatz using Principal Component Analysis (PCA)~\cite{sorzano2014surveydimensionalityreductiontechniques} in Fig.~\ref{fig:space_analysis_PCA}.This validates previous observations that, in shallow PQCs, the geometric features of the optimization landscape degrade with increasing dimensions~\cite{Huembeli_2021}. Low-dimensional problems exhibit highly textured surfaces with clearly defined lower and upper regions. However, in the 7-node problem, we observe the emergence of repeated valleys and peaks, leading to more intricate structures and potential local minima.

An interesting technique for visualizing high-dimensional functions involves projecting higher-dimensional objects into random lower-dimensional constructs, which preserves distances from the original high-dimensional space~\cite{bingham_ella_random_projection}. In this case, we proceed by picking two random directions in parameter space in a 5 qubit ansatz solving the 5 nodes standard problem (Sec.~\ref{sec:problem_def}). We set a 300 steps from $[-1, 1]$ and explore the parameter space along multiple random directions. This method shows even in shallow QAOA complex interference patterns and hyperbolic structures Fig.~\ref{fig:space_analysis_random_direction}. Those figures represent a visual motivation for natural gradient methods: we see that in parameter space, a more suited metric, i.e. the Fubini-Study metric tensor, which describes a Riemannian geometry Eq.~(\ref{equ:study_tensor}).

\subsection{Convergence analysis}
\paragraph*{Convergence robustness} is assessed by compiling results from runs on the same 3-node regular graph MaxCut problem, initiated from a random set of initial points. In these evaluations, we exclude non-converging runs, defined as those that do not achieve a 1\% tolerance, to better highlight the overall convergence behavior. The stopping condition at 1\% allows us to observe the dynamic behavior of the methods while ensuring that there is no influence from stagnated trajectories in local minima or excessive iterations. Thus, the Lipschitz property (LP) serves as an effective descriptor of how the methods adapt to the same objective function geometry. To avoid statistical biases, we only consider experiments where at least 50\% of the runs achieved the specified tolerance. We present four statistical metrics representing the LP estimation for the given problem in Sec.~\ref{sec:local_lip_theory}.

\begin{table}[htbp]
  \centering
  \small 
  \begin{tabular}{c c c c c}
      \hline
      \textbf{Methods} & \textbf{av} & \textbf{std}  & \textbf{med} & \textbf{iqr}\\
      \hline
      \hline
      BFGS & 10.736 & 18.855 & 3.353 & 7.737\\
      SR1 & 12.0144 & 19.190 & 4.055 & 7.564\\
      NCG & 60.643 & 116.662 & 2.679 & 31.514\\
      SP-BFGS & 90.571 & 43.590 & 95.860 & \textbf{51.84}\\
      DFP & \textbf{93.568} & 30.862 & \textbf{103.362} & 33.448\\
      \hline
      QNG (diag) & 221.578 & 66.464 & 210.936 & 77.135\\
      QNG (block) & 404.684 & 119.77 & 381.91 & 129.793\\
      qBroyden & 411.238 & 138.408 & 386.249 & 149.343\\
      qBang & 232.709 & 74.829 & 217.482 & 61.351\\
      m-QNG & \textbf{632.674} & 150.492 & 611.474 & 191.993\\
      \hline
  \end{tabular}
  \caption{LP constant estimation for each method with full gradient estimation, 3 nodes, Bayesian defaults in Tab.~\ref{tab:table_quasi_newton_quantum_gradient}, 1000 experiments per method. seed=32, 20 max iterations}
  \label{table:lpestimator}
\end{table}



\begin{figure}[htbp]
  \centering
  \includegraphics[scale=0.3]{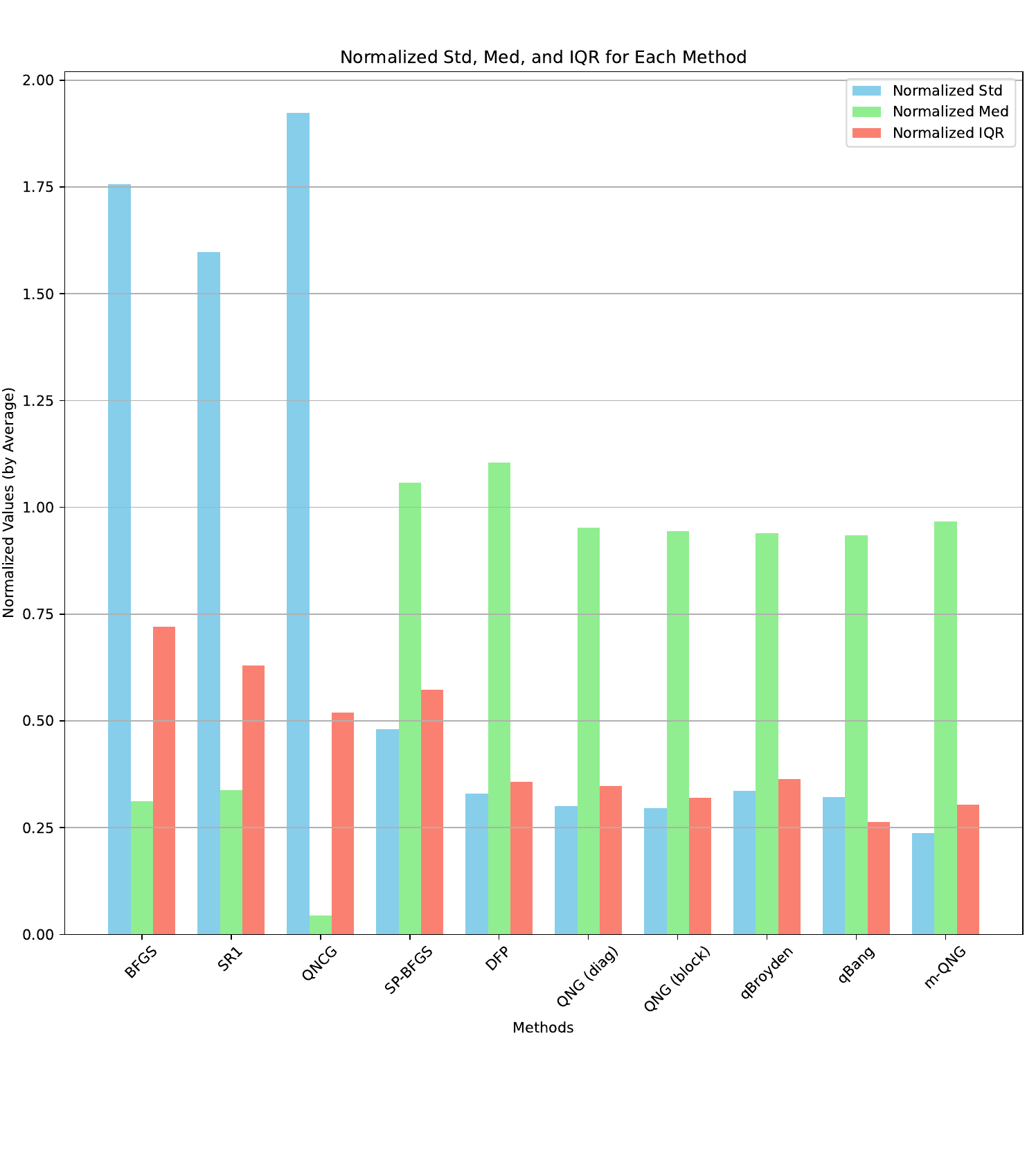}
  \caption{Each method normalized on their averages. We find three families of convergence behaviors: first are the skewd distributions: BFGS, SR1 and NCG. Well-behaved leptokurtic distributions (peaked) with SP-BFGS, DFP, QNG(block and diagonal), qBroyden and m-QNG. Last behavior are platykurtic distributions (flat) with qBang.}
  \label{fig:lp_normalized}
\end{figure}

Raw results in Tab.~\ref{table:lpestimator} shows natural gradient and quasi-Newton methods have significantly different LP estimations. On average, natural gradients provide much higher LP estimations, this means that the natural gradient preconditioning technically allows for larger step size, this validates of why natural gradients are introduced in the first place: providing preconditioning from the true direction of greater change in the optimization space according to information geometry. This is expected from the definition of the natural gradient, which indicates the true gradient in the quantum parametric space. High average LP means rapid convergence, Tab.~\ref{table:lpestimator} shows that SP-BFGS and DFP dominate quasi-Newton methods, where m-QNG and qBroyden show the stronger convergence. Those raw points however, do not show the full picture. Fig.~\ref{fig:lp_normalized} represents normalized distributions of the LP estimations on the mean. As explained on distribution in Sec.~\ref{sec:local_lip_theory}, BFGS, SR1 and NCG show very skewed distributions: they are not well behaved in addition to provide slower convergence than the two other quasi-Newton methods. Tab.~\ref{tab:lips_metrics} helps interpreting the data: the methods to the left are -as seen by the optimizers- behaving as flat landscape, provide poor convergence and due to low median provide low sensitivity to updates. The latter is not a sought after feature for non-convex geometries, complex geometries require high exploration. Furthermore DFP and SP-BFGS provides well rounded, regular and peaked distributions, with the only difference being a slightly higher scaling for SP-BFGS with slightly lower median, indicating slightly more rigid updates for SP-BFGS, but with more exploration and dispertion at each runs. Overall, natural gradient methods provide similar performances against each other, to the exception to m-QNG showing the lower variation overall.

These results demonstrate the behavior under ideal conditions, characterized by the lowest geometric complexity, as the 3-node system represents the simplest dimensional problem for MaxCut, using the most shallow QAOA ansatz. The remainder of this analysis will examine varying behaviors as we scale up the problems.

\paragraph*{Convergence behavior} can be investigated through an evaluation of convergence trajectories. While the previous paragraph offers a geometric perspective on a specific problem, it does not provide insights into how the methods perform as the complexity of the objective function increases, specifically with larger parameter spaces and higher dimensionality. Sequencially we investigate quasi-Newton preconditioners, natural gradient based methods, then we give some comments about the introduction of second order estimation in stochastic methods.

\begin{figure}[htbp]
    \centering\includegraphics[scale=0.5]{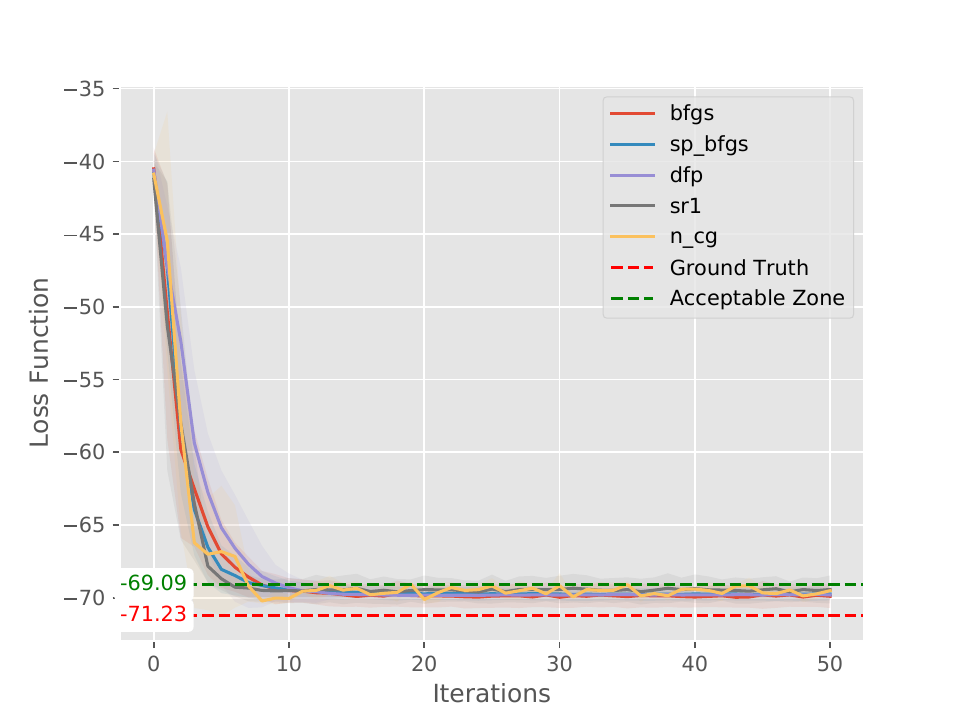}
    \centering\includegraphics[scale=0.5]{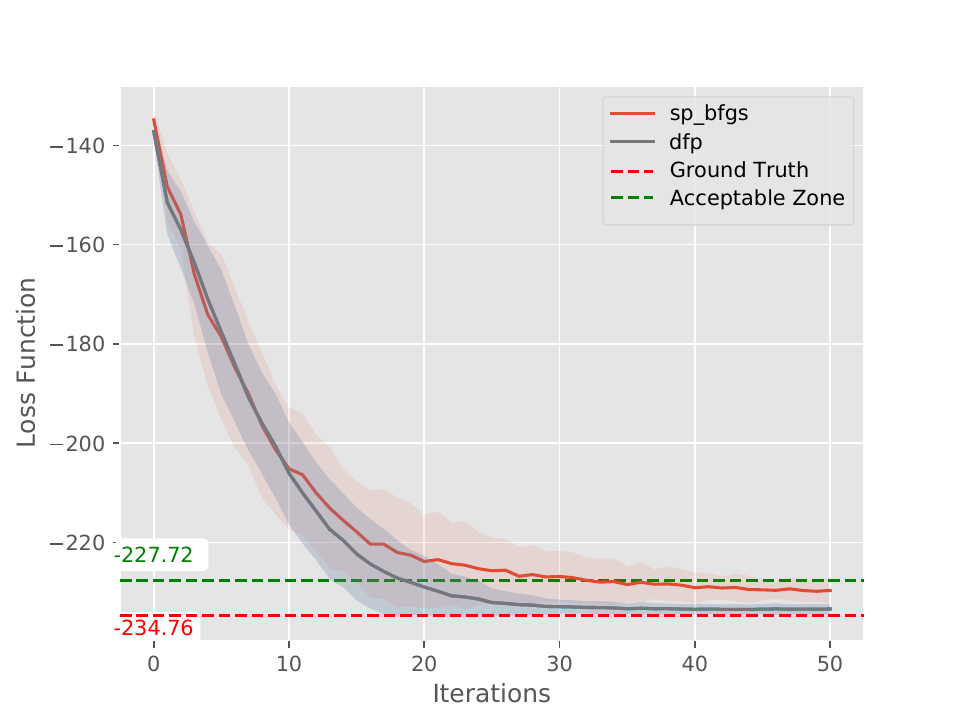}
    \caption{Convergence analysis, quasi-Newton types methods for benchmark problem, \textit{top} figure problem with 3 nodes, \textit{low} figure problem with 5 nodes of seed=32, 50 experiments over randomly sampled initial conditions. 512 shots, 60 iterations}
    \label{fig:3_5_nodes_conv_second_order}
\end{figure}

\begin{figure}[htbp]
    \centering\includegraphics[scale=0.5]{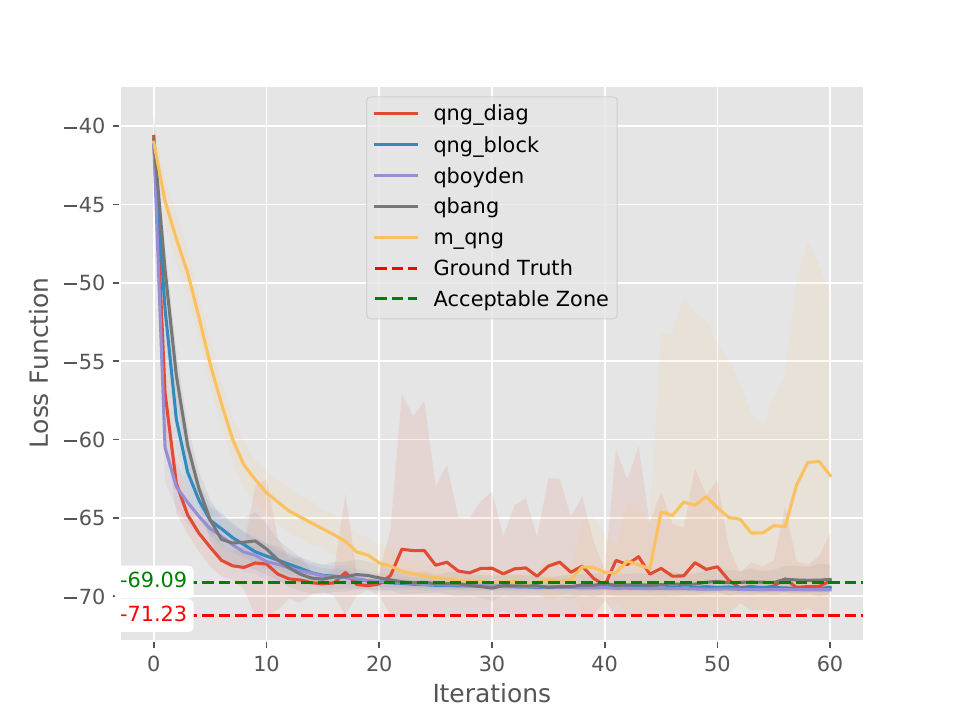}
    \centering\includegraphics[scale=0.5]{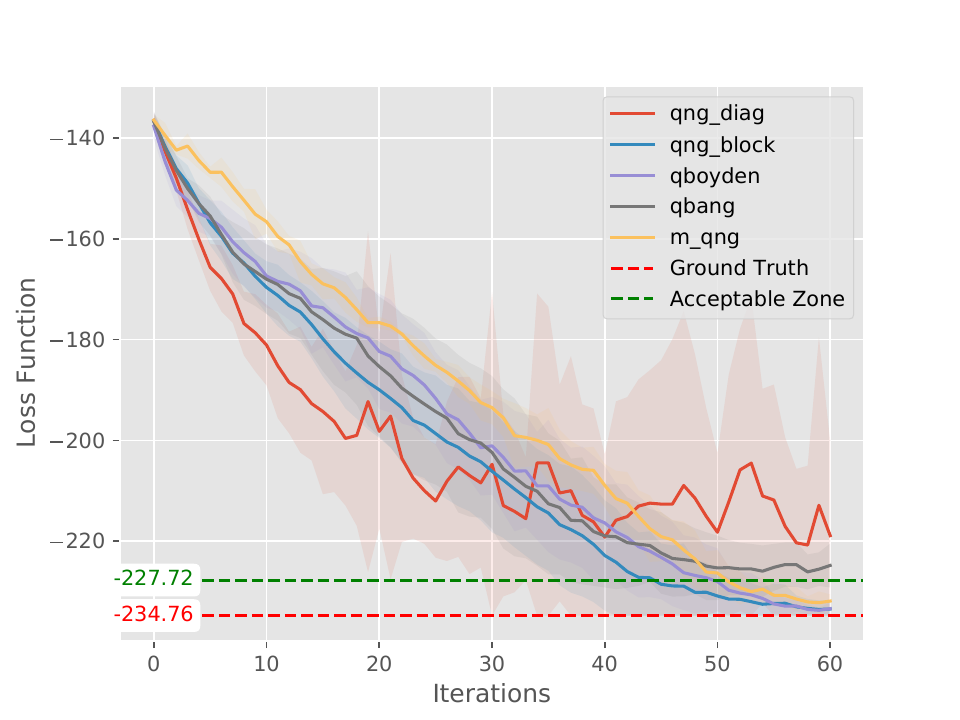}
    \caption{Convergence analysis, QNG types methods for benchmark problem, \textit{top} figure problem with 3 nodes, \textit{low} figure problem with 5 nodes of seed=32, 50 experiments over randomly sampled initial conditions. 512 shots, 60 iterations}
    \label{fig:3_5_nodes_conv_natural_gradient}
\end{figure}

\begin{figure}[htbp]
  \centering\includegraphics[scale=0.5]{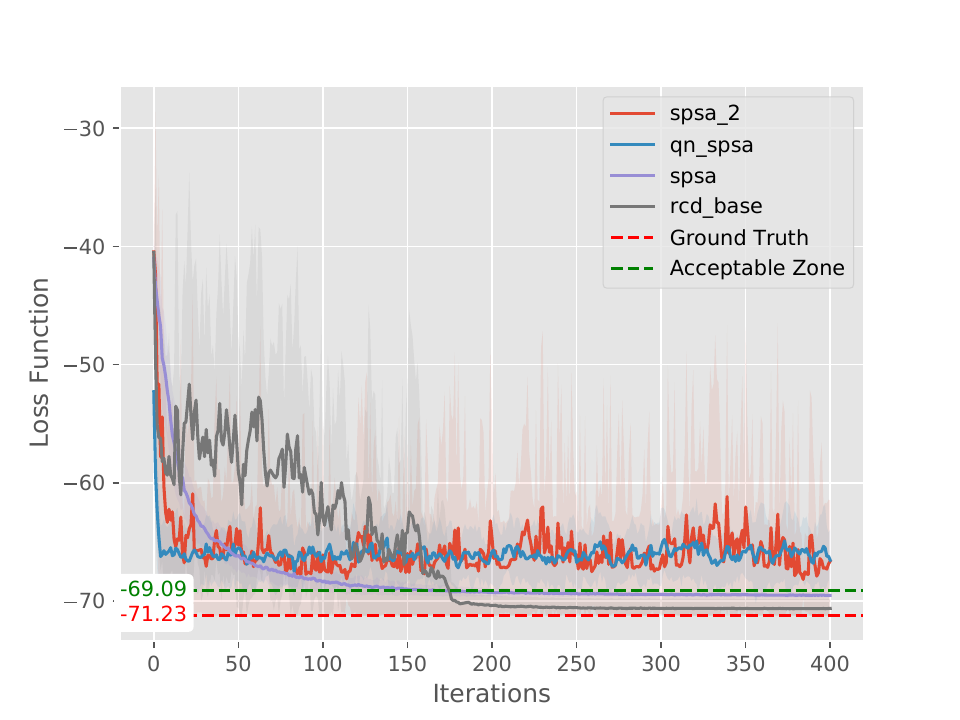}
  \centering\includegraphics[scale=0.5]{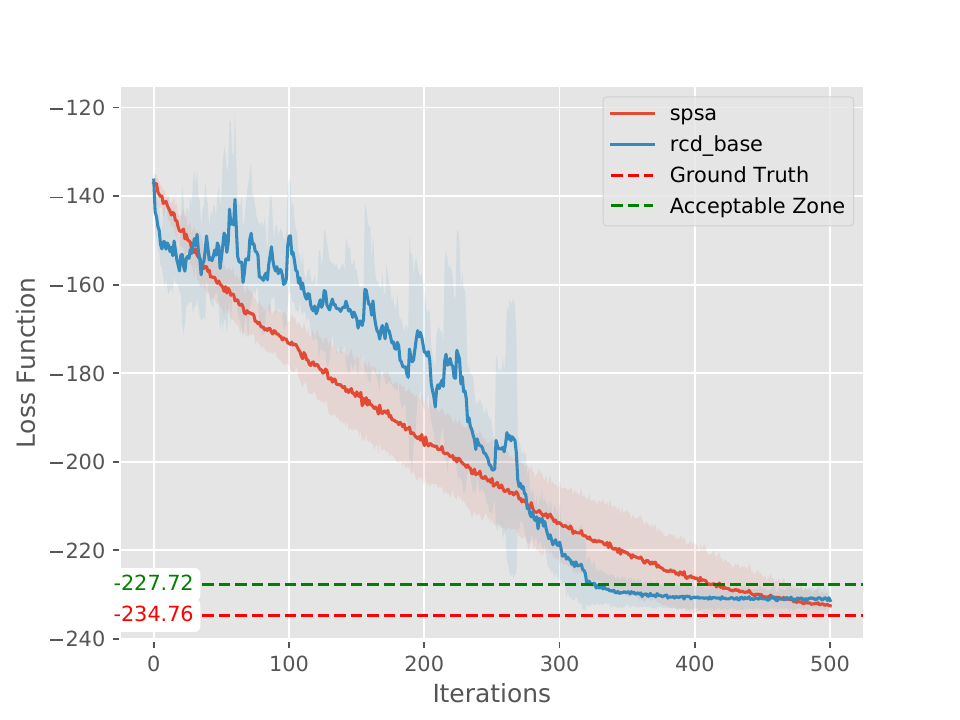}
  \caption{Convergence analysis, stochastic methods types methods for benchmark problem, \textit{top} figure problem with 3 nodes, \textit{low} figure problem with 5 nodes of seed=32, 50 experiments over randomly sampled initial conditions. 512 shots, 400 max iterations}
  \label{fig:3_5_nodes_stochastic}
\end{figure}

\paragraph*{Discussion}
The classic preconditionners SR1 and BFGS and NCG seems from Fig.~\ref{fig:3_5_nodes_conv_second_order} and Tab.~\ref{table:lpestimator} to be entirely unsuitable for a QAOA solver applied to MaxCut. They exhibit low average LP values, suggesting slow convergence overall, and demonstrate a null convergence ratio at 3\% tolerance for 5-node problems. In contrast, DFP and SP-BFGS present superior characteristics, offering quality solutions, strong convergence, and a favorable trade-off for exploring the parameter space.

On the side of the natural gradient experiments, although natural gradient promises better robustness in Tab.~\ref{table:lpestimator}, Fig.~\ref{fig:3_5_nodes_conv_natural_gradient}, demonstrate that this does not necessarily translate to improved performance in practice. This is likely due to the QNG algorithms suffering from additional statistical noise in the QFIM matrix evaluation. This seems to remain true regardless of the QFIM approximation choice however, as qBang and qBroyden rely on approximations of the QFIM through a learned filter and moving average.

Finally, in Fig.~\ref{fig:3_5_nodes_stochastic}, we compare performances between the 3-node and 5-node configurations. We observe significant improvements for QNSPSA and 2SPSA at 3 nodes, though this comes with increased statistical variations across runs. However, at 5 nodes, the translation of hyperparameters from the 3-node case fails to maintain consistency in capturing second-order information. This suggests that retuning is likely necessary for second-order stochastic methods, even with minor changes in problem dimensions and parameters.

In the next section, we go into more details and explore the full capabilities of those optimizers through the exploration of their hyperparameter space using Bayesian optimization, allowing for an in-depth cost to performance analysis.

\subsection{Benchmarks}
In this section, we explore significative results from overall benchmarks in appendix App.~\ref{app:all_benchmarks}, highlighting specific sections providing commentary. During those runs made on a 10-CPU node we provide two bayesian sweeps for each optimizer through all test problems in Sec.~\ref{sec:problem_def} from 3 nodes to 8 nodes. The first sweep is set without stopping condition to maximum iteration, the second with stopping condition at 3\% tolerance. Measure falls into performance and cost metrics such as last objective function value, walltime(s), Hamming distance to solution (binary distance from true binary solution), convergence ratio, Calls to QPU (qCalls), and with exit condition to convergence: average time per iteration, average qCall, qCall per iteration, average iteration to convergence

The averaged objective function with and without exit condition are used as cost function of the Bayesian search. In the following results, each points should be interpreted as an averaged value over 20 runs with initial conditions picked randomly from an uniform distribution. we note that NCG is absent to computation over 3 nodes, as the method fails to converge further at all. This is most likely due to the fact that CG-type methods perform best in locally-convex geometries, which is not the case for MaxCut QAOA as shown in Sec.~\ref{sec:landscape_analysis}.

\paragraph*{Performances}
\begin{figure*}[htbp]
  \centering
  \includegraphics[scale=0.35]{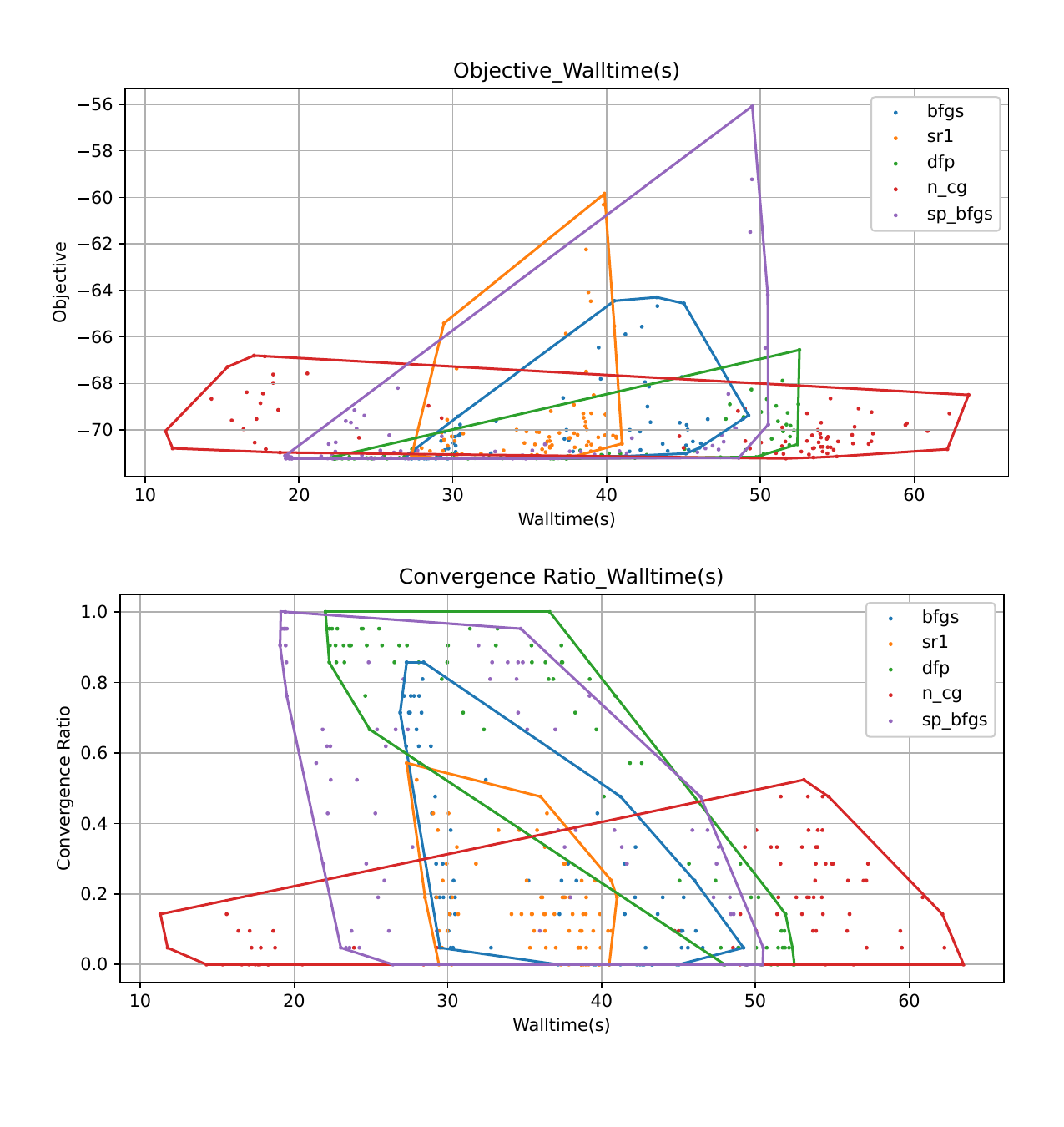}
  \raisebox{0cm}{\includegraphics[scale=0.35]{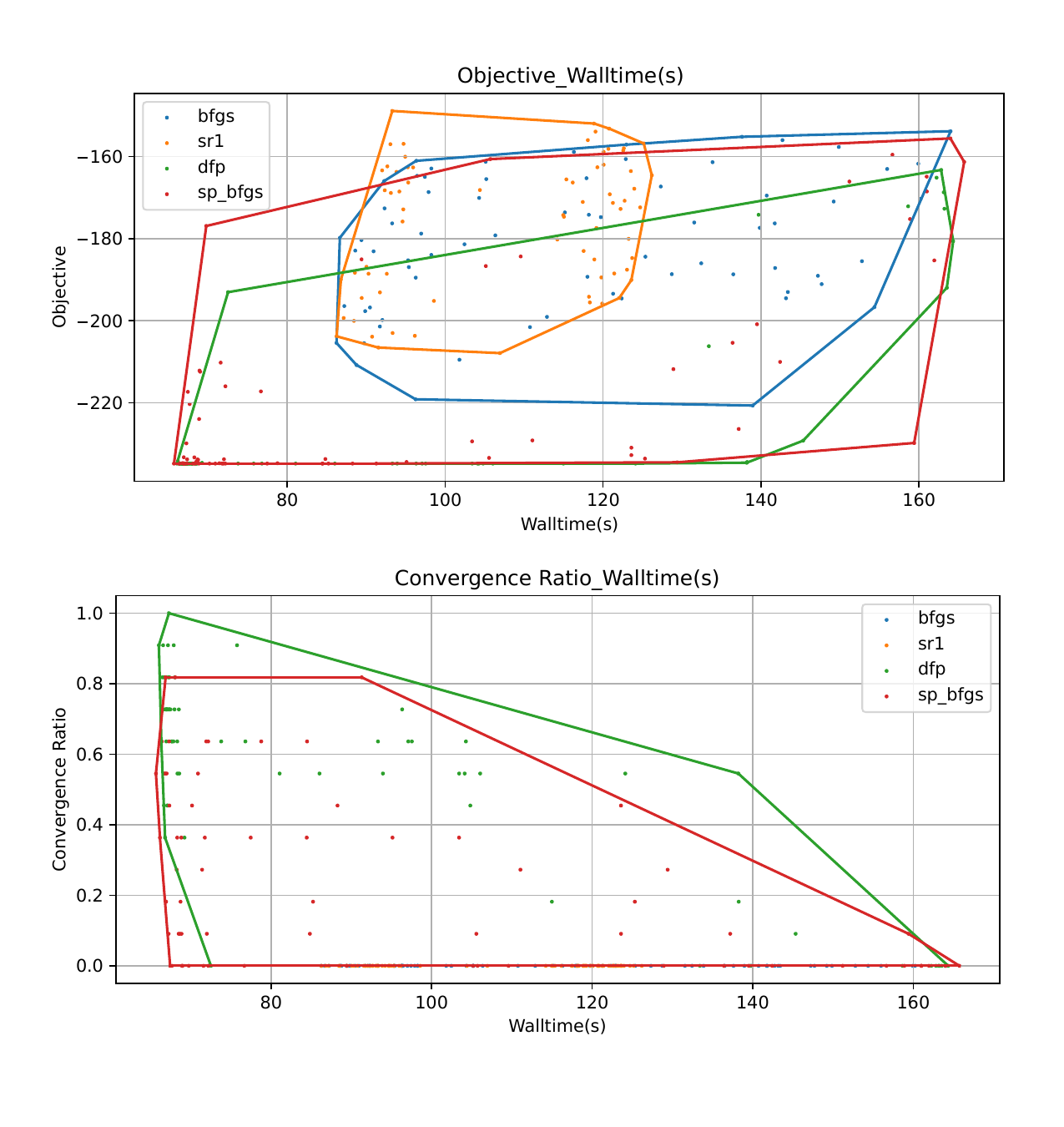}} 
  \caption{3 nodes problem (left) and 5 nodes problem (right) on the quasi-Newton methods, 70 bayesian sweeps on second order methods, stopping condition reached at maxit 60 iterations. Objective function are last objective values (averaged over 20 runs), Convergence ratio is the ratio of runs arriving to 3\% tolerance (20 runs), walltime (s) is measured from call of first iteration to stopping condition reached.}
  \label{fig:performance_3_5_second_order}
\end{figure*}

By incrementally increasing the problem size and constructing \verb|convex-hull| graphs, we can expand upon previous observations regarding the full expressible behavior set of the optimizers. The shapes of these graphs define the boundaries within a specifically chosen behavior space. Optimizers that share the same surface area on the graph across multiple metrics should be expected to perform similarly on a given QAOA MaxCut problem.

In Fig.~\ref{fig:performance_3_5_second_order} representing 3 and 5 nodes problems, we observe the failure of BFGS, NCG and SR1 methods to adapt to the optimization landscape as it scales. Already for 3 nodes, only DFP and SP-BFGS find hyper-parameter set with 100\% runs converging to 3\%. Furthermore, DFP allows for better scaling reaching close to 100\% convergence, this trend is validated up to 8 nodes in Fig.~\ref{fig:8_nodes_performances}.

\begin{figure*}[htbp]
  \centering
  \includegraphics[scale=0.35]{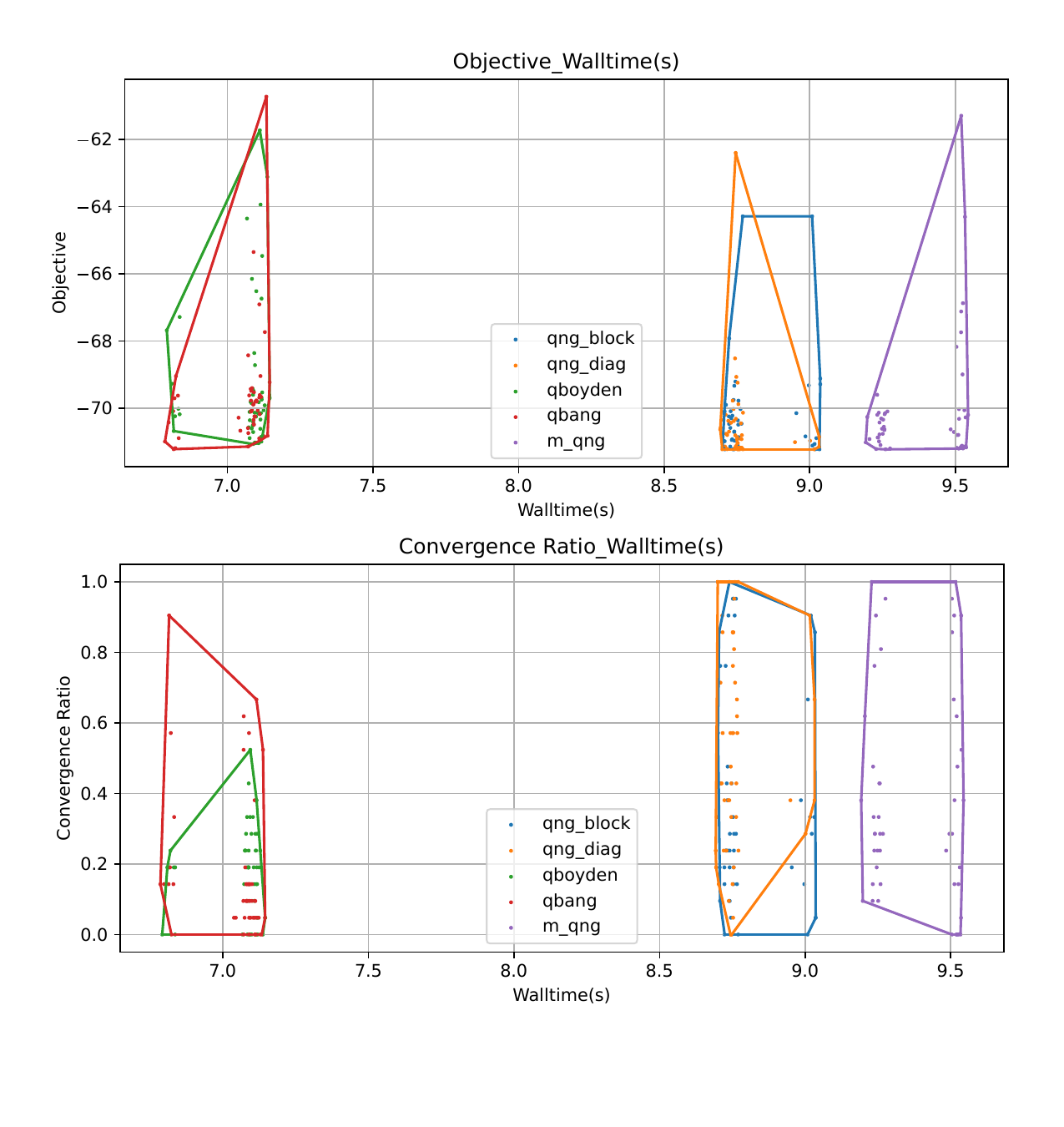}
  \raisebox{0.1cm}{\includegraphics[scale=0.35]{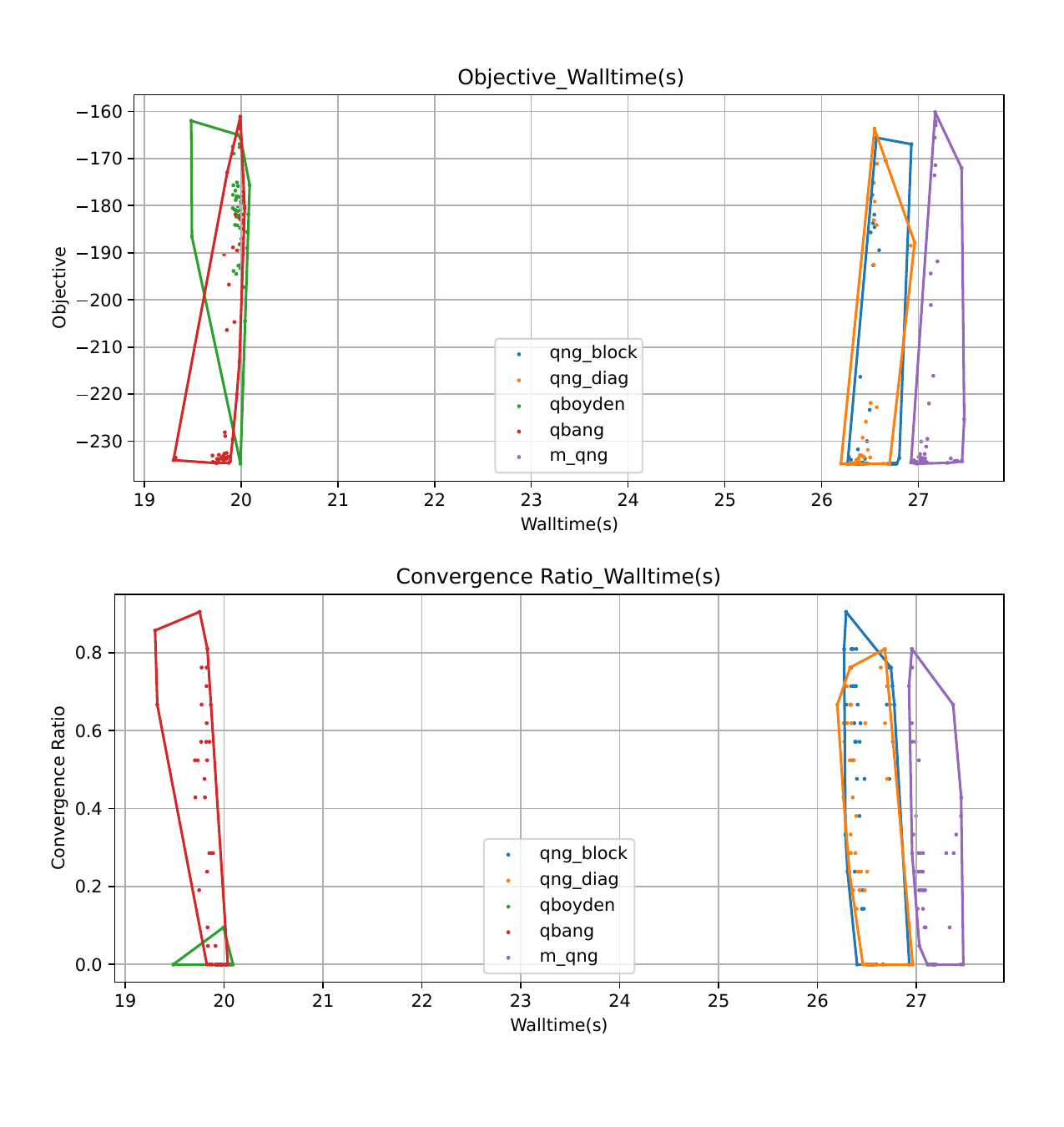}} 
  \caption{3 nodes problem (left) and 5 nodes problem (right), 70 bayesian sweeps Natural gradient methods, stopping condition reached at maxit 60 iterations. Objective function are last objective values (averaged over 20 runs), Convergence ratio is the ratio of runs arriving to 3\% tolerance (20 runs), walltime (s) is measured from call of first iteration to stopping condition reached.}
  \label{fig:performance_3_5_qng}
\end{figure*}

Scaling again from 3 nodes to 5 nodes shows a rapid loss of convergence ratios for qBroyen in Fig.~\ref{fig:performance_3_5_qng}, showing poor performances compared to the other methods. One can see however that qBang and qBroyden methods provide significatively faster operations for similar performances, depending on initial conditions.

\begin{figure}[t]
  \centering
  \includegraphics[scale=0.32]{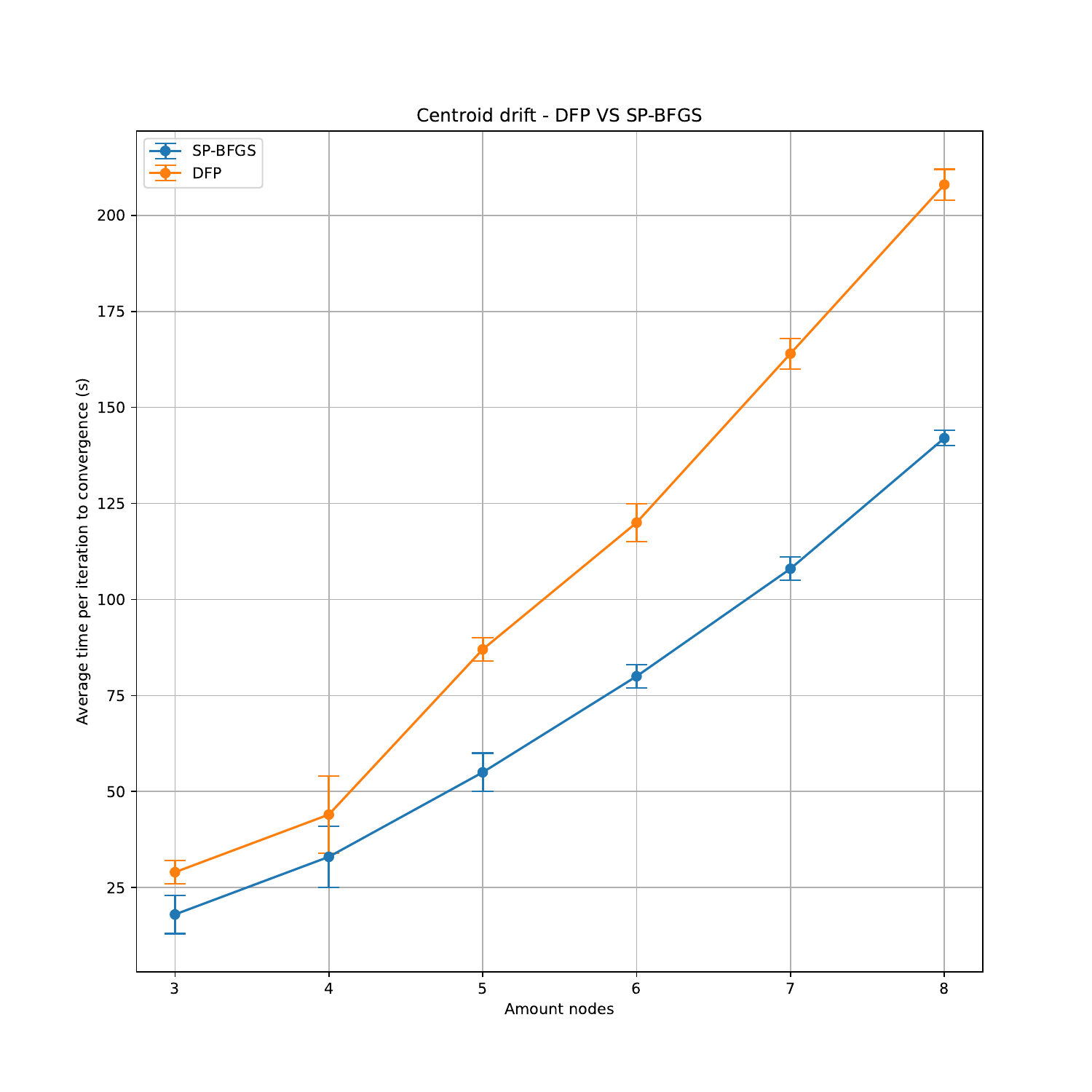}
  \caption{Distance between SP-BFGS and DFP centroids in time per iterations to convergence. We observe from a well-behaved quasi-Newton methods such as DFP form a secant-penalized update, the performance gain in time per iteration is increasing with the problem size.}
  \label{fig:centroid_improvement}
\end{figure}

Introducing secant penalization facilitates a significantly lower cost while maintaining a useful convergence ratio. The average improvements associated with this approach increase with problem size, as illustrated in Fig.~\ref{fig:centroid_improvement}.

\begin{figure*}[t]
  \centering
  \includegraphics[scale=0.35]{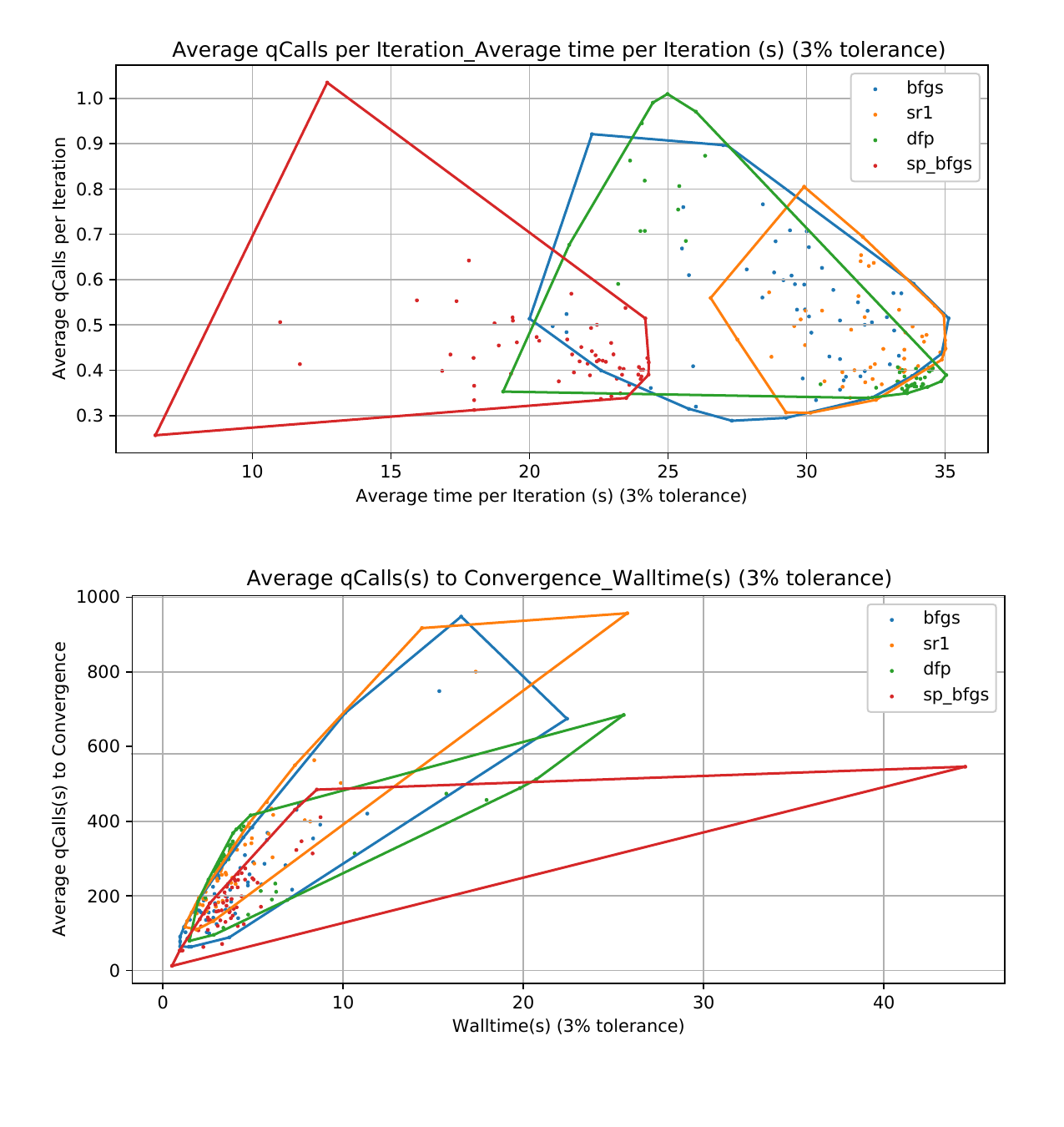}
  \raisebox{-0.09cm}{\includegraphics[scale=0.355]{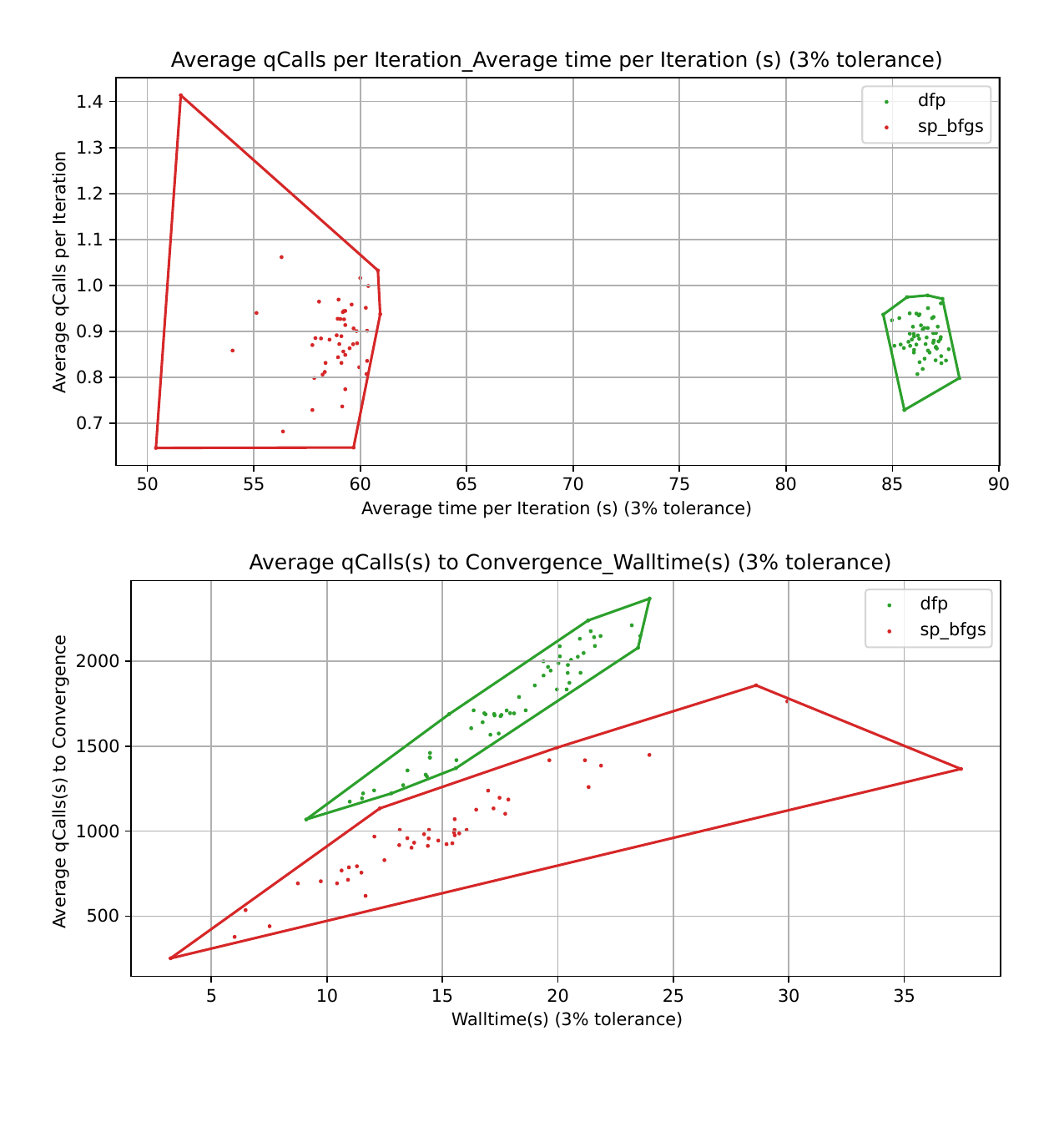}} 
  \caption{3 nodes problem (left) and 5 nodes problem (right), 70 bayesian sweeps on second order methods, stopping condition reached when 3\% tolerance is reached. Objective function are last objective values (averaged over 20 runs), Convergence ratio is the ratio of runs arriving to 3\% tolerance (20 runs), walltime (s) is measured from call of first iteration to stopping condition reached.}
  \label{fig:cost_3_5_second_order}
\end{figure*}

\paragraph*{Cost analysis}
The cost analysis is done through a similar procedure, with added a with a 3\% stopping condition, hence measuring cost metrics to convergence. To reach tolerance, we observe validating trends concerning SP-BFGS capabilities in finding quality solutions for less resources in time and QPU quieries in Fig.~\ref{fig:cost_3_5_second_order}. Those trends are validated by scaling further the tests up to 8 nodes in App.~\ref{app:all_benchmarks}. I normal use, observations on poor convergence quality for SR1, NCG and BFGS from 3 nodes problem to 5 nodes problem in Fig.~\ref{fig:cost_3_5_second_order} shows complete failure in finding optimal solutions. On the other hand, SP-BFGS provides significatively better cost per iteration than even DFP. This is not only true for overall walltime, but also the average quantum calls to convergence still in Fig.~\ref{fig:cost_3_5_second_order}.

The improvement in generalization and cost for SP-BFGS is likely attributable to the secant-penalization mechanism, which updates the Hessian matrix only when sufficient information is gathered from sampling the objective function. The cost advantage from DFP to SP-BFGS appears to increase as the problem dimension grows; this can be confirmed by examining Fig.~\ref{fig:centroid_improvement}, which illustrates the differences in centroid distances and time per iteration (to convergence) between DFP and SP-BFGS. Overall, these figures indicate that DFP is the best-performing quasi-Newton method; however, there are hyperparameter configurations in which SP-BFGS significantly outperforms DFP in terms of execution time.

Introducing the secant-penalization allows for a significantly lower cost while keeping useful convergence ratio, whose average improvements increases with the problem size~Fig.\ref{fig:centroid_improvement}

At convergence, qBang qBang is significantly more cost-effective in terms of quantum calls compared to QNG methods, resulting in a drastic reduction in average time per iteration. The introducton of momentum in QNG pushes this convergence speed from QNG (block or diagonal approximation of the QFIM) to the same convergence speed than qBang. Note that convergence speed per iteration allows to compare overall behavior: QNG provide stronger convergence, but at the cost of the block matrix QFIM estimation, which qBroyden and qBang are avoiding. This result however do show experimentally that qBang approximates the QFIM through a learned parametric filter over the optimization history. At scale however to 8 nodes, QNG block and diagonal only remain capable of producing 3\% solutions App.~\ref{app:all_benchmarks}.

\begin{figure*}[htbp]
  \centering
  \includegraphics[scale=0.35]{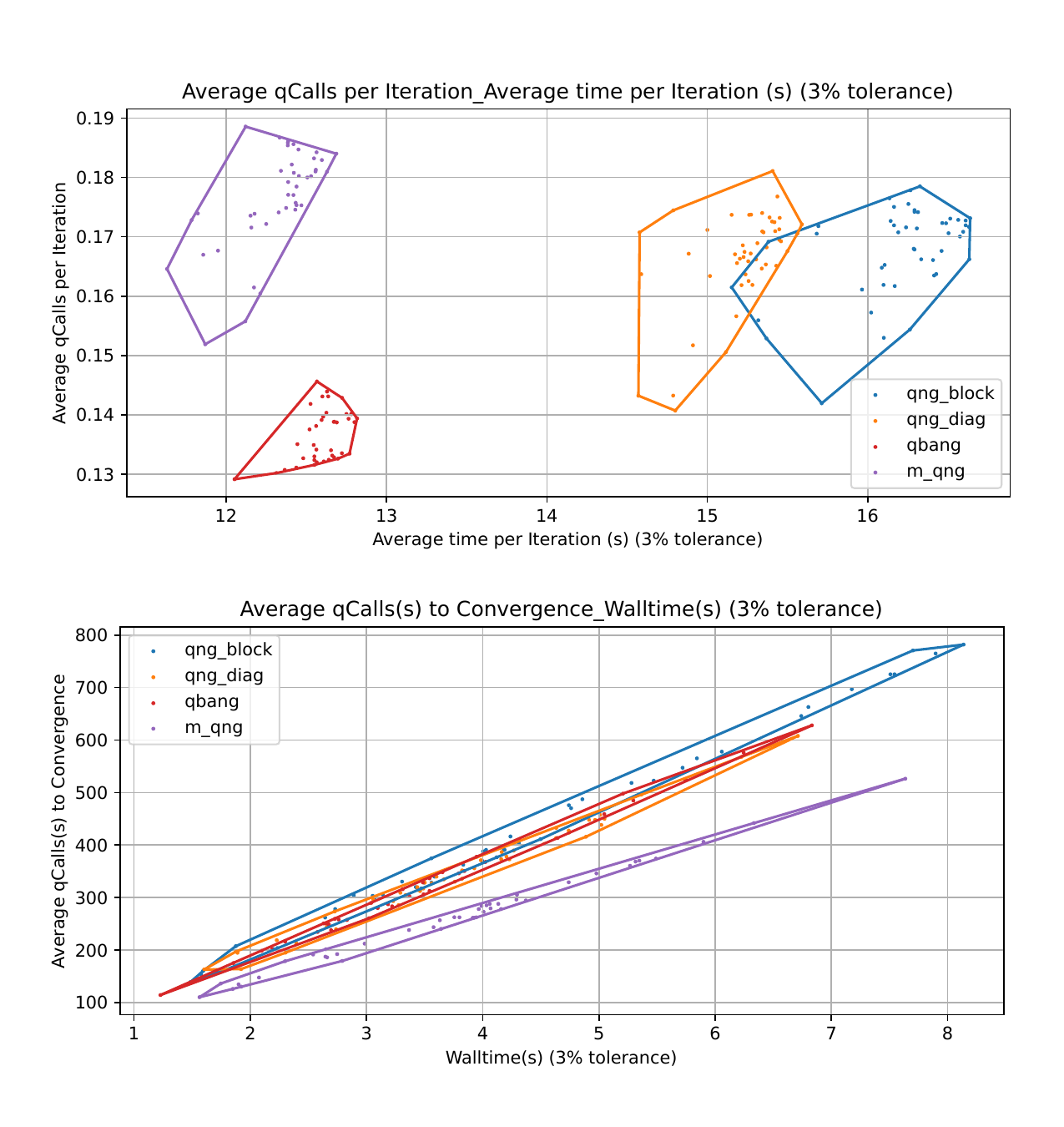}
  \raisebox{-0.15cm}{\includegraphics[scale=0.35]{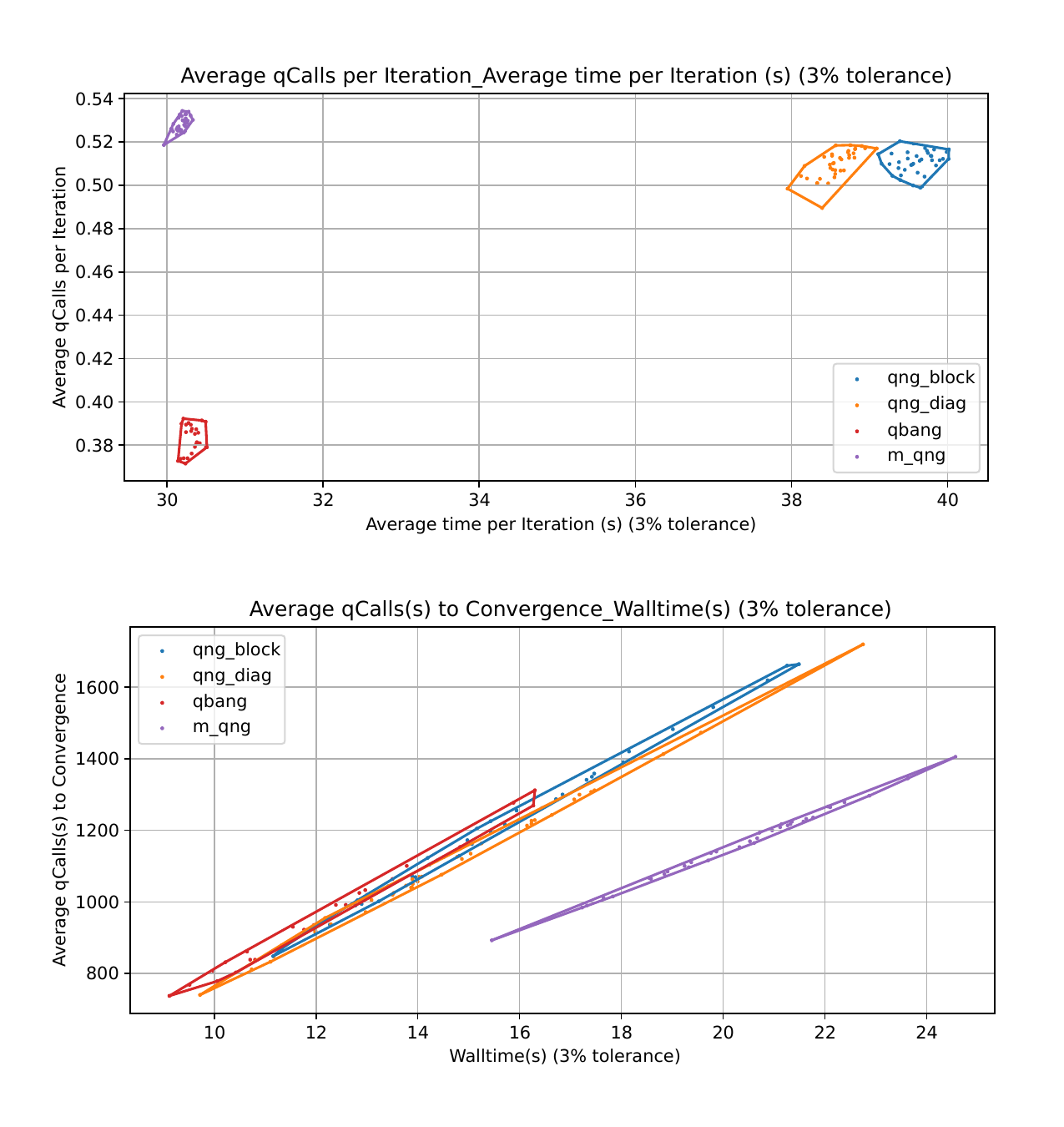}} 
  \caption{3 nodes problem (left) and 5 nodes problem (right), 70 bayesian sweeps on natural gradient methods, stopping condition reached when 3\% tolerance is reached. Objective function are last objective values (averaged over 20 runs), Convergence ratio is the ratio of runs arriving to 3\% tolerance (20 runs), walltime (s) is measured from call of first iteration to stopping condition reached.}
  \label{fig:cost_3_5_qng}
\end{figure*}

\subsection{Second order information in stochastic methods}
\begin{figure*}[htbp]
  \centering
  \includegraphics[scale=0.35]{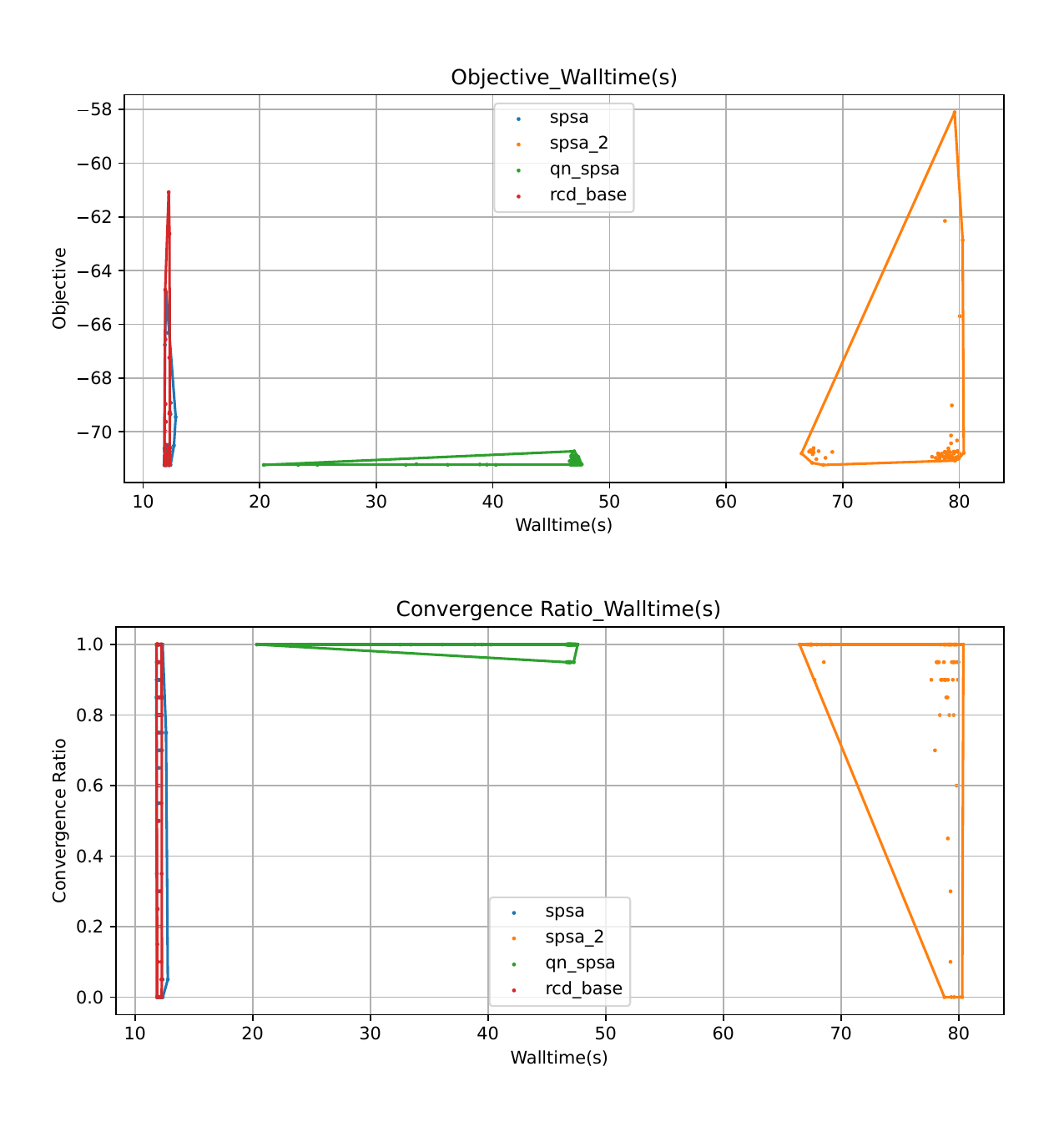}
  \raisebox{-0.45cm}{\includegraphics[scale=0.36]{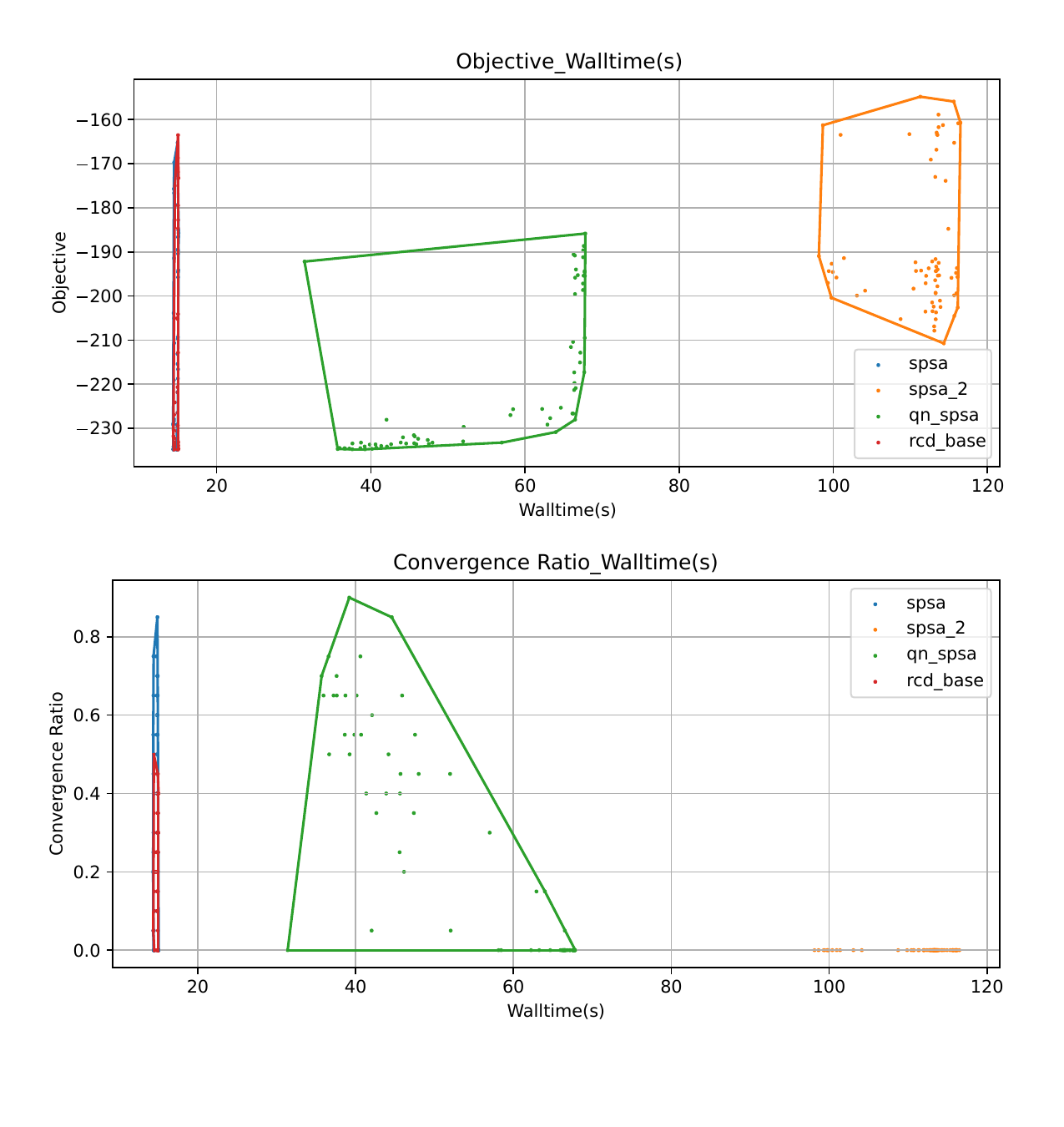}} 
  \caption{3 nodes problem (left) and 5 nodes problem (right), 70 bayesian sweeps on stochastic methods, stopping condition reached when 400 max iteration is reached. Objective function are last objective values (averaged over 20 runs), Convergence ratio is the ratio of runs arriving to 3\% tolerance (20 runs), walltime (s) is measured from call of first iteration to stopping condition reached.}
  \label{fig:cost_3_5_stochastic}
\end{figure*}

Introducing second-order estimation in SPSA necessitates additional hyperparameters that must be fine-tuned for optimal performance. The quality of the second-order information estimation is contingent upon these parameters, which must either be pre-tuned or leveraged from past experiences.Overall, we see the first advantage of first order stochastic methods is to provide solutions at constant speed in ~Fig.\ref{fig:cost_3_5_stochastic} again from 3 nodes to 5 nodes, which shows increase difficulty in fine-tuning parameters. QNSPSA however provides an improvement on SPSA2 at 400 iterations, at the cost of more QPU calls~\ref{app:all_benchmarks}.

Overall, QNSPSA significantly outperforms SPSA2, particularly due to its advantage of having fewer hyperparameters to tune, which allows for more efficient use of second-order information compared to SPSA2. However, for the chosen parameters, second-order methods such as SPSA and RCD, especially SPSA, still provide the last non-zero convergence ratios for larger problems, as detailed in App.~\ref{app:all_benchmarks}. In practice, both QNSPSA and SPSA2 should only be utilized when the statistical error on samples can be minimized, for instance, through high shot counts, to facilitate an unbiased estimation of the Hessian or QFIM.

\section{Conclusion}
\label{sec:Conclusion}
In this paper, we demonstrated that the choice of preconditioner in second-order methods significantly influences optimization outcomes in shallow QAOA ansatz. The complex structure of the optimization landscape of the PQC objective function led us to conclude that SR1, BFGS, and NCG are unsuitable for these types of problems. In contrast, the DFP method exhibits the best performance in this context.

Our contribution introduces a simple secant-penalization rule that stabilizes BFGS when confronted with non-convexities and noisy gradient sampling. Notably, SP-BFGS is less computationally expensive in terms of QPU calls and overall computational time compared to DFP, which we identified as the best-behaved quasi-Newton preconditioner. Furthermore, the tuning of the additional phenomenological parameters $(N_0, N_1)$ does not present significant challenges.

Additionally, we found that for small dimensions, QNG-type methods provide much faster convergence than second-order methods, albeit at a higher cost due to QFIM estimation. Although more economical methods like qBang and qBroyden still offer faster convergence than quasi-Newton methods, their QFIM approximations—based on a learned mechanism over the optimization history—tend to degrade as the complexity of the landscape increases.

Finally, incorporating second-order information in stochastic problems necessitates careful tuning of hyperparameters, without guaranteeing improvements over first-order methods such as SPSA or RCD. While QNSPSA outperforms SPSA2, it is likely to require a high shot count to prevent the QFIM statistical estimator from becoming counterproductive.

Future research should focus on integrating a secant-penalization rule into the QFIM estimation of qBang to bypass uninformative gradient measurements during optimization and avoid cumulative errors. Additionally, further investigation into the penalization rule within SP-BFGS is warranted. We demonstrated that a simple linear model with an interception rule was sufficient to enhance preconditioner behavior; however, more sophisticated rules, problem-specific schedules, phenomenological models, or machine learning approaches could be introduced in Eq.~\ref{eq:mitigation_interception}, with their formulation depending on the uniform gradient bounded noise hypothesis.

\section{Acknowledgements}
This research was made possible by the financing of the BMWK-Project »EniQmA« for the Systematic Development of Hybrid Quantum Computing Applications.

\newpage
\subsection{References and footnotes}
\bibliographystyle{unsrt}


\appendix
\onecolumn
\section{Algorithmic complement}
\label{sec:algorithm_complement}
\begin{algorithm}[H]
    \begin{algorithmic}[1]
        \STATE \textbf{Input:} $f$, $\beta\_compute\_linear$, $x_\text{init}$, $\alpha$, $\beta_{reduce}$, $c_0$, $c_1$, $N_0$, $N_s$, MAXIT
        \STATE \textbf{Output:} $x$, best\_value

        \STATE $x \gets x_{\text{init}}$
        \STATE $i \gets 0$
        \STATE $H_i \gets \text{Identity Matrix}$

        \WHILE{$i < \text{MAXIT}$}
            \STATE $\nabla_i \gets \nabla_\theta f(x)$
            \IF{$\lVert \nabla_i \rVert < \text{tol}$}
                \STATE \textbf{break}
            \ENDIF

            \STATE $p_i \gets -H_i \cdot \nabla_i$

            \WHILE{not satisfying Armijo-Wolfe condition}
                \IF{$f(x + \alpha p_i) \leq f(x) + c_0 \alpha \cdot \nabla_i^T p_i$}
                    \STATE \textbf{break}
                \ENDIF
                \IF{$-p_i^T \nabla_\theta f(x + \alpha p_i) \leq -c_1 p_i^T \nabla_i$}
                    \STATE \textbf{break}
                \ENDIF
                \STATE $\alpha \gets \beta_{reduce} \cdot \alpha$
            \ENDWHILE

            \STATE $x_i \gets x + \alpha p_i$
            \STATE $s_i \gets x_i - x$
            \STATE $\nabla_{i+} \gets \nabla f(x_i)$
            \STATE $y_i \gets \nabla_{i+} - \nabla_i$

            \STATE $\beta_i \gets \beta\_compute\_linear(N_0, N_s, s_i)$
            \STATE proj $\gets s_i^T \cdot y_i$
            \STATE $\gamma_i \gets \frac{1}{proj + \frac{1}{\beta_i}}, \quad \omega_i \gets \frac{1}{proj + \frac{2}{\beta_i}}$
            \STATE $H_{ns} \gets (I - \omega_i (s_i \odot y_i)) \cdot H_i (I - \omega_i (y_i \odot s_i))$
            \STATE $H_{nc} \gets \omega_i \cdot \left(\frac{\gamma_i}{\omega_i} + (\gamma_i - \omega_i) \cdot y_i^T \cdot H_i y_i\right) \cdot s_i s_i^T$
            \STATE $H_n \gets H_{ns} + H_{nc}$

            \STATE $x \gets x_i$
            \STATE $H_i \gets H_n$
            \STATE $i \gets i + 1$
        \ENDWHILE

        \RETURN {$x$}
    \end{algorithmic}
    \label{alg:sp_bfgs}
    \caption{SP-BFGS implementation, integration with the Amijo-Wolf backtracking implementation. \textit{compute\_linear} function is the expression in adapting $\beta_i$ as a function of the measured gradient at iteration $i$, for example with interception or noise model in~\ref{eq:mitigation_interception}}
\end{algorithm}

\section{Problem definition}
\label{sec:problem_def}

\hfill
\begin{figure}[H]
    \centering
    \begin{minipage}{0.3\textwidth}
    \centering
    \includegraphics[width=\linewidth]{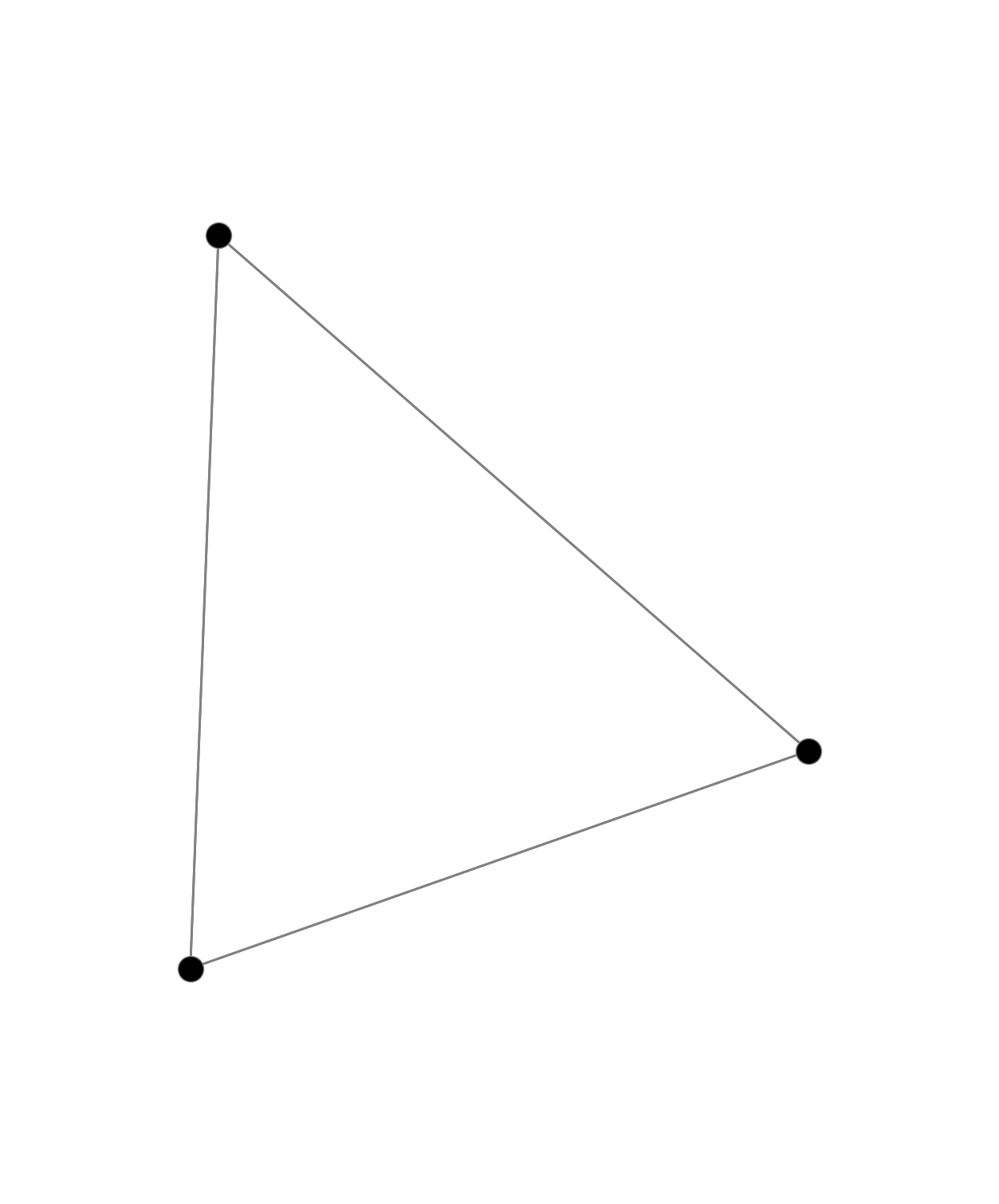}
    \caption{3 nodes, ground solution to MaxCut: (-71.23)}
    \label{fig:3_nodes_problem}
    \end{minipage}%
    \hfill
    \begin{minipage}{0.3\textwidth}
    \centering
    \includegraphics[width=\linewidth]{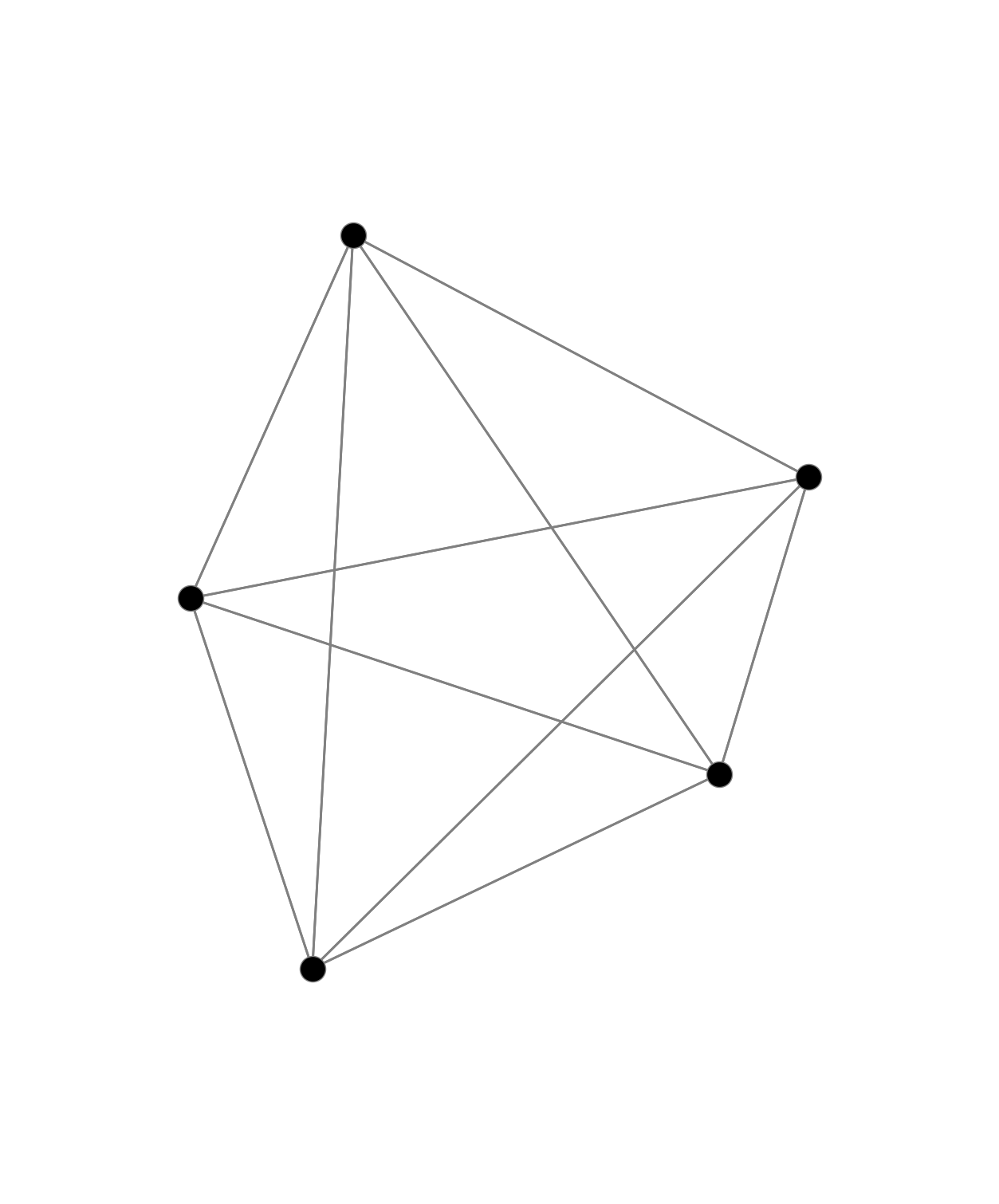}
    \caption{5 nodes, ground solution to MaxCut: (-234.76)}
    \label{fig:5_nodes_problem}
    \end{minipage}%
    \hfill
    \begin{minipage}{0.3\textwidth}
    \centering
    \includegraphics[width=\linewidth]{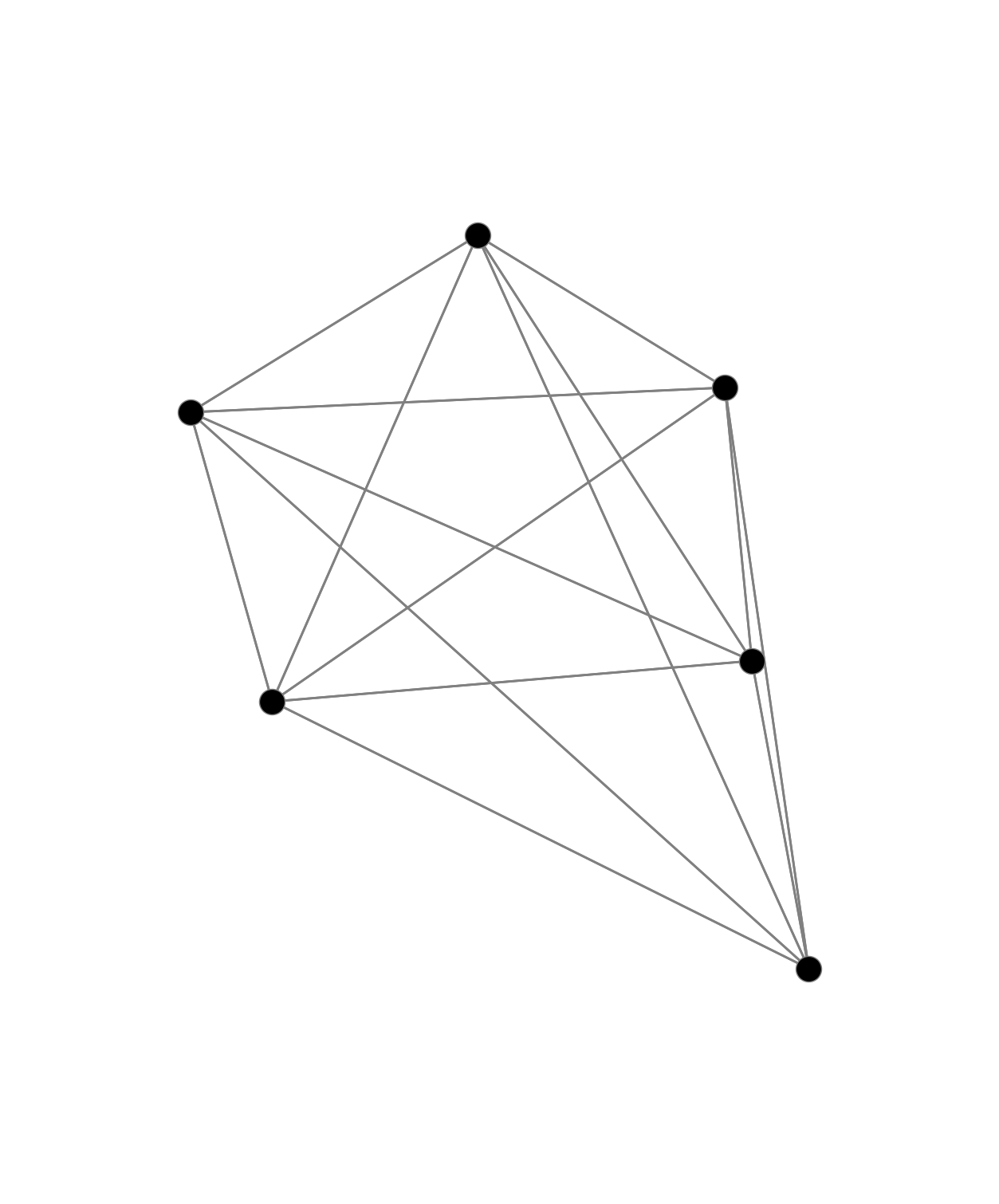}
    \label{fig:6_nodes_problem}
    \caption{6 nodes, ground solution to MaxCut: (-293.26)}
    \end{minipage}

    \vspace{1cm}

    \begin{minipage}{0.45\textwidth}
    \centering
    \includegraphics[width=\linewidth]{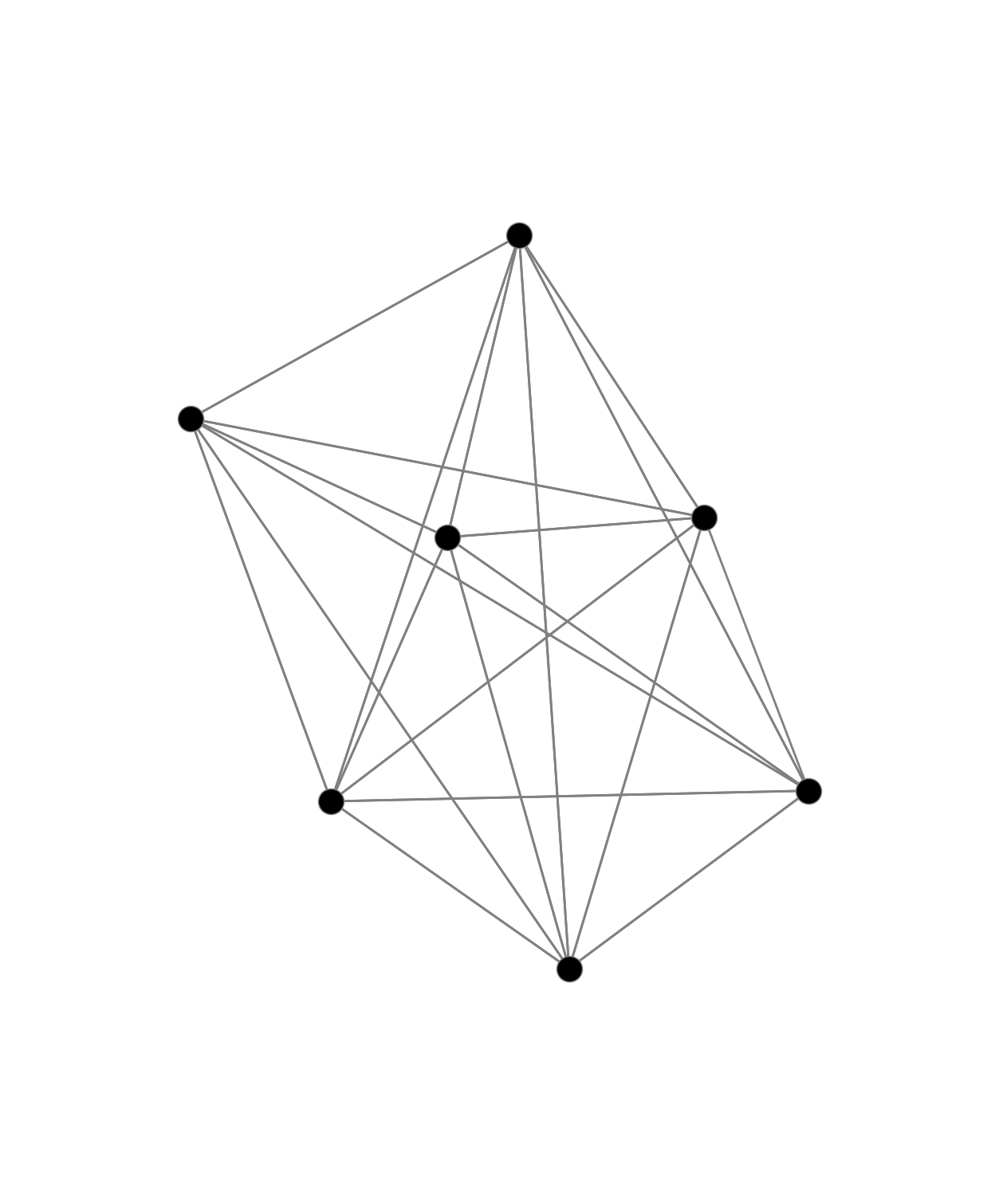}
    \caption{7 nodes, ground solution to MaxCut: (-458.63)}
    \end{minipage}%
    \hfill
    \begin{minipage}{0.45\textwidth}
    \centering
    \includegraphics[width=\linewidth]{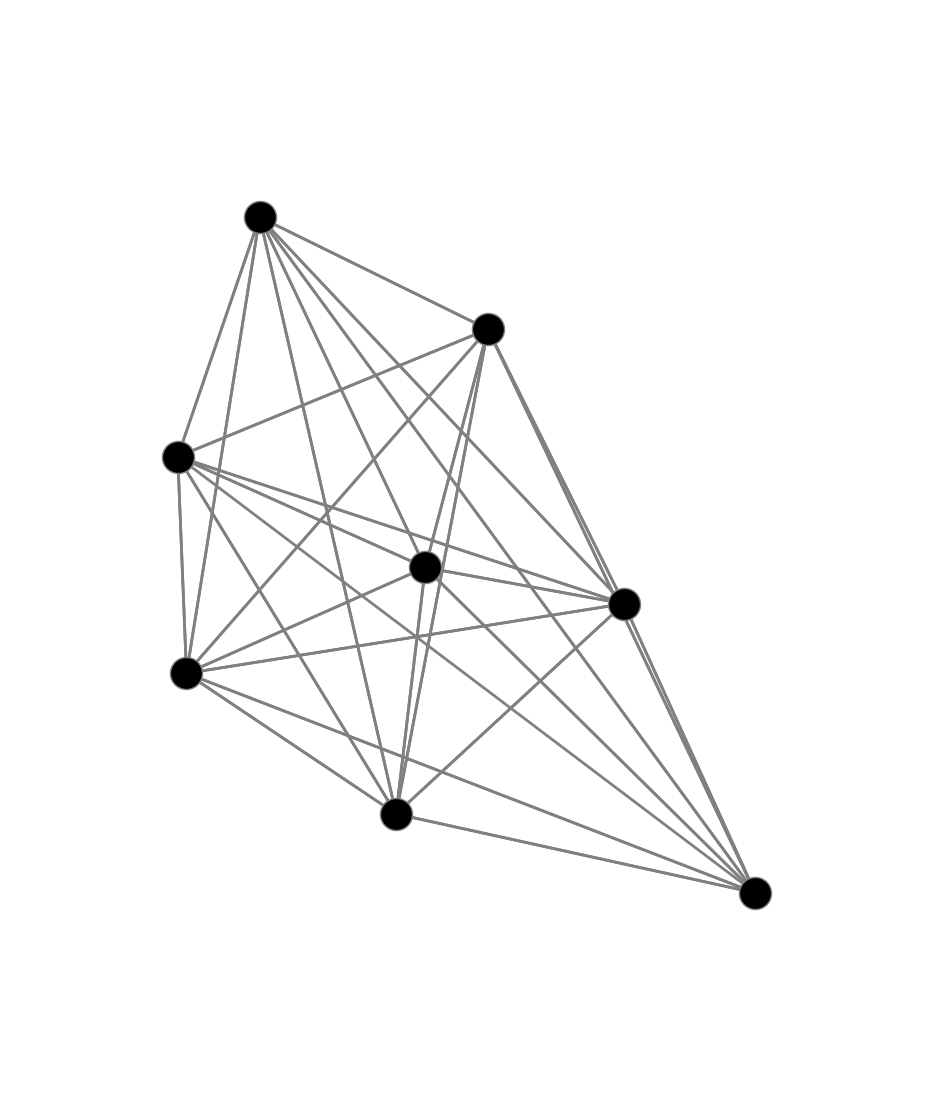}
    \caption{8 nodes, ground solution to MaxCut: (-519.13)}
    \label{fig:8_nodes_problem}
    \end{minipage}%


    \caption{Selected problems for benchmarking}
\end{figure}

\section{Hyperparameters table}
\label{app:default_parameters}

\begin{table}[H]
    \centering
    \small 
    \begin{tabular}{| c | c c c c|}
        \hline
        \textbf{Family} & \textbf{Methods} & \textbf{Params} & \textbf{Search space} & \textbf{Defaults}\\
        \hline
        Quasi-Newton & BFGS & $\alpha$ & (1e-5, 0.99) & 0.70\\
        & & $\beta$ & (0.8, 0.9) & 0.8\\
        & & $c_1$ & (1e-5, 5) & 0.0001\\
        & & $c_2$ & (0.1, 1) & 1.0\\
        \hline
        & DFP & $\alpha$ & (1e-5, 0.99) & 0.37\\
        & & $\beta$ & (0.8, 0.9) & 0.89\\
        & & $c_1$ & (1e-5, 5) & 0.0017\\
        & & $c_2$ & (0.1, 1) & 1.0\\
        \hline
        & N-CG & $\alpha$ & (1e-5, 0.99) & 0.99\\
        & & $\beta$ & (0.8, 0.9) & 0.83\\
        & & $c_1$ & (1e-5, 5) & 0.34\\
        & & $c_2$ & (0.1, 1) & 0.59\\
        \hline
        & SR1 & $\alpha$ & (1e-5, 0.99) & 0.48\\
        & & $\beta$ & (0.8, 0.9) & 0.83\\
        & & $c_1$ & (1e-5, 5) & 1e-5\\
        & & $c_2$ & (0.1, 1) & 1.0\\
        \hline
        & SP-BFGS & $\alpha$ & (1e-5, 0.99) & 0.049\\
        & & $\beta$ & (0.8, 0.9) & 0.82\\
        & & $N_0$ & (1e-5, 1) & 1e-5\\
        & & $N_i$ & (1e-5, 1) & 1e-5\\
        & & $c_1$ & (1e-5, 1) & 1.67e-5\\
        & & $c_2$ & (0.1, 1) & 1.0\\
        \hline
        \hline
        Quantum N. Gradient & QNG (block, diag) & $\alpha$ & (1e-5, 0.99) & 0.0016, 0.0026\\
        \hline
        & qBroyden & $\alpha$ & (1e-5, 0.99) & 0.0088\\
        & & $\epsilon$ & (1e-5, 0.99) & 0.0003\\
        \hline
        & qBang & $\alpha$ & (1e-5, 0.99) & 0.14\\
        & & $\epsilon$ & (1e-5, 0.98) & 5.06e-5\\
        & & $\beta_1$ & (1e-5, 0.99) & 0.0078\\
        & & $\beta_2$ & (1e-5, 0.99) & 0.0001\\
        \hline
        & m-QNG & $\alpha$ & (1e-5, 0.99) & 0.14\\
        & & $\epsilon$ & (1e-5, 0.99) & 5.06e-5\\
        & & $\beta_1$ & (1e-5, 0.99) & 0.0078\\
        & & $\beta_2$ & (1e-5, 0.99) & 0.0001\\
        \hline
    \end{tabular}
    \caption{Selected hyperparameters for Quasi-Newton and Quantum Natural Gradient methods.}
    \label{tab:table_quasi_newton_quantum_gradient}
\end{table}



\subsection{Full benchmark}

\label{app:all_benchmarks}
\subsubsection{Second order quasi-Newton methods}
\label{app:subsection_second_order_benchmarks}
\begin{figure}[H]
    \centering
    \includegraphics[scale=0.4]{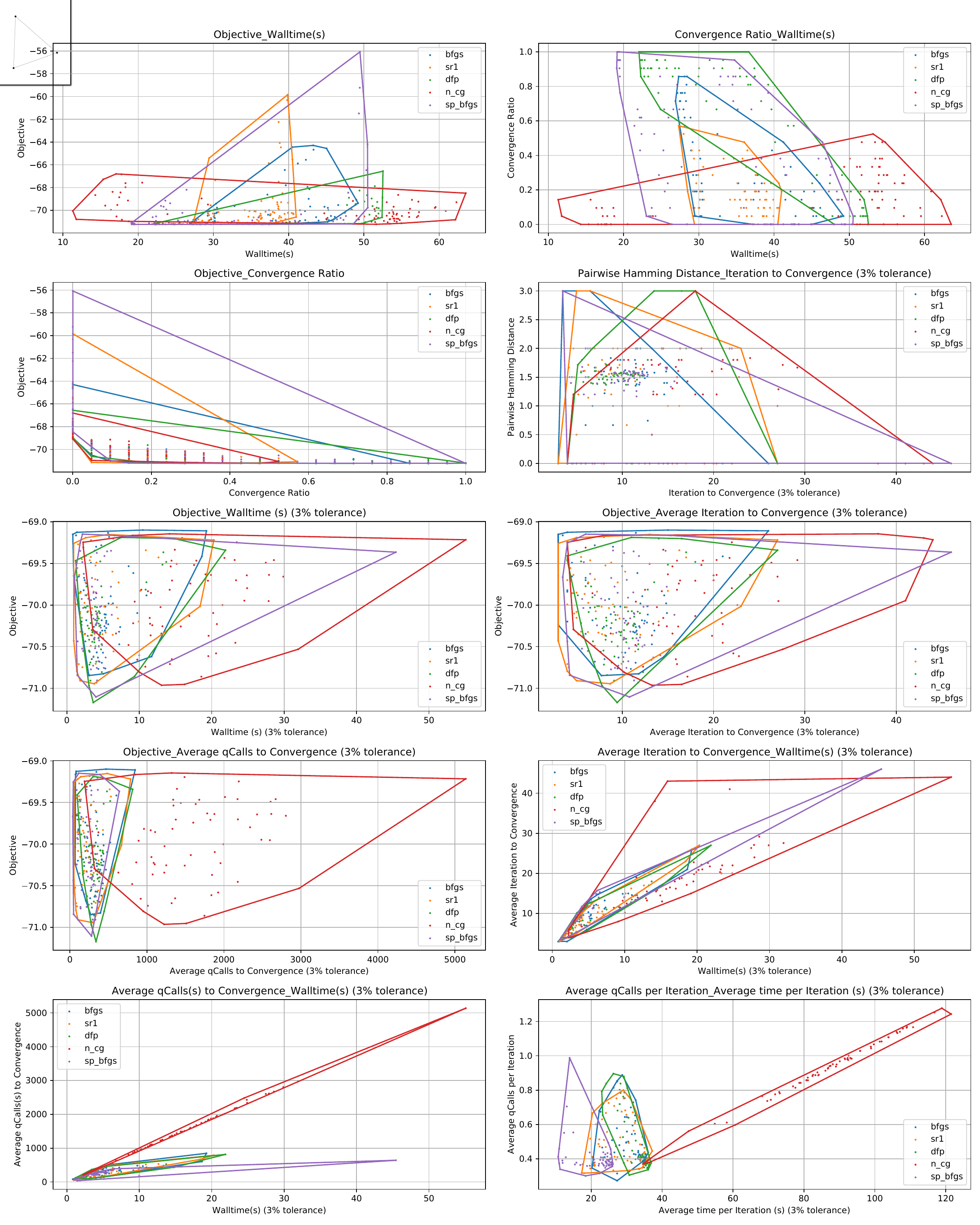}
    \caption{3 nodes problems, 50 iterations, second order methods noiseless, 70 Bayesian sweeps}
\end{figure}

\begin{figure}[H]
    \centering
    \includegraphics[scale=0.4]{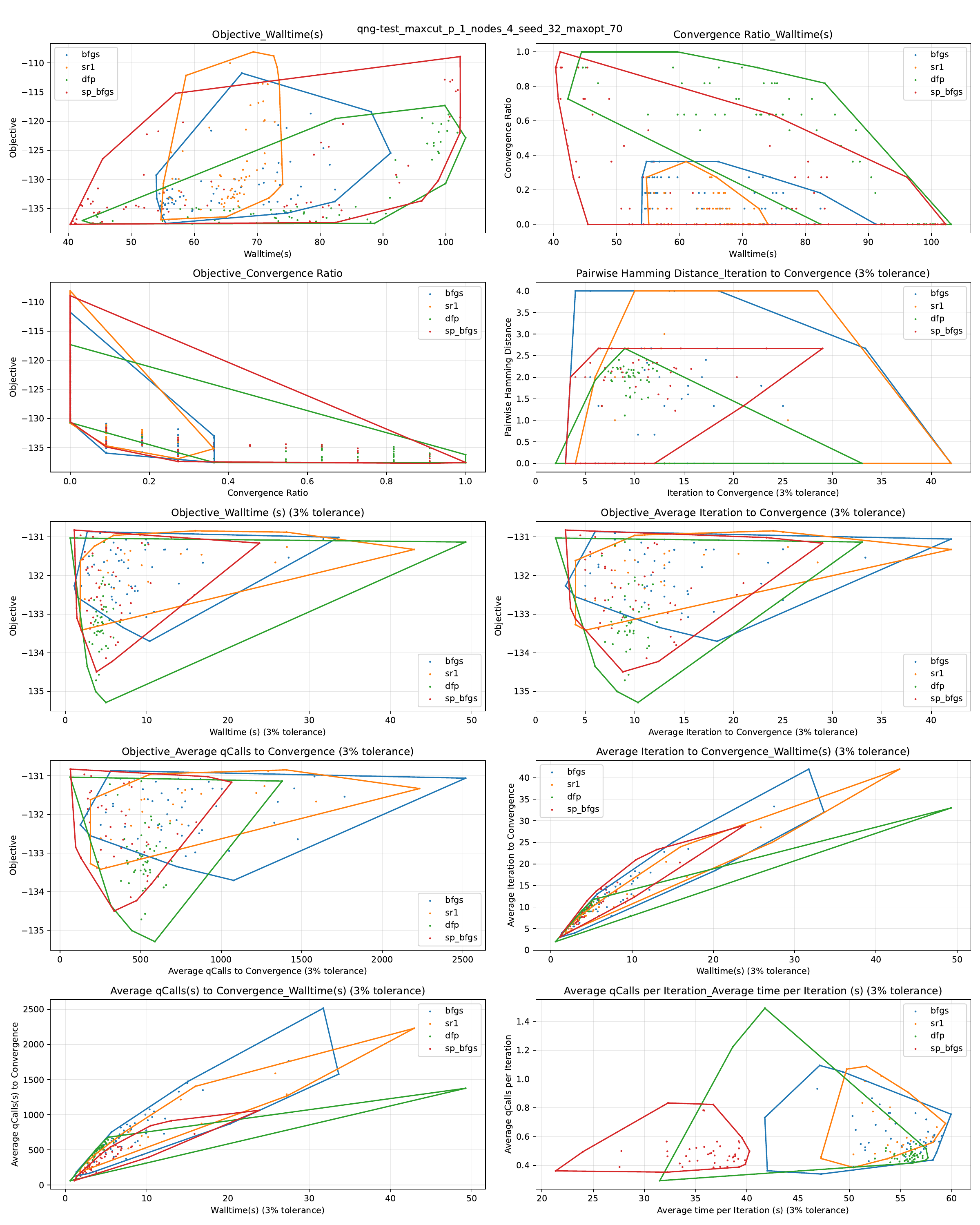}
    \caption{4 nodes problems, 70 iterations, second order methods noiseless, 70 Bayesian sweeps}
\end{figure}

\begin{figure}[H]
    \centering
    \includegraphics[scale=0.4]{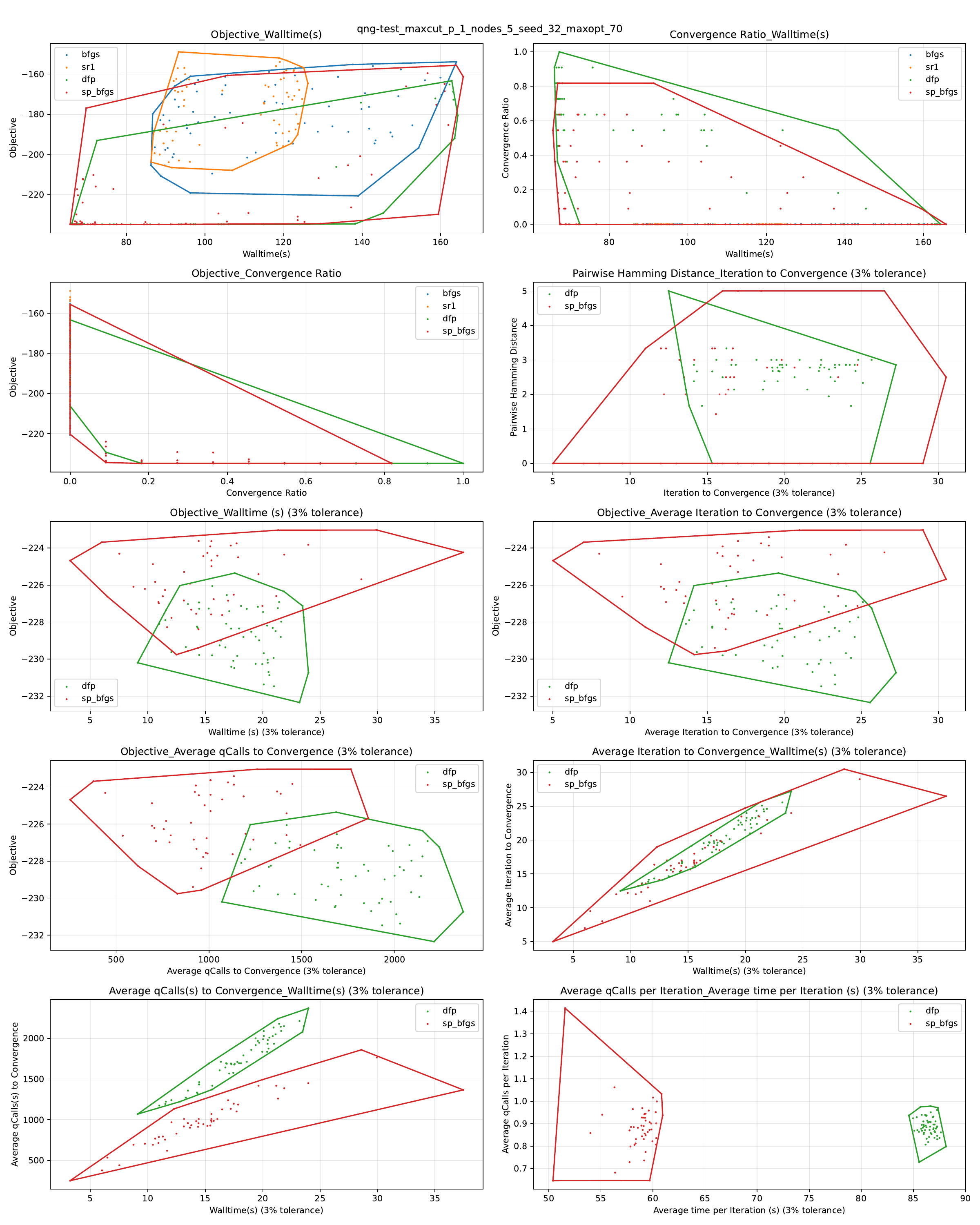}
    \caption{5 nodes problems, 70 iterations, second order methods noiseless, 70 Bayesian sweeps}
\end{figure}

\begin{figure}[H]
    \centering
    \includegraphics[scale=0.4]{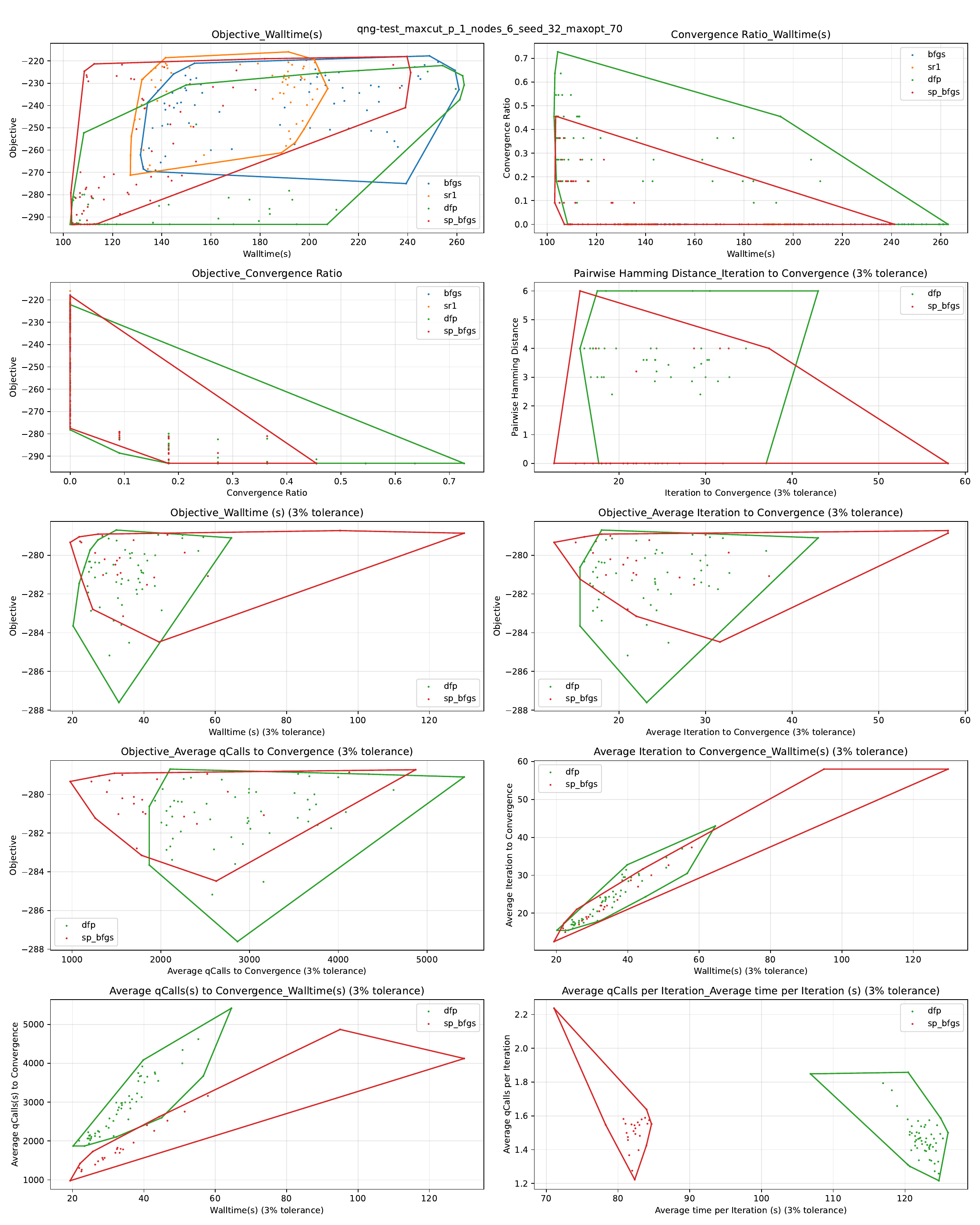}
    \caption{6 nodes problems, 70 iterations, second order methods noiseless, 70 Bayesian sweeps}
\end{figure}

\begin{figure}[H]
    \centering
    \includegraphics[scale=0.4]{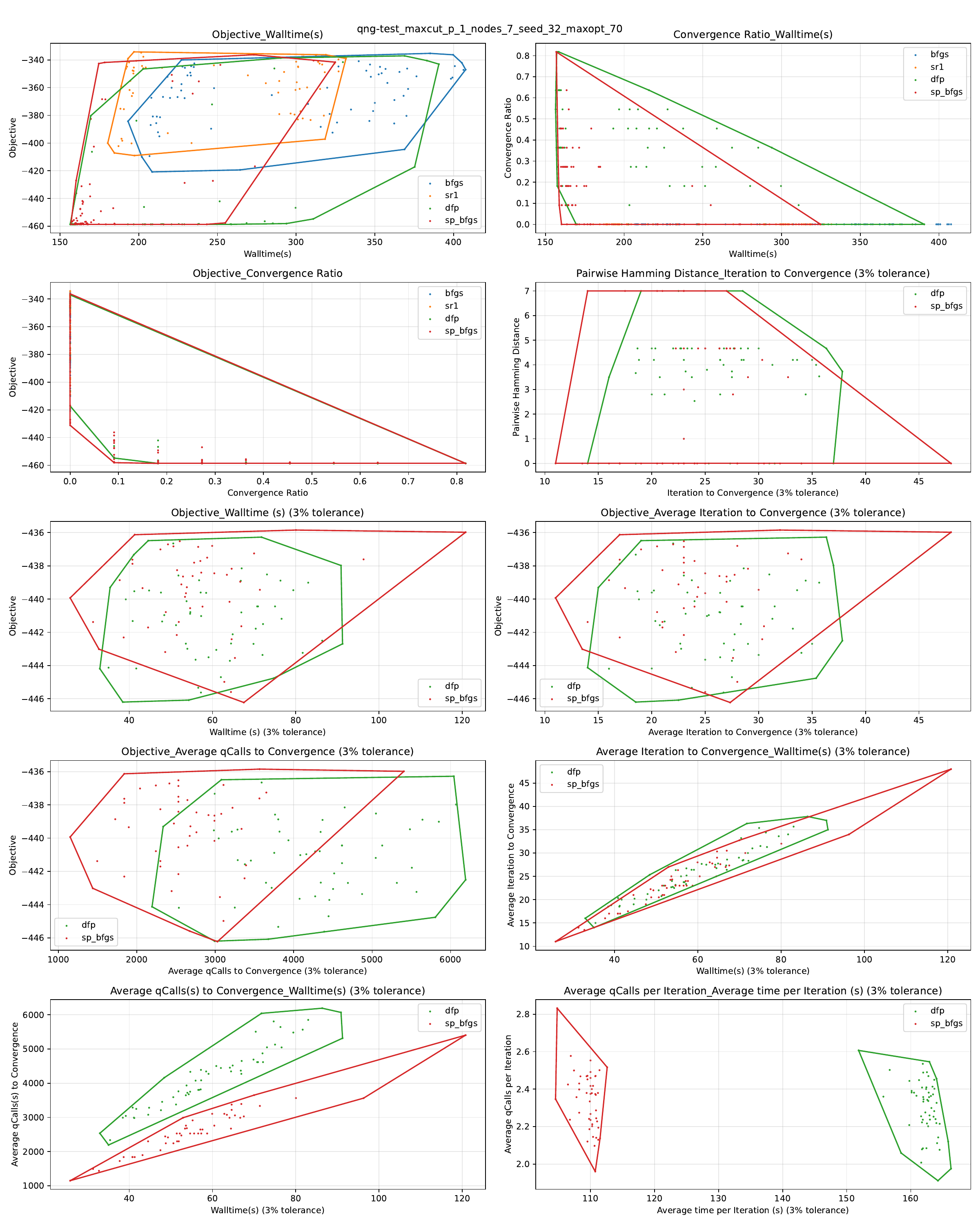}
    \caption{7 nodes problems, 70 iterations, second order methods noiseless, 70 Bayesian sweeps}
\end{figure}

\begin{figure}[H]
    \centering
    \includegraphics[scale=0.4]{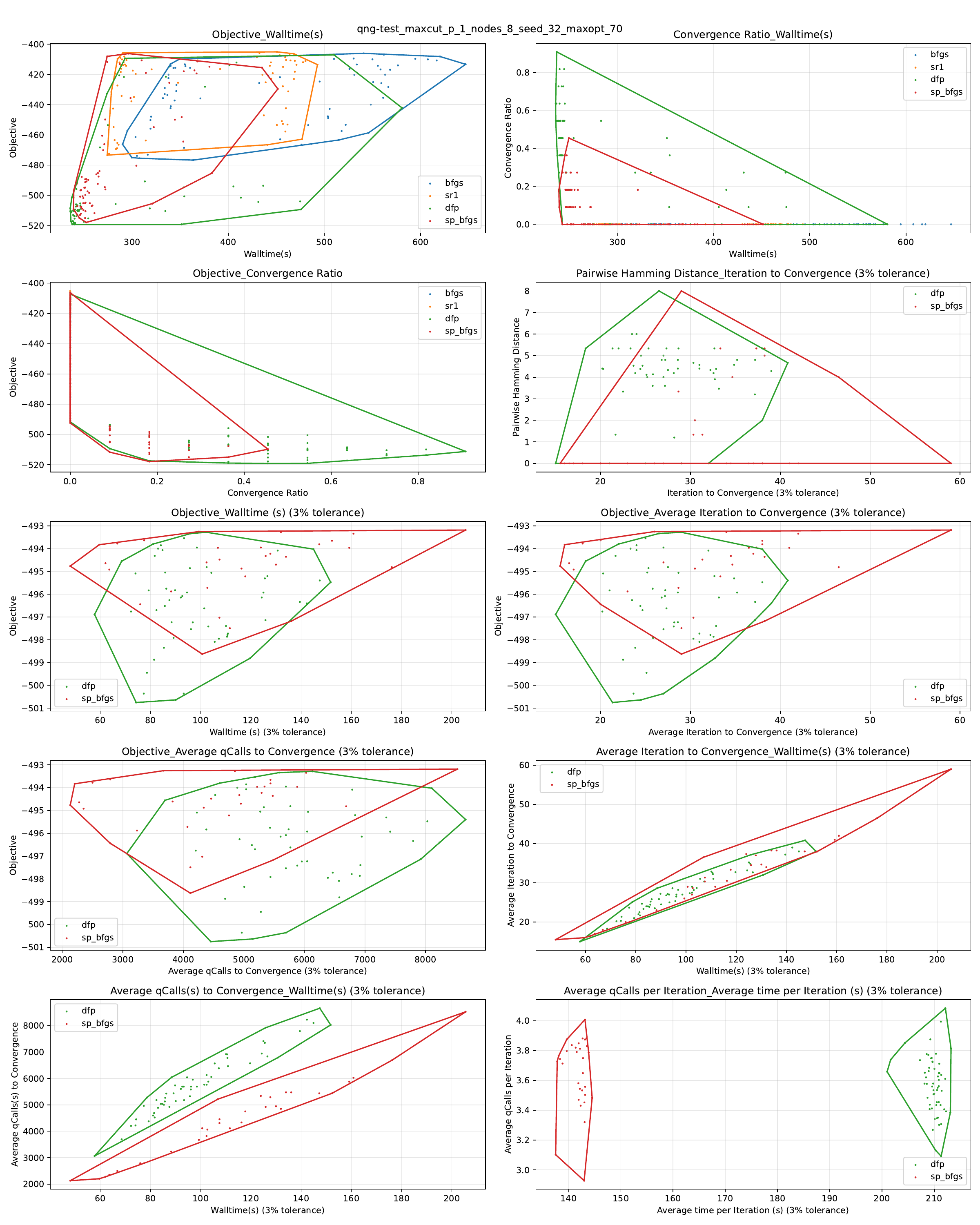}
    \caption{8 nodes problems, 70 iterations, second order methods noiseless, 70 Bayesian sweeps}
    \label{fig:8_nodes_performances}
\end{figure}

\subsubsection{Natural gradient methods}
\begin{figure}[H]
    \centering
    \includegraphics[scale=0.4]{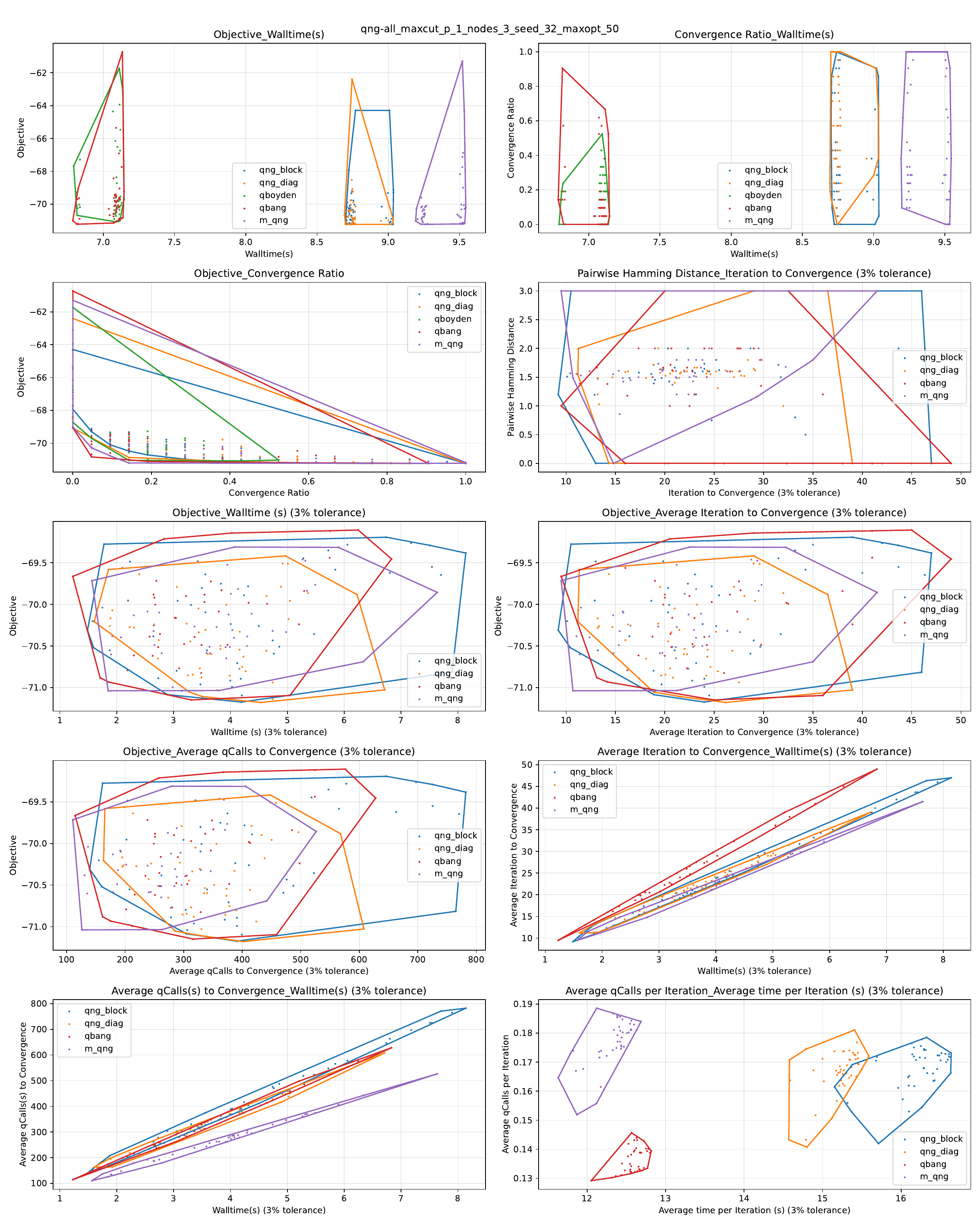}
    \caption{3 nodes problems, 50 iterations, qng-type noiseless, 70 Bayesian runs}
\end{figure}

\begin{figure}[H]
    \centering
    \includegraphics[scale=0.4]{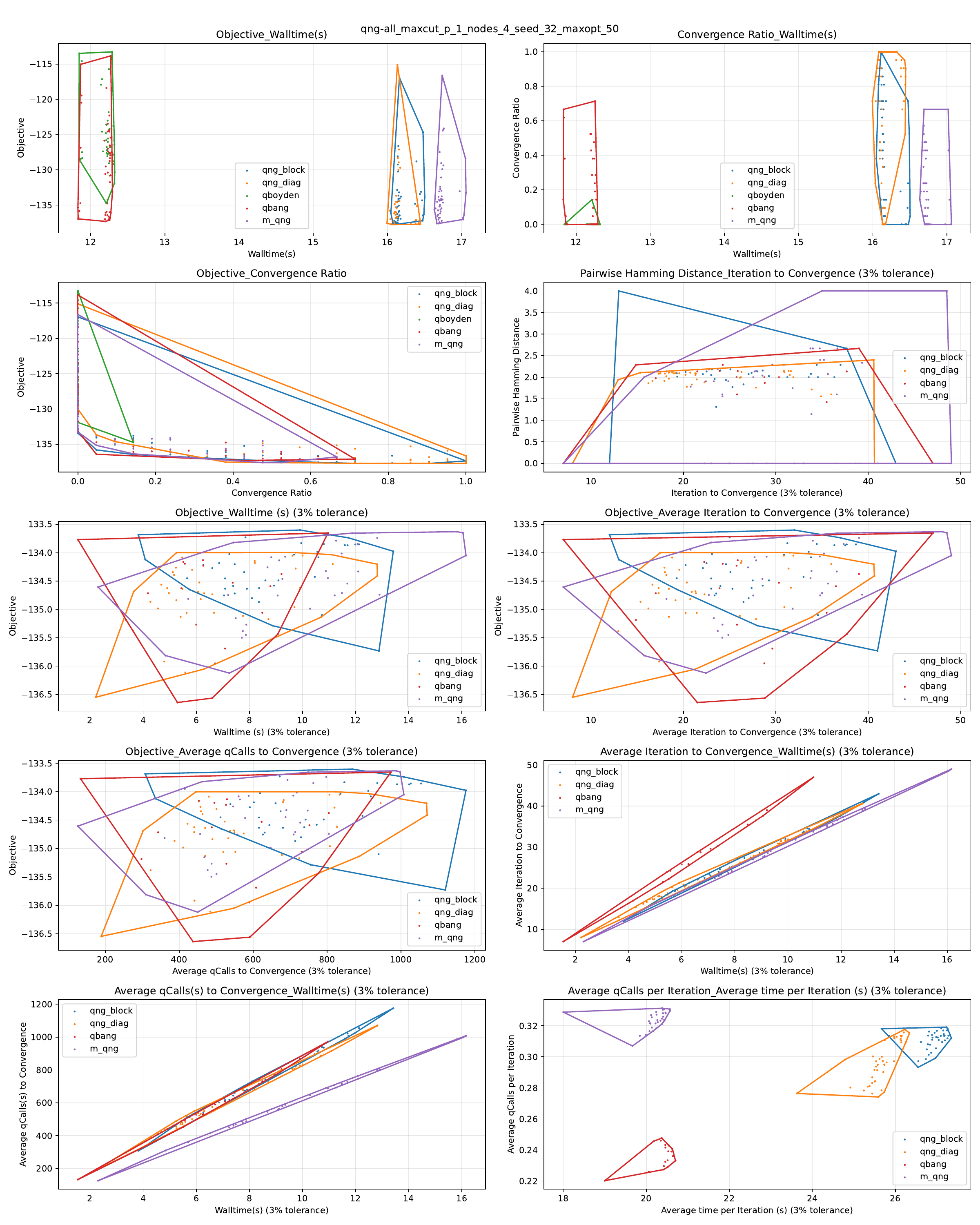}
    \caption{4 nodes problems, 50 iterations, qng-type noiseless, 70 Bayesian runs}
\end{figure}

\begin{figure}[H]
    \centering
    \includegraphics[scale=0.4]{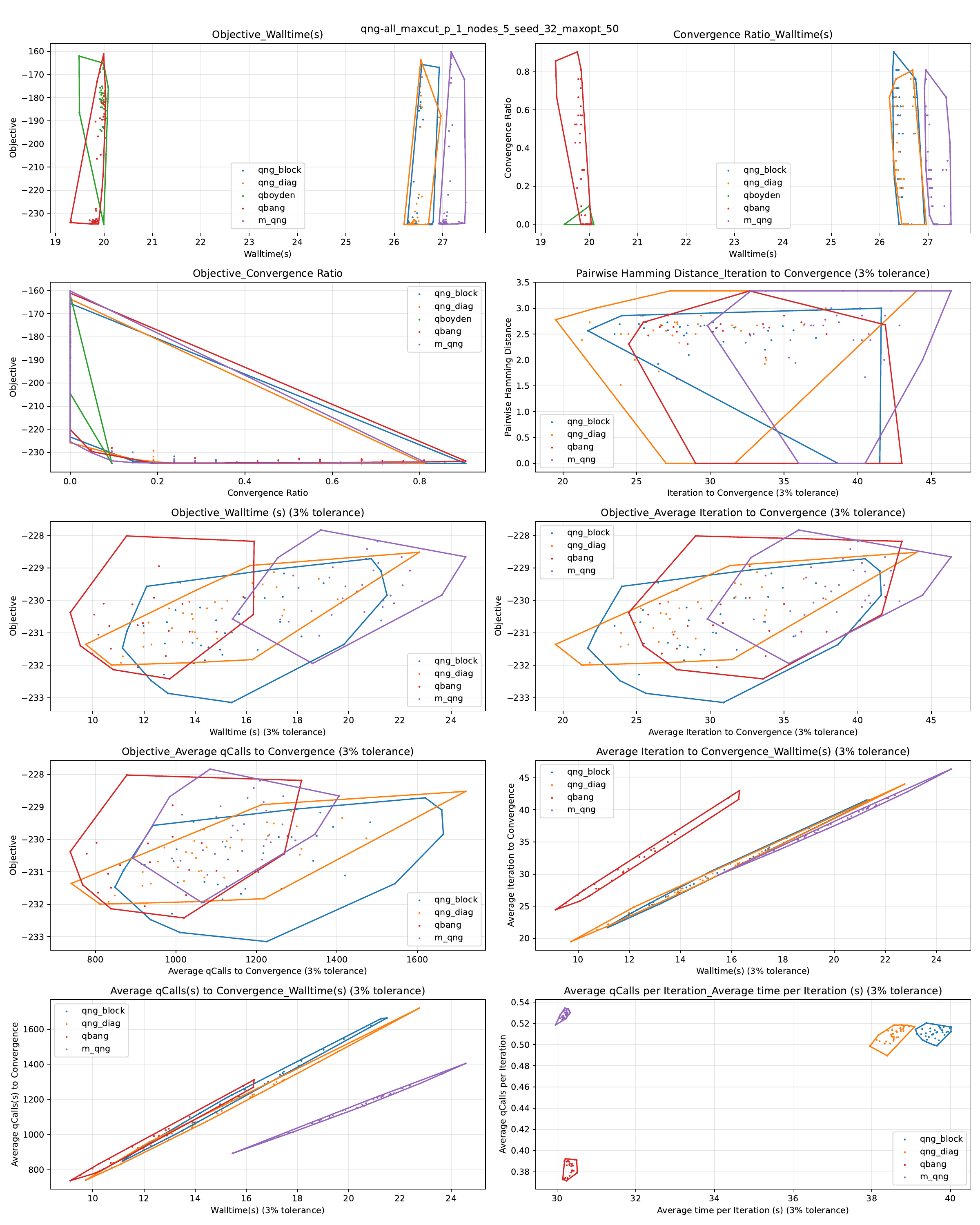}
    \caption{5 nodes problems, 50 iterations, qng-type noiseless, 70 Bayesian runs}
\end{figure}

\begin{figure}[H]
    \centering
    \includegraphics[scale=0.4]{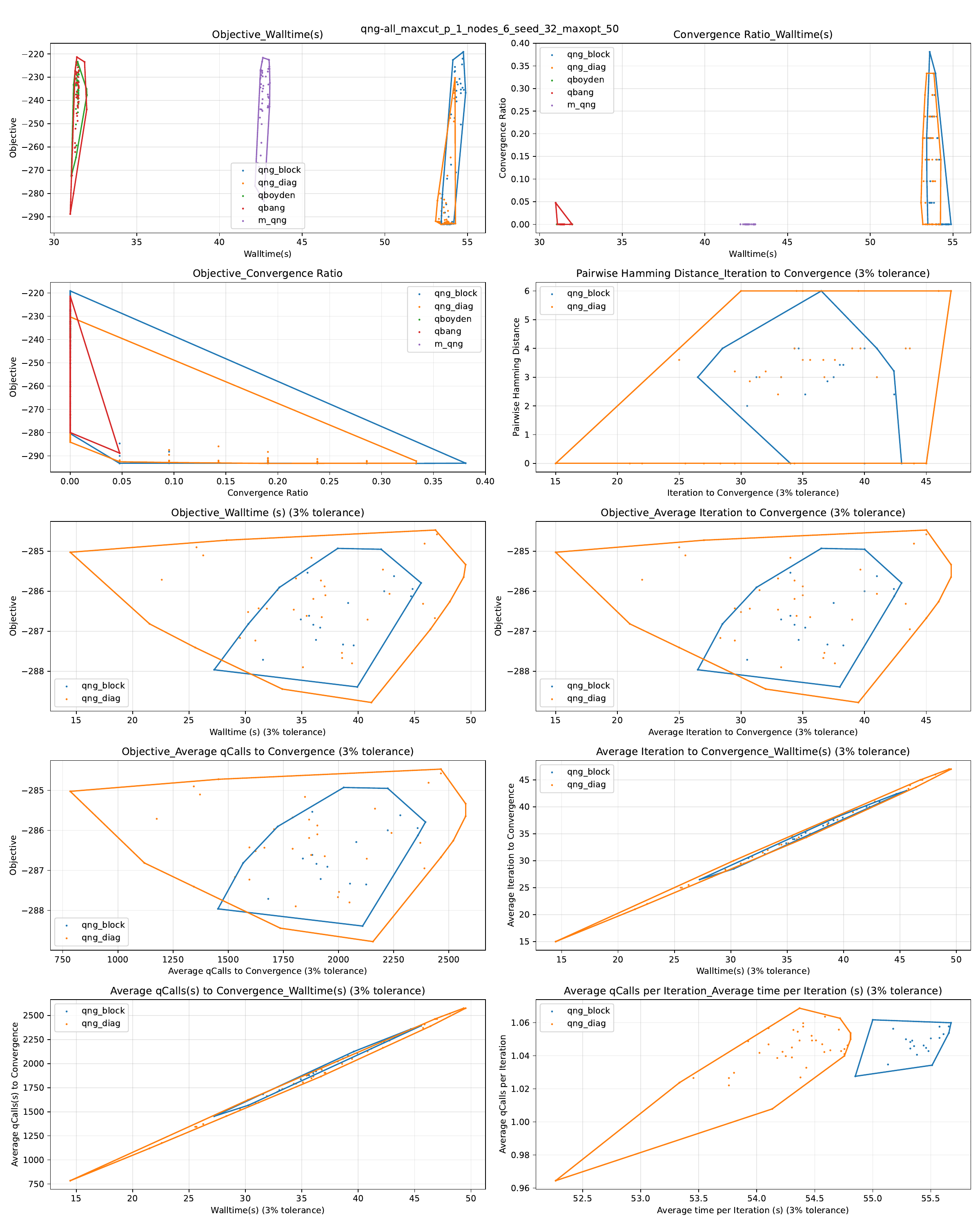}
    \caption{6 nodes problems, 50 iterations, qng-type noiseless, 70 Bayesian runs}
\end{figure}

\begin{figure}[H]
    \centering
    \includegraphics[scale=0.4]{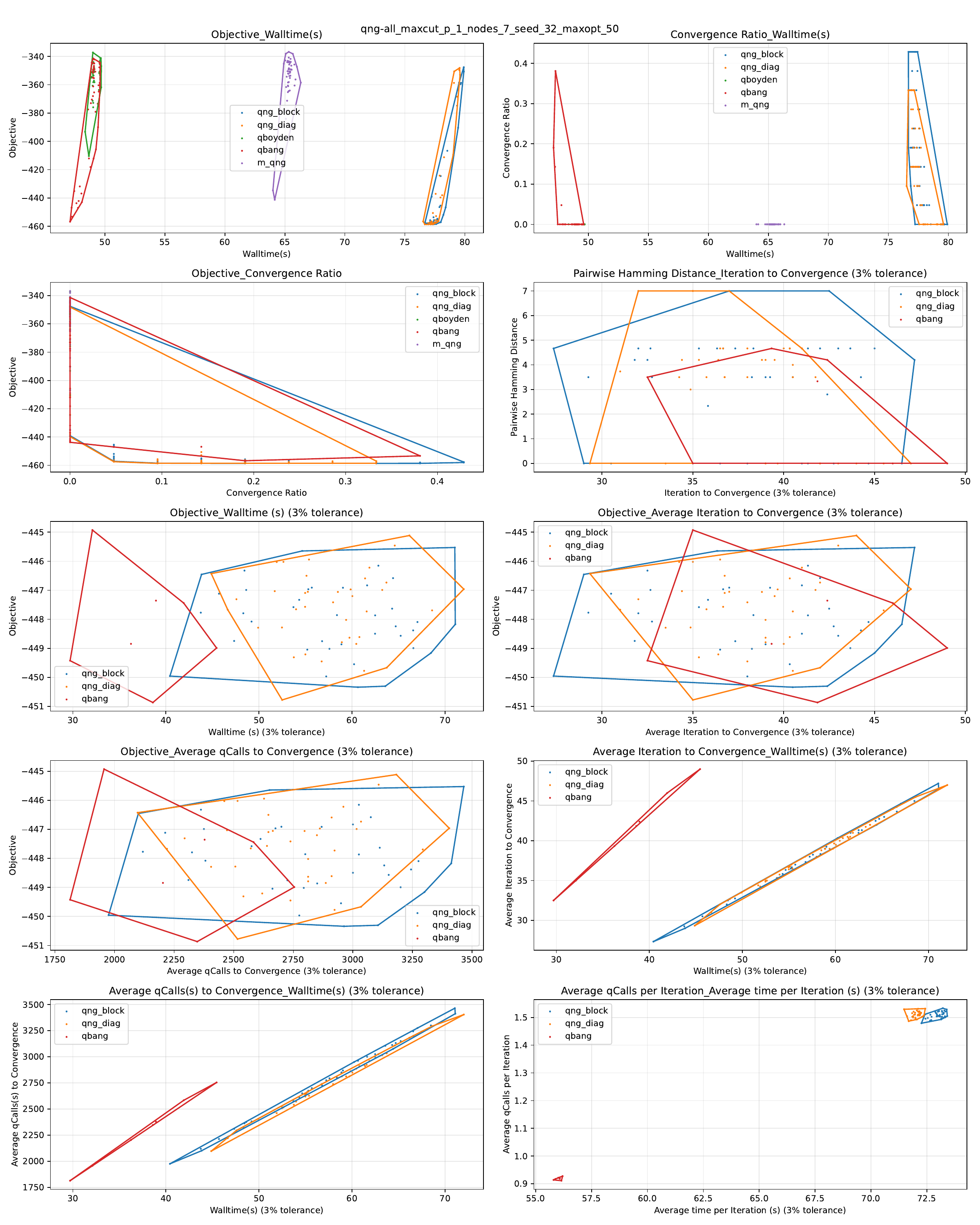}
    \caption{7 nodes problems, 50 iterations, qng-type noiseless, 70 Bayesian runs}
\end{figure}

\begin{figure}[H]
    \centering
    \includegraphics[scale=0.4]{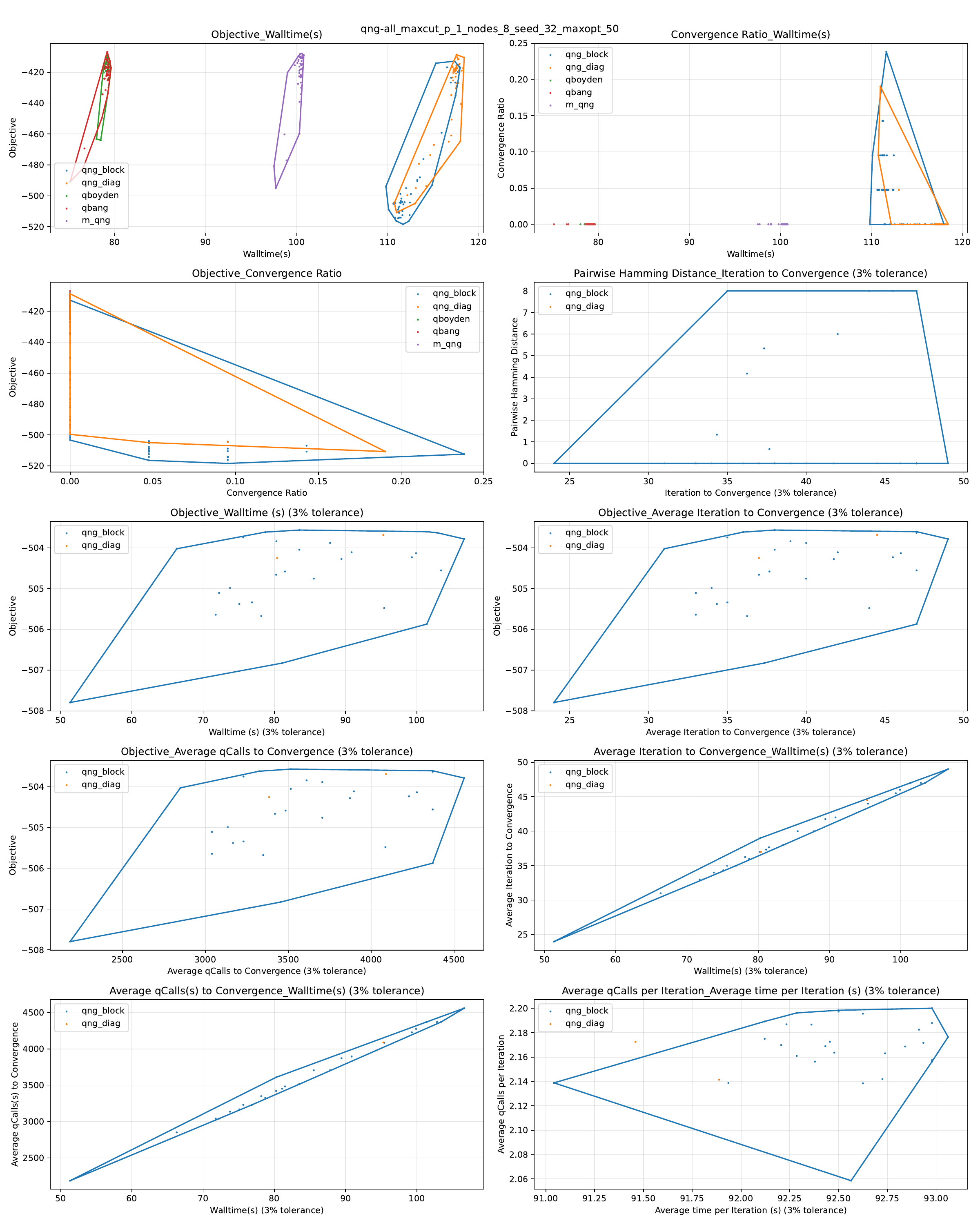}
    \caption{8 nodes problems, 50 iterations, qng-type noiseless, 70 Bayesian runs}
\end{figure}

\subsubsection{Stochastic methods}
\begin{figure}[H]
    \centering
    \includegraphics[scale=0.4]{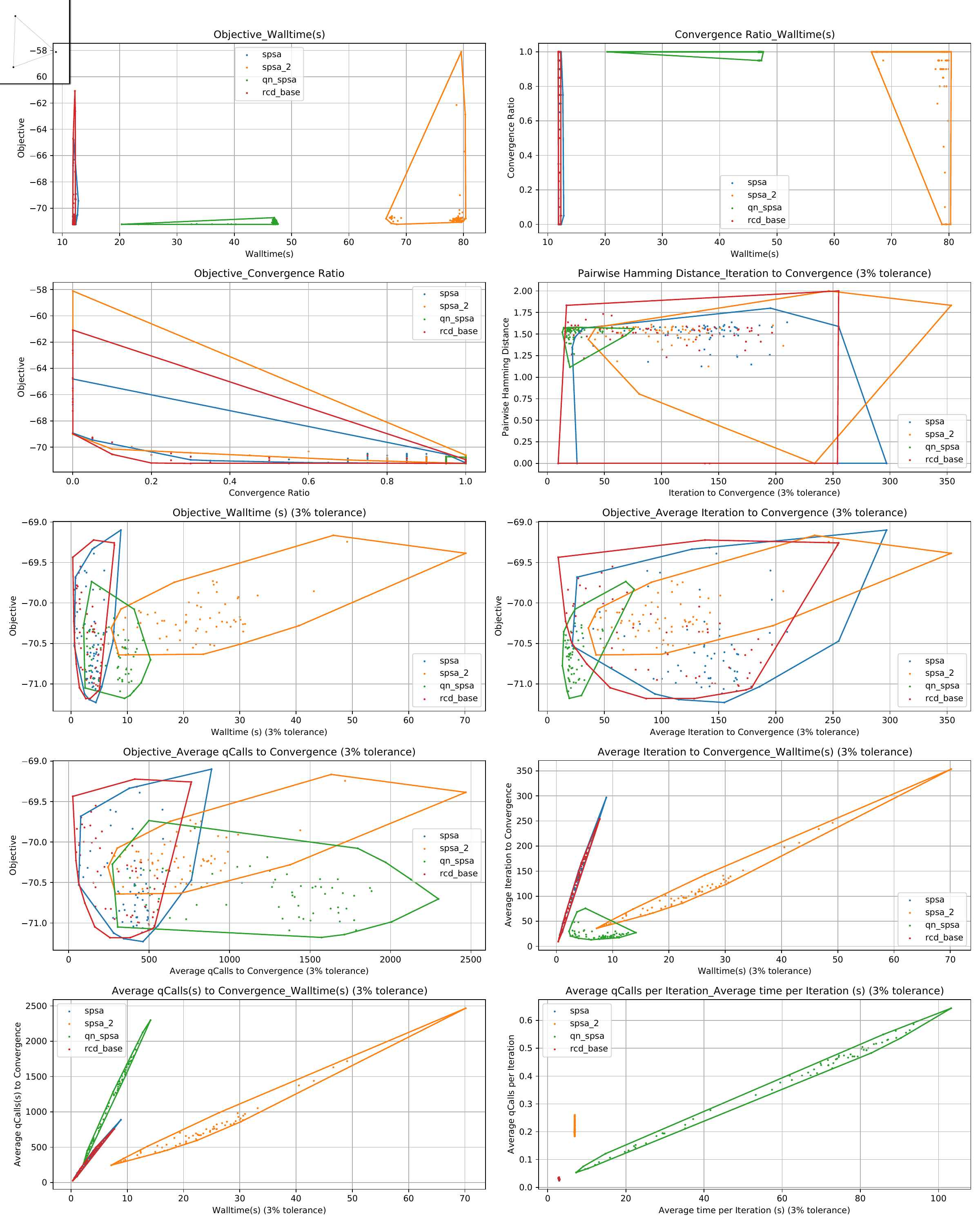}
    \caption{3 nodes problems, 400 iterations max, stochastic methods on shot-noise, 70 Bayesian sweeps}
\end{figure}

\begin{figure}[H]
    \centering
    \includegraphics[scale=0.4]{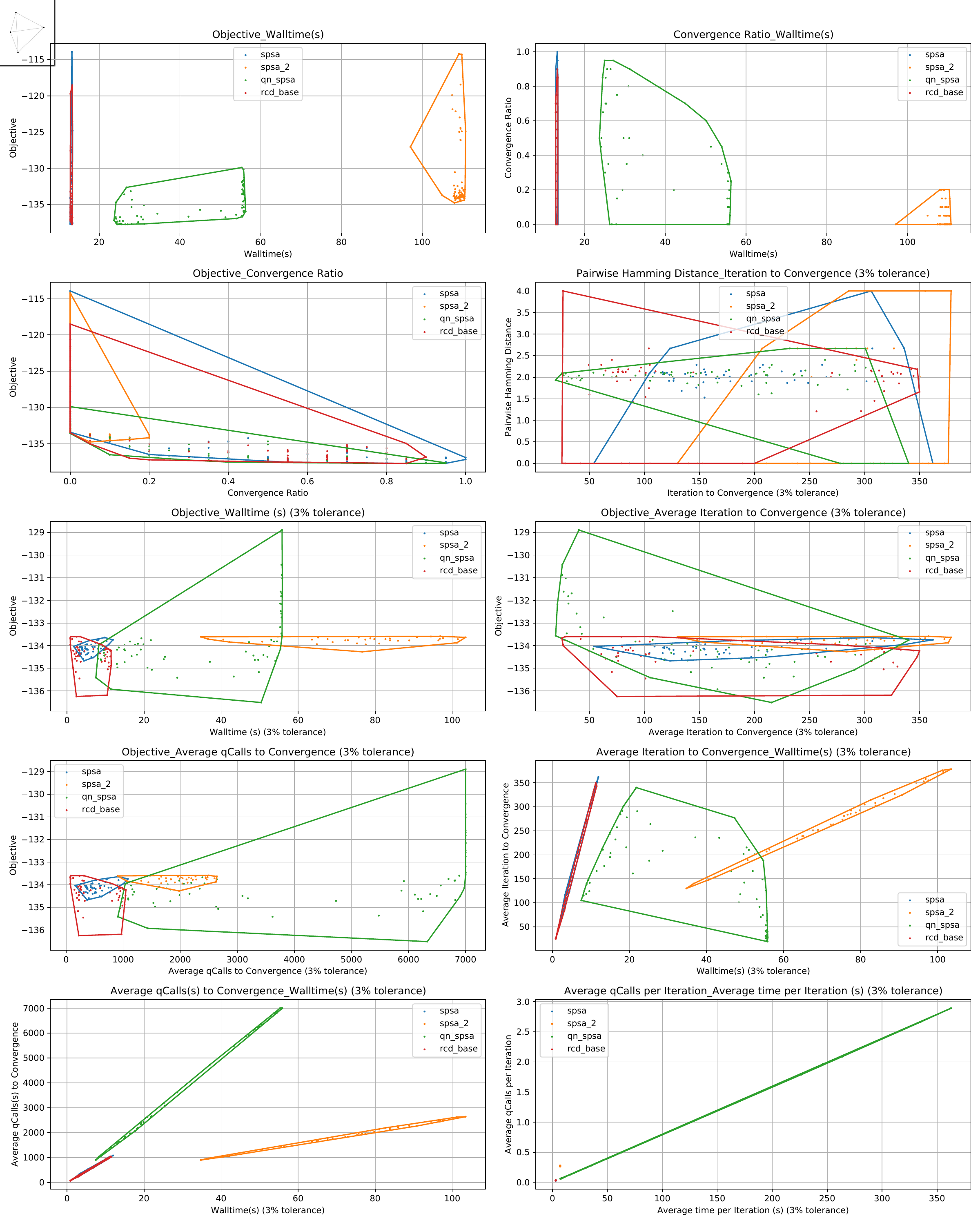}
    \caption{4 nodes problems, 70 iterations, second order methods noiseless, 70 Bayesian sweeps}
\end{figure}

\begin{figure}[H]
    \centering
    \includegraphics[scale=0.4]{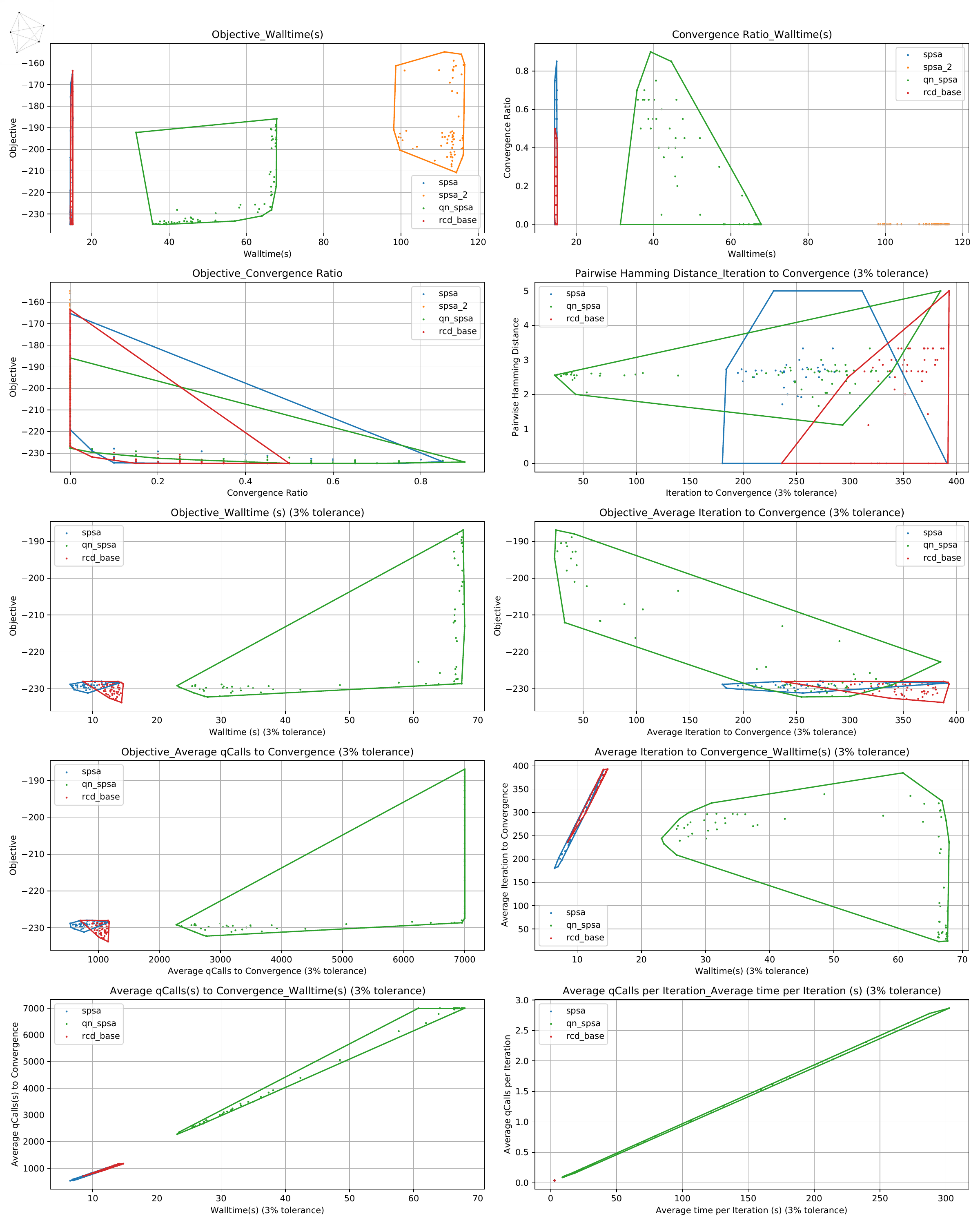}
    \caption{5 nodes problems, 70 iterations, second order methods noiseless, 70 Bayesian sweeps}
\end{figure}

\begin{figure}[H]
    \centering
    \includegraphics[scale=0.4]{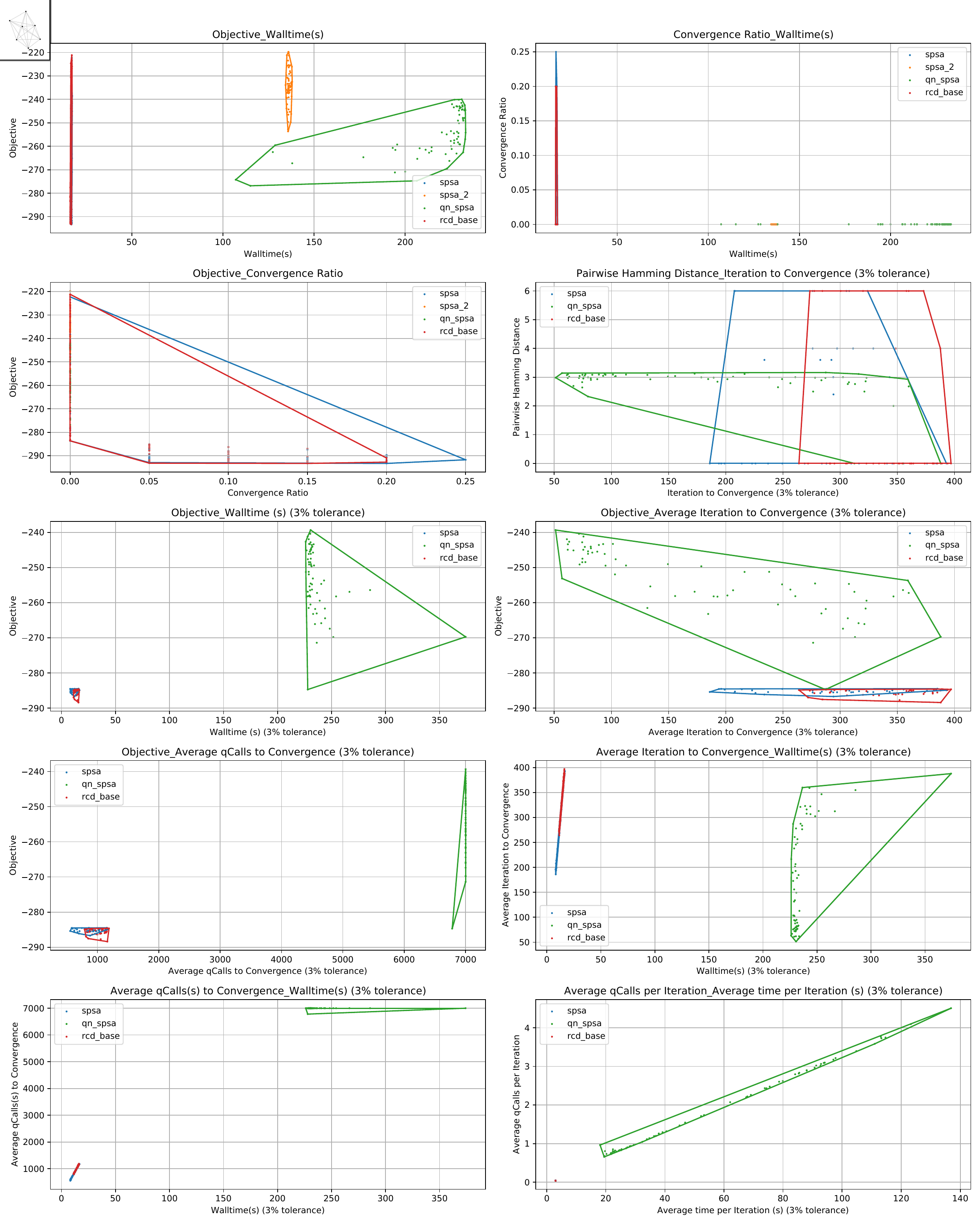}
    \caption{6 nodes problems, 70 iterations, second order methods noiseless, 70 Bayesian sweeps}
\end{figure}

\begin{figure}[H]
    \centering
    \includegraphics[scale=0.4]{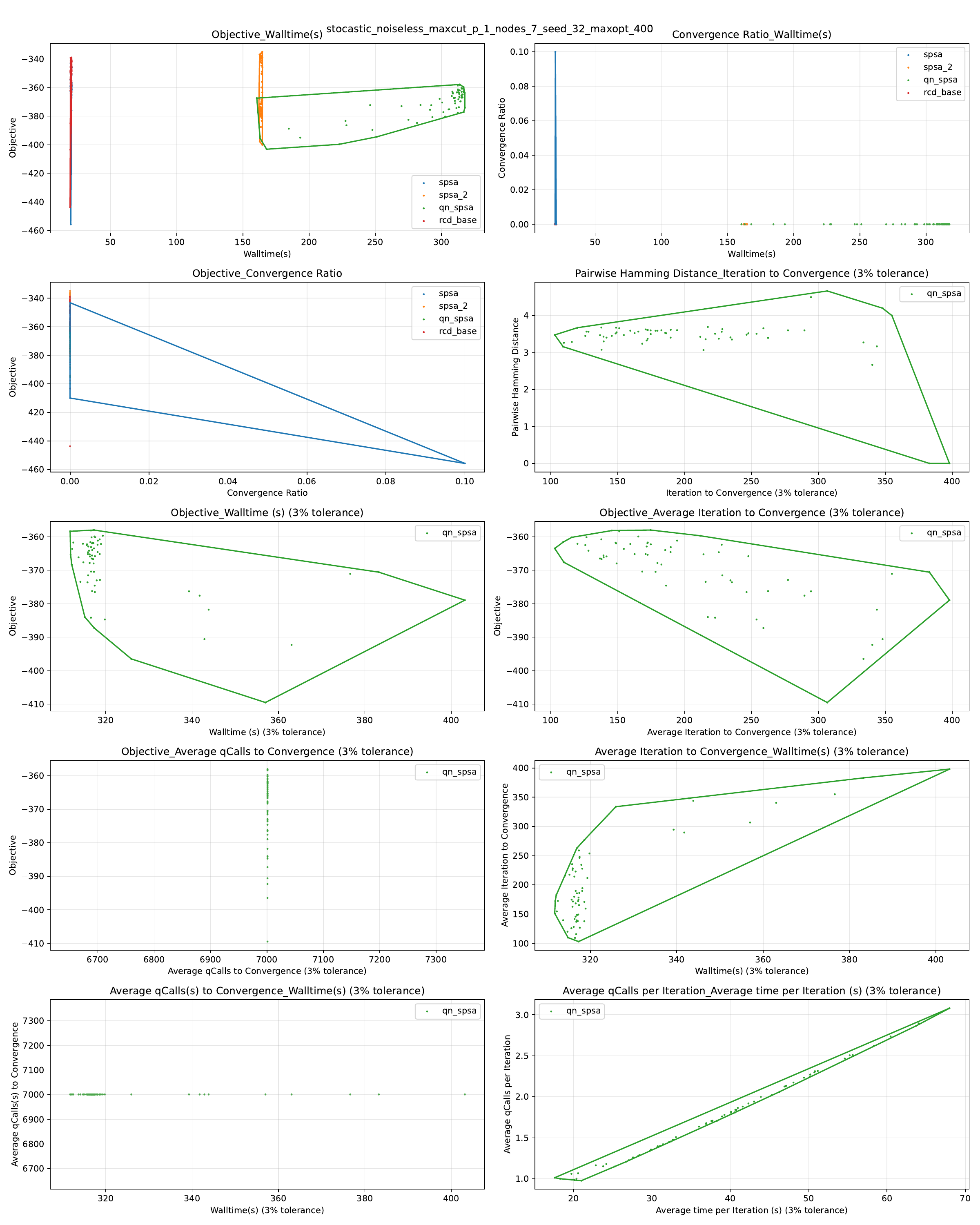}
    \caption{7 nodes problems, 70 iterations, second order methods noiseless, 70 Bayesian sweeps}
\end{figure}

\end{document}